\documentclass{article} 
\usepackage{iclr2027_conference,times}

\usepackage{geometry}
\usepackage{amsmath,amsfonts,bm}

\def\eqref#1{equation~\ref{#1}}

\def\1{\bm{1}}

\DeclareMathAlphabet{\mathsfit}{\encodingdefault}{\sfdefault}{m}{sl}
\SetMathAlphabet{\mathsfit}{bold}{\encodingdefault}{\sfdefault}{bx}{n}

\usepackage{hyperref}
\usepackage{url}
\usepackage{titletoc}
\usepackage{amsmath,amssymb,amsfonts,amsthm}
\usepackage{mathtools}
\usepackage{bm}
\usepackage{booktabs}
\usepackage{multirow}
\usepackage{graphicx}
\usepackage{ragged2e}
\usepackage{mathrsfs}
\usepackage{textcomp}
\usepackage{wrapfig}
\usepackage[table]{xcolor}
\usepackage{enumitem}
\usepackage{microtype}
\usepackage{algorithm}
\usepackage{algpseudocode}
\usepackage{adjustbox}
\usepackage{subcaption}
\usepackage[section]{placeins}
\usepackage{tabularx}
\definecolor{AlignedAccent}{HTML}{A84852}
\definecolor{AlignedFill}{HTML}{FCF4F4}
\definecolor{OrthogonalAccent}{HTML}{405FA8}
\definecolor{OrthogonalFill}{HTML}{F3F5FC}
\definecolor{SecondaryText}{HTML}{62666D}

\newtheorem{proposition}{Proposition}

\theoremstyle{definition}

\theoremstyle{remark}

\newcommand{\RR}{\mathbb{R}}

\newcommand{\norm}[1]{\left\lVert #1 \right\rVert}

\definecolor{mycyan}{RGB}{0,204,204}
\definecolor{mygreen}       {rgb} {0.10,0.50,0.30}
\definecolor{mylightgreen}  {RGB} {50,168,82}
\definecolor{lightred}      {rgb} {1.00,0.80,0.80}
\definecolor{lightblue}     {rgb} {0.80,0.80,1.00}
\definecolor{mediumred}     {rgb} {1.00,0.60,0.60}
\definecolor{mediumblue}    {rgb} {0.40,0.50,0.70}
\definecolor{pennblue}{cmyk}{1,0.65,0,0.30}
\definecolor{pennred}{cmyk}{0,1,0.65,0.34}
\definecolor{myred}{RGB}{199,107,102}
\definecolor{myindigo}{RGB}{75, 0, 130}

\usepackage{array}
\usepackage{tikz}
\usepackage[most]{tcolorbox}

\definecolor{ArxivCardGray}{HTML}{F2F2F2}
\definecolor{ArxivMuted}{HTML}{5F6368}
\definecolor{ArxivRule}{HTML}{D9D9D9}
\definecolor{ArxivLink}{HTML}{1769E0}

\newtcolorbox{arxivtitlecard}{
  enhanced,
  breakable=false,
  width=\textwidth,
  left skip=0mm,
  colback=ArxivCardGray,
  colframe=ArxivCardGray,
  boxrule=0pt,
  arc=15pt,
  outer arc=15pt,
  left=8mm,
  right=8mm,
  top=7mm,
  bottom=7mm,
  boxsep=0pt,
  before skip=2mm,
  after skip=7mm,
}

\newcommand{\arxivmetarow}[2]{%
  \noindent\textbf{#1}\hspace{0.45em}#2\par
}

\definecolor{HGAlignedBg}{HTML}{E8F4FF}
\definecolor{HGAlignedInk}{HTML}{174F75}
\definecolor{HGOrthBg}{HTML}{E8F5E9}
\definecolor{HGOrthInk}{HTML}{176444}
\definecolor{HGHeaderBg}{HTML}{EDF2F8}
\definecolor{HGMuted}{HTML}{555555}

\newcommand{\HGBlueArrow}{%
  \textcolor{HGAlignedInk}{$\rightarrow$}%
}
\newcommand{\HGGreenArrow}{%
  \textcolor{HGOrthInk}{$\rightarrow$}%
}

\newcommand{\HGMetric}[3]{%
  \tikz[baseline=(metric.base)]{%
    \node[
      fill=#1,
      rounded corners=2pt,
      inner xsep=3pt,
      inner ysep=2pt,
      outer sep=0pt,
      text width=\dimexpr\linewidth-8pt\relax,
      align=center
    ] (metric) {%
      \color{#2}\bfseries
      \fontsize{6}{7.6}\selectfont #3%
    };%
  }%
}

\newcommand{\HGChatIcon}{%
  \tikz[baseline=-.2ex,scale=.09]{%
    \fill[HGAlignedInk,rounded corners=1.5pt]
      (0,0) rectangle (2.8,2);
    \fill[HGAlignedInk]
      (.5,0)--(.35,-.6)--(1.15,0);
    \foreach \x in {.6,1.4,2.2}
      {\fill[white] (\x,1) circle (.12);}
  }%
}

\newcommand{\HGShieldIcon}{%
  \tikz[baseline=-.2ex,scale=.09,yscale=.6]{%
    \draw[
      HGAlignedInk,
      fill=HGAlignedBg,
      line width=0.5pt
    ]
      (0,1.5)--(1.3,2)--(2.6,1.5)--(2.6,0)
      .. controls (2.5,-1.3) and (1.7,-2.2) ..
      (1.3,-2.5)
      .. controls (.9,-2.2) and (.1,-1.3) ..
      (0,0)--cycle;
  }%
}

\newcommand{\HGGlobeIcon}{%
  \tikz[baseline=-.2ex,scale=.09]{%
    \draw[HGOrthInk,line width=0.5pt]
      (0,0) circle (1.35);
    \draw[HGOrthInk,line width=0.5pt]
      (-1.35,0)--(1.35,0)
      (0,-1.35)--(0,1.35);
    \draw[HGOrthInk,line width=0.5pt]
      (-.75,-1.1)
      .. controls (.1,-.3) and (.1,.3) ..
      (-.75,1.1)
      (.75,-1.1)
      .. controls (-.1,-.3) and (-.1,.3) ..
      (.75,1.1);
  }%
}

\newcommand{\HGFileIcon}{%
  \tikz[baseline=-.2ex,scale=.09,yscale=.77]{%
    \draw[HGOrthInk,line width=0.5pt]
      (0,-1.5)--(0,2)--(1.7,2)--(2.5,1.2)
      --(2.5,-1.5)--cycle;
    \draw[HGOrthInk,line width=0.5pt]
      (1.7,2)--(1.7,1.2)--(2.5,1.2)
      (.4,.5)--(2,.5)
      (.4,0)--(2,0)
      (.4,-.5)--(2,-.5);
  }%
}

\newcommand{\HGBench}[3]{%
  \makebox[\linewidth][c]{%
    #1\hspace{.4em}%
    \begin{minipage}[c]{\dimexpr\linewidth-1.8em\relax}
      \raggedright
      \textbf{#2}\\[-1pt]
      {\normalfont\color{HGMuted}%
       \fontsize{6.5}{7.5}\selectfont
       \citep{#3}}
    \end{minipage}%
  }%
}

\definecolor{StateTeal}{HTML}{1AA3A3}
\definecolor{SteerRed}{HTML}{C4453F}

\newcommand{\statevar}[1]{\textcolor{StateTeal}{#1}}
\newcommand{\steervar}[1]{\textcolor{SteerRed}{#1}}

\newcommand{\hk}{\statevar{h_k}}
\newcommand{\hrefk}{\statevar{h_k^{\mathrm{ref}}}}

\newcommand{\xk}{\statevar{x_k}}
\newcommand{\xkp}{\statevar{x_{k+1}}}

\newcommand{\Ak}{\statevar{A_k}}
\newcommand{\phik}{\statevar{\phi_k}}

\newcommand{\xik}{\statevar{\xi_k}}
\newcommand{\Dk}{\statevar{D_k}}
\newcommand{\wk}{\statevar{w_k}}

\newcommand{\uk}{\steervar{u_k}}
\newcommand{\urefk}{\steervar{u_k^{\mathrm{ref}}}}
\newcommand{\duk}{\steervar{\delta u_k}}

\newcommand{\Bk}{\steervar{B_k}}
\newcommand{\Kk}{\steervar{K_k}}

\newcommand{\Ctar}{\steervar{C_k^{\mathrm{tar}}}}
\newcommand{\Coff}{\steervar{C_k^{\mathrm{off}}}}

\newcommand{\zk}{\steervar{z_k}}

\newcommand{\gammastar}{\steervar{\gamma^\star}}
\newcommand{\Srob}{\steervar{S_{\mathrm{rob}}}}

\newif\ifsemanticcolors
\semanticcolorstrue       

\ifsemanticcolors
    \renewcommand{\statevar}[1]{\textcolor{StateTeal}{#1}}
    \renewcommand{\steervar}[1]{\textcolor{SteerRed}{#1}}
\else
    \renewcommand{\statevar}[1]{#1}
    \renewcommand{\steervar}[1]{#1}
\fi

\semanticcolorstrue

\title{Mind Control of LLMs via Closed-Loop Stimulation: Out-of-Distribution Robustness}

\author{Kasra Fallah\thanks{Equal contribution.} \\
Columbia University \\
\texttt{kasra.fallah@columbia.edu}
\And
Rudramani Singha\footnotemark[1] \\
UT Southwestern\\
\texttt{rgs2151@columbia.edu}
\And
Hannah Ghanei \\
University of Surrey \\
\texttt{hg00859@surrey.ac.uk}
\And
Paul Sajda\\
Columbia University \\
\texttt{ps629@columbia.edu}
\And
James Anderson \\
Columbia University \\
\texttt{ja3451@columbia.edu}
\And
Erfan Zabeh \\
Columbia Zuckerman Institute \\
\texttt{ez2290@columbia.edu}
}

\iclrfinalcopy 
\begin{document}
\pagestyle{plain}

\begin{arxivtitlecard}


{\fontsize{18}{25.5}\selectfont\bfseries
Mind Control of LLMs via Closed-Loop Stimulation:\\ Out-of-Distribution Robustness\par}

\vspace{3.5mm}

{\fontsize{10}{13}\selectfont\bfseries
Kasra Fallah\textsuperscript{1,*},
Rudramani Singha\textsuperscript{2,*},
Hannah Ghanei\textsuperscript{3},
Paul Sajda\textsuperscript{1,4},
James Anderson\textsuperscript{1},
Erfan Zabeh\textsuperscript{2,4,5,$\dagger$}\par}

\vspace{2mm}

{\fontsize{8.6}{10.6}\selectfont\color{ArxivMuted}
\textsuperscript{1}Department of Electrical Engineering, Columbia University, New York, NY, USA\\
\textsuperscript{2}Peter O'Donnell Jr. Brain Institute, UT Southwestern Medical Center, Dallas, TX, USA\\
\textsuperscript{3}School of Computer Science and Electronic Engineering, University of Surrey, Guildford, UK\\
\textsuperscript{4}Department of Biomedical Engineering, Columbia University, New York, NY, USA\\
\textsuperscript{5}Center for Theoretical Neuroscience, Zuckerman Institute, Columbia University, New York, NY, USA
\par}

\vspace{1.2mm}

{\fontsize{8.7}{10.5}\selectfont\color{ArxivMuted}
\textsuperscript{*}Equal contribution
\par
\textsuperscript{$\dagger$}Correspondence:
ez2290@columbia.edu\par}

\vspace{4mm}
{\fontsize{9.4}{12.3}\selectfont\justifying
Brains and large language models need not be considered equivalent systems to reveal an interpretable commonality: both exhibit high-dimensional, nonlinear dynamics whose behavior can only be modeled imperfectly. This creates a shared challenge whenever we seek to intervene in their dynamics. A controller may either rely on a model identified around a particular operating regime and risk failure as the system moves away from that regime, or explicitly account for uncertainty and a changing internal state. Neurostimulation has confronted an analogous problem for decades. The limitations of fixed, open-loop stimulation have motivated closed-loop and state-dependent approaches that condition intervention on the ongoing state of the neural system. Inspired by this principle, we formulate LLM steering as a feedback-control problem in which the network's internal state, rather than a fixed intervention alone, determines how the system is perturbed. We further account for the inevitable mismatch between tractable dynamical models and the nonlinear transformer they approximate by adopting a robust control approach. As one principled realization of this framework, we develop an $H_\infty$-based steering controller that explicitly limits sensitivity to disturbances and linearization mismatches. Across multiple steering benchmarks, the resulting controller improves performance over existing steering baselines, with particularly pronounced gains under out-of-distribution conditions and varying length complexities---regimes in which the assumptions underlying locally identified dynamics are increasingly challenged. Beyond controller performance, we introduce robust steerability, $S_{\mathrm{rob}}$, a pre-deployment measure of tolerance to dynamics mismatch; across model--task pairs, it exhibits task-specific structure and varies systematically with model scale. Beyond robust LLM steering, our measure asks not merely whether control is possible, but how much uncertainty it can survive---a principle that may span artificial and biological neural systems.\par}

\vspace{3mm}
{\fontsize{8.7}{10.5}\selectfont
\arxivmetarow{Keywords:}{closed-loop steering, robust control, distribution shift, activation steering, $H_\infty$ control}
}
\vspace{-0.8cm}
\begin{minipage}[r]{0.48\linewidth}
    \RaggedRight
\end{minipage}
\hfill
\begin{minipage}[c]{0.48\linewidth}
    \raggedleft
    \includegraphics[height=9mm]{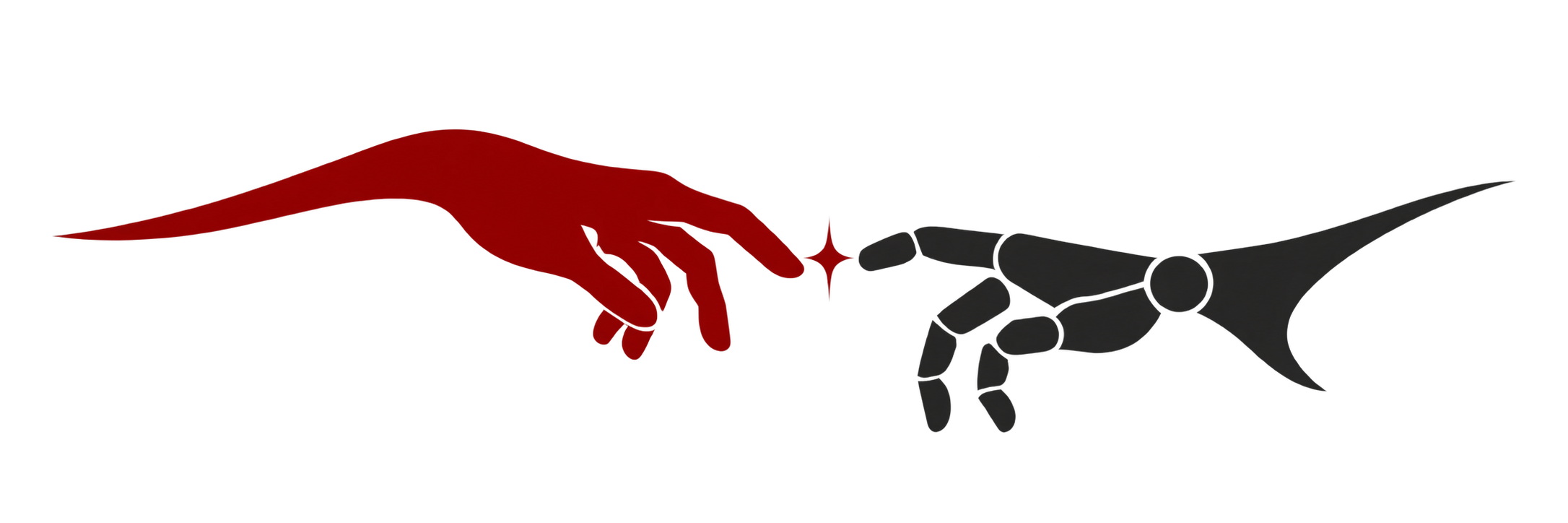}
\end{minipage}
\end{arxivtitlecard}

\begin{figure}[h]
    \centering
    \includegraphics[width=0.95\linewidth]{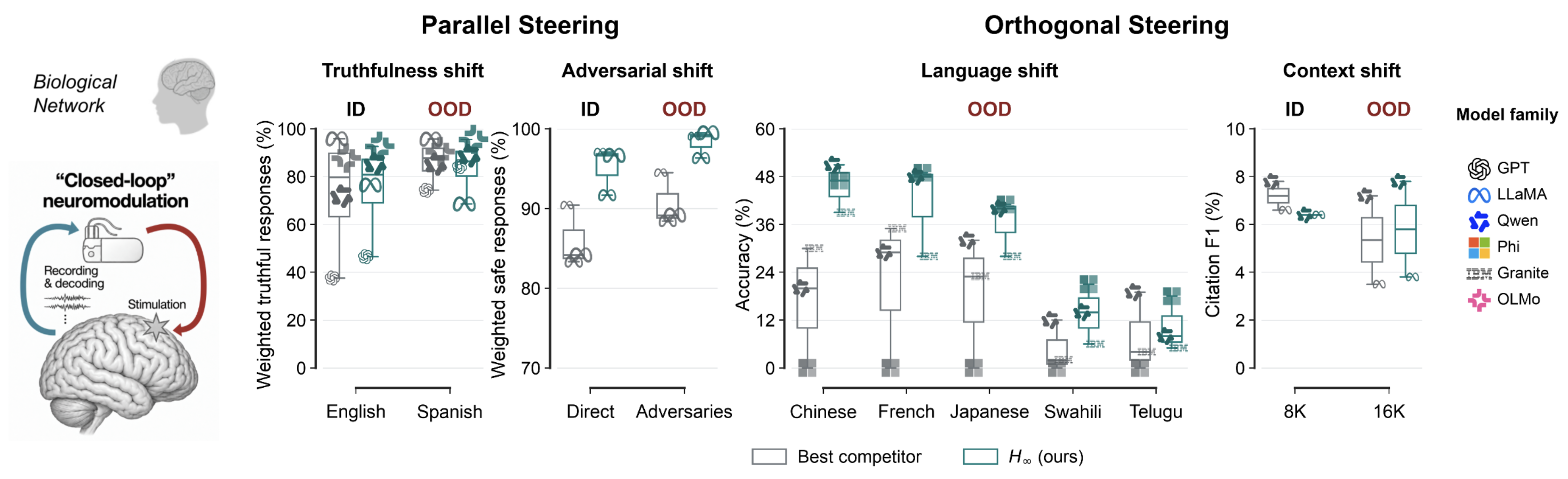}
    \caption{
        {Borrowing from closed-loop neurostimulation, we steer LLMs with robust $H_\infty$ feedback and derive a score characterizing steering reliability under distribution shift.}
    }
    \label{fig:graphical_abstract}
\end{figure}


\section{Introduction}
\label{sec:intro}

Activation steering provides a partial control surface for large language
models: residual-stream directions added or removed at inference can increase
truthfulness, sustain refusal, or change language without weight updates
\citep{li2023iti,zou2023representation,turner2024activation,rimsky2024caa,arditi2024refusal}.
Yet these interventions---a direction, layer, and strength---are calibrated
under one prompt distribution and deployed on unseen inputs. Calibration
therefore measures success under construction conditions, not reliability in
use. A vector that improves truthfulness on English calibration questions may
weaken in Spanish or as the context length increases, while its calibration
metrics give no warning
\citep{tan2024analysing,hsu2026contextual,maraia2026can}. \textit{A successful
perturbation is not the same as reliable control.}

\begin{wrapfigure}[26]{r}{0.40\textwidth}
    \setlength{\columnsep}{8pt}
    \setlength{\intextsep}{1pt}
    \vspace{-10pt}

    \centering
    \includegraphics[width=0.9\linewidth]
    {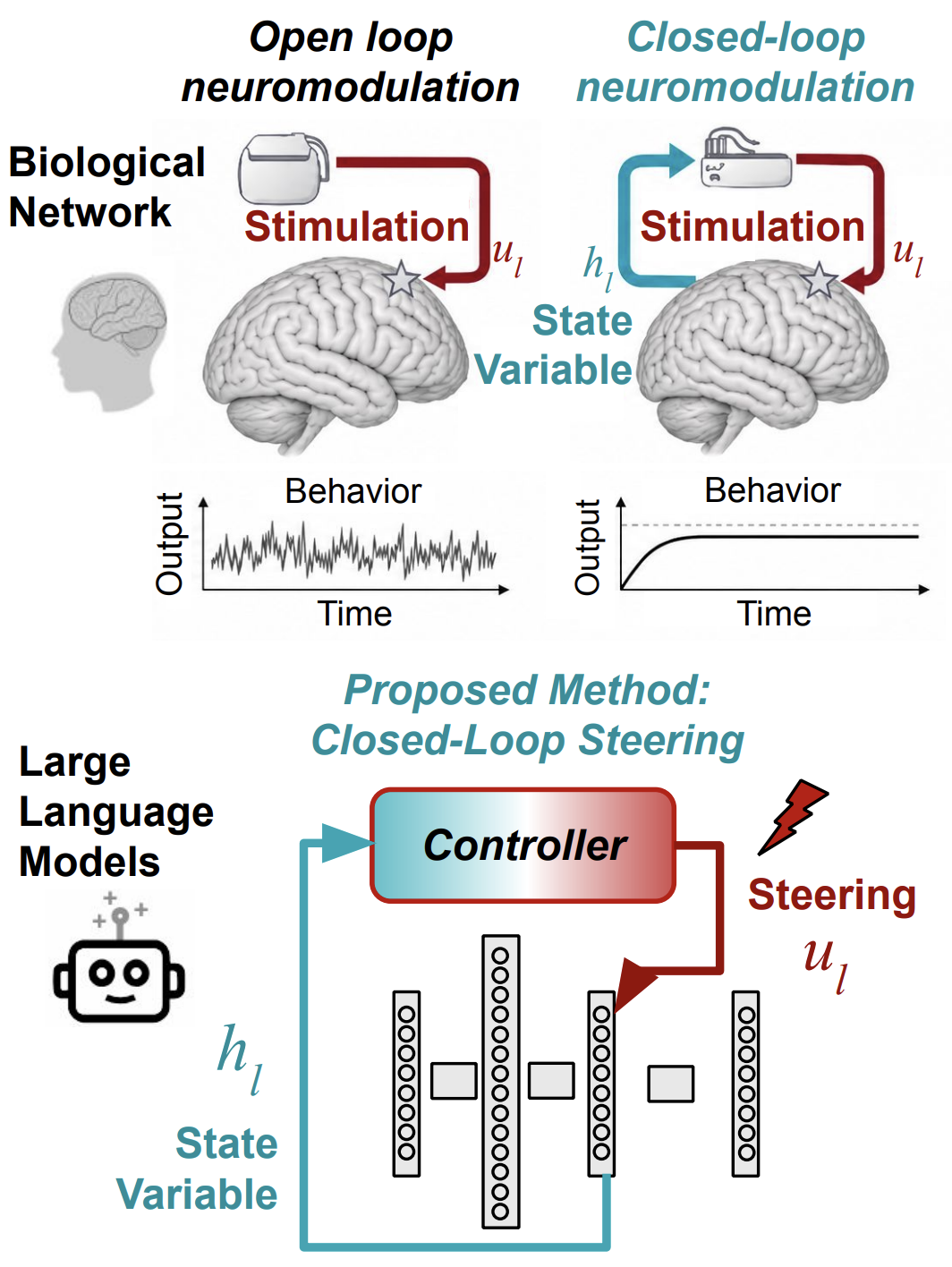}

    \vspace{-0.3cm}

    \caption{\textbf{Method Schematic.}
    Closed-loop stimulation enables controlled modulation of the system's
    behavioral dynamics.}
    \label{fig:brain_control}

    \vspace{-10pt}
\end{wrapfigure}

Neuroscience has confronted this distinction directly, and its response has two
parts. First, the effect of stimulation depends on the state in which it is
delivered: across invasive and non-invasive paradigms, state alters the
magnitude, direction and behavioral consequence of otherwise identical
stimulation
\citep{faller2022daily,silvanto2008state,bergmann2018brain,bradley2022state}.
Response variability is not noise around a fixed input--output map but a
signature of the dynamics through which perturbations propagate, and this
motivated a shift from open-loop stimulation to state-aware closed-loop
intervention
\citep{zrenner2016closedloop,zrenner2024closedloop,hebb2014creating,topalovic2023wearable,davila2022control,kao2015singletrial,natarajan2026multimodalphysiologicaldecodingreveals,shanechi2019bmi}.
Second, reliable control of an imperfectly known system requires designing
against that imperfection. Optimal feedback-control accounts of movement
explain behavioral robustness as correction according to task consequence
rather than tracking of a fixed trajectory
\citep{todorov2002optimal,scott2004optimal}, and human responses to
unpredictable disturbances have been described by robust-control formulations
related to $\mathcal{H}_\infty$ \citep{crevecoeur2019robust}: intervention
reliability depends on how uncertainty is amplified from perturbation to
behavior. Language models and biological circuits are not equivalent systems,
but both are high-dimensional, nonlinear and imperfectly modeled, and
intervention in both inherits the same reliability problem.

In activation steering, this problem takes a concrete form. Recent work has cast
transformer depth as a finite control horizon, hidden representations as
states, and activation interventions as control inputs whose effects propagate
through subsequent layers \citep{nguyen2025pid,skifstad2026activationlqr}.
These methods address how to steer a representation under an identified nominal
model, but not whether it remains steerable when those dynamics change. At
deployment, shifts in prompt, context, or activation trajectory alter the
network's local Jacobians and therefore how the same intervention propagates.
Distribution shift thus changes not only the state being controlled but also
the dynamics of control itself. Consequently, two representations with similar
decodability and nominal steering performance can differ sharply in reliability
under shift. We therefore ask:
\vspace{-0.35cm}
\begin{quote}
\textit{Can a language model's internal dynamics tell us, before deployment,
whether a representation will remain reliably steerable as those dynamics
change?}
\end{quote}
\vspace{-0.35cm}
We study this question through robust control. Around a nominal steered
trajectory, we model local representation dynamics as
$\xkp=\Ak\xk+\Bk\uk+\Dk\wk,$ where $\Ak$ represents the identified layer
dynamics, $\Bk\uk$ the admissible activation intervention, and $\Dk\wk$ the
deviation from the nominal model. Crucially, $\Dk$ is estimated from empirical
representation-dynamics residuals, capturing the directions in which mismatch
actually occurs rather than imposing isotropic uncertainty. We define robust
steerability as $\Srob=1/\gamma^\star$, where
$\gamma^\star=\inf_K\|T_{w\to z}(K)\|_\infty$ is the minimum achievable
closed-loop amplification from disturbance to steering-relevant output. We use robust control as a diagnostic of intervention
reliability rather than as a new controller-synthesis method, and ask what
$\Srob$, computed from calibration activations alone, reveals about which
model--behavior pairs admit reliable intervention. Beyond the benchmark gains,
synthesis returns a measure for whether or not the controller is successful:
$\Srob$ measures how much deviation from the identified dynamics an
intervention can absorb before it stops producing the intended behavior, and it
varies systematically across models and behaviors (Sec.~\ref{sec:S_rob&Dconclusion}). 
The question is not whether an intervention works, but how careful it has to
be.

Our formulation has two ingredients; our pre-deployment analysis (Sec.~\ref{sec:prblm_formulation}) 
earns each separately. \textit{Robustness}: in both intracranial recordings and
transformer hidden states, local linearization holds near the identification
regime and leaves a residual that grows as the trajectory departs from
it---structured and state-dependent, not incidental noise. \textit{Feedback}:
in the neural recordings, where open- and closed-loop stimulation can be
compared \citep{basu2023closed,bush2006multi}, conditioning on the evolving
state lowers output-level mismatch and narrows the out-of-distribution gap; no
language model affords that comparison, since there closed-loop efficacy is the
claim under test. The intervention must therefore track the state, and the
model that shapes it cannot be trusted beyond its fitting regime. An
$\mathcal{H}_\infty$ feedback law is exactly that pair
(Sec.~\ref{sec:prblm_formulation}). 

\begin{wrapfigure}[47]{l}{0.65\textwidth}
    \vspace{-0.3cm}
    \centering
    \includegraphics[width=0.64\textwidth]{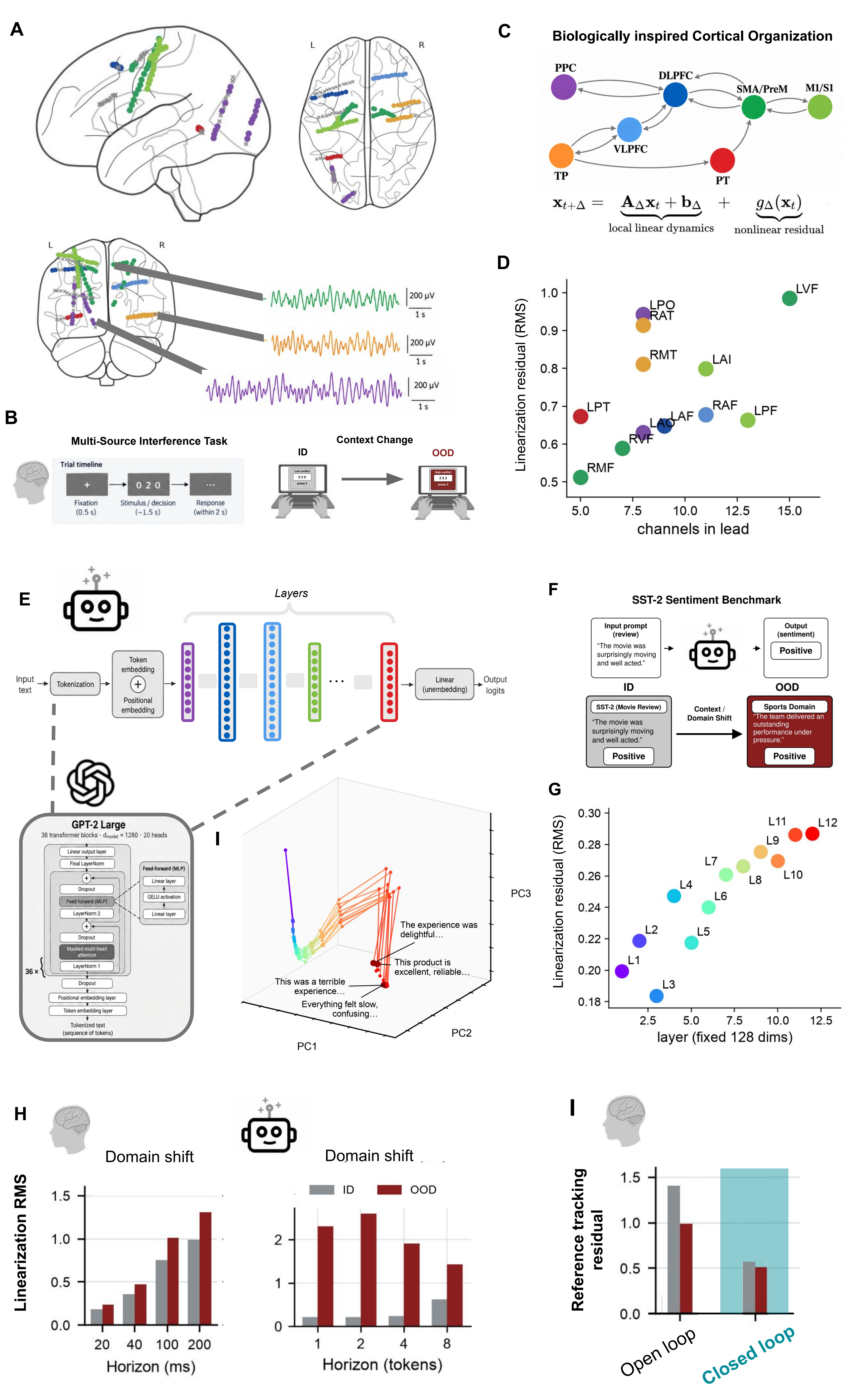}
    \vspace{-0.3cm}
\end{wrapfigure}

\refstepcounter{figure}
\label{fig:human_linearization}

\noindent
{\small
Figure~\thefigure:\;
\textbf{Linearization residuals under distribution shift in biological and artificial neural systems motivate closed-loop control.}
\textbf{(A--D)} Human intracranial recordings during MSIT show that neural dynamics admit useful short-horizon linear prediction, while residual model mismatch increases when a frozen model is evaluated under task-context shift.
\textbf{(E--G)} We apply the same local-dynamics abstraction to GPT-2 hidden states under an analogous context shift.
\textbf{(H)} In both systems, linearization residuals increase from in-distribution (ID) to out-of-distribution (OOD) conditions, indicating greater mismatch between locally identified and realized dynamics under shift.
\text bf{(I)} In the human data, closed-loop stimulation reduces \emph{output-level mismatch} and narrows the ID--OOD gap relative to open-loop stimulation. Together, these results motivate treating model mismatch as a disturbance and LLM steering as a state-dependent, robust closed-loop control problem.
\par}

The exchange runs both ways. A language model is not a model of \textit{the
brain}, but it is a large nonlinear network whose states can be measured,
perturbed and evaluated at will---a \textbf{model organism} for intervention in complex
neural systems. Sec.~\ref{sec:S_rob&Dconclusion} 
returns the measure to that setting, where it could separate stimulation
targets that tolerate coarse models from those that need precise closed-loop
control. What follows begins with those dynamics: what a local model captures,
and what it leaves behind.

\section{Problem Formulation}
\label{sec:prblm_formulation}
\subsection{Neuronal and Transformer Based Dynamics}
Neural population activity can be viewed as the evolution of a high-dimensional nonlinear dynamical system. Let $\statevar{x}_t$ denote the population state and $\statevar{c}_t$ the task and physiological context; over an interval $\Delta$, we write
\vspace{-0.15cm}
$$
\statevar{x}_{t+\Delta}
=
\mathcal{F}_{\Delta}(\statevar{x}_t,\statevar{c}_t)
+
\statevar{\epsilon}_t,
\vspace{-0.1cm}
$$
where $\mathcal{F}_{\Delta}$ captures recurrent nonlinear interactions and $\statevar{\epsilon}_t$ collects unresolved dynamics and measurement variability. Although the underlying dynamics are nonlinear, population trajectories can remain locally predictable over short horizons, while both the dynamics and the effect of intervention depend on the operating state and context \citep{vyas2020computation}. This motivates a local dynamical description rather than a globally fixed input--output model.

We adopt the same dynamical viewpoint for transformer based architectures, with network depth replacing physical time. Given a token sequence $q=(q_1,\ldots,q_T)$, a decoder-only transformer initializes the residual stream as
$\statevar{H_1}=\operatorname{Embed}(q)\in\RR^{T\times d}$ and propagates it through self-attention and an MLP \citep{vaswani2017attention,xiong2020layernorm}:

$$
\begin{aligned}
\statevar{H_k^{\mathrm{attn}}}
&=
\statevar{H_k}
+
\operatorname{Attn}_k
\left(
\operatorname{Norm}_k^{\mathrm{attn}}(\statevar{H_k})
\right),\\
\statevar{H_{k+1}}
&=
\statevar{H_k^{\mathrm{attn}}}
+
\operatorname{MLP}_k
\left(
\operatorname{Norm}_k^{\mathrm{mlp}}
(\statevar{H_k^{\mathrm{attn}}})
\right)
=: \statevar{\Phi_k}(\statevar{H_k}).
\end{aligned}
$$
Activation steering perturbs intermediate representations without modifying model weights
\citep{li2023iti,zou2023representation,turner2024activation,
rimsky2024caa,arditi2024refusal}. Following recent control-theoretic formulations
\citep{nguyen2025pid,skifstad2026activationlqr}, we control the final-token state
$\hk:=\statevar{H_k}[T,:]$. Under causal attention, the remaining token context can be absorbed into a layer-dependent nonlinear map $\phik$, giving the post-block intervention model
$$
\statevar{h_{k+1}}
=
\phik(\hk)+\Bk\uk,
\qquad
k=1,\ldots,L.
$$
Here, $\uk$ denotes the intervention coordinates and $\Bk$ their embedding into the residual stream. We interpret $k$ as a finite time axis \citep{haber2017stable,chen2018neuralode} and reapply the depth-wise controller as the autoregressive context evolves. We assume post-block injection in the main text; Appendix~\ref{app:intervention-placement} considers alternative intervention locations.

\vspace{-0.3cm}
\subsection{Local Dynamics and Structured Model Mismatch}
\vspace{-0.25cm}

Let $\{(\hrefk,\urefk)\}_{k=1}^{L}$ denote a reference trajectory realizing the desired behavior, with deviations
$\xk=\hk-\hrefk$ and $\uk=\uk-\urefk$. Linearizing the controlled dynamics about this trajectory gives
$\Ak:=\left.\partial\phik(\statevar h)/\partial\statevar h\right|_{\hrefk}$.
Rather than treating the approximation error as negligible or isotropic, we represent the remaining mismatch through a structured disturbance channel:
\begin{equation}
\tag{REP-Lin}
\label{eq:lin-llm}
\xkp=\Ak\xk+\Bk\uk+\Dk\wk.
\end{equation}
Here, $\Dk\wk$ captures deviations between the locally identified dynamics and the realized trajectory; $\Dk$ is estimated from calibration residuals as described in Sec.~\ref{sec:method}.In addition to the dynamics~\eqref{eq:lin-llm} we introduce a regulated output \(z_k\) that captures target-tracking error, intervention effort, and undesired changes to off-target representations. Specifically, we define
\begin{equation}\label{eq:out}
z_k =
\begin{bmatrix}
(Q_k^{1/2} C_k^{\mathrm{tar}} x_k)^\top &
(R_k^{1/2} u_k)^\top &
(S_k^{1/2} C_k^{\mathrm{off}} x_k)^\top
\end{bmatrix}^{\top}.
\end{equation}

Figure~\ref{fig:human_linearization} motivates this modeling choice empirically. In human intracranial recordings, short-horizon neural dynamics admit useful local prediction, yet the residual persists and increases when a frozen model is evaluated under changed task or stimulation context. Applying the same analysis to transformer hidden states yields the same qualitative pattern: local linearization is most accurate near the identification regime, while mismatch grows as the deployment trajectory moves away from it. Thus, in both systems, the residual is structured and state dependent rather than an incidental perturbation around a fixed model.

The stimulation analysis in Fig.~\ref{fig:human_linearization}I provides the complementary control observation: conditioning the intervention on the evolving neural state reduces output-level mismatch and narrows the ID--OOD gap relative to a fixed open-loop intervention. Together, these observations motivate the two ingredients of our LLM formulation: \emph{feedback}, because intervention effects depend on the current state, and \emph{robustness}, because the local dynamics used to design that feedback are necessarily imperfect. We therefore design the steering controller against the amplification of the empirically observed mismatch $\Dk\wk$ through the network.
\begin{figure}[t]
    \centering
    \includegraphics[width=\textwidth]{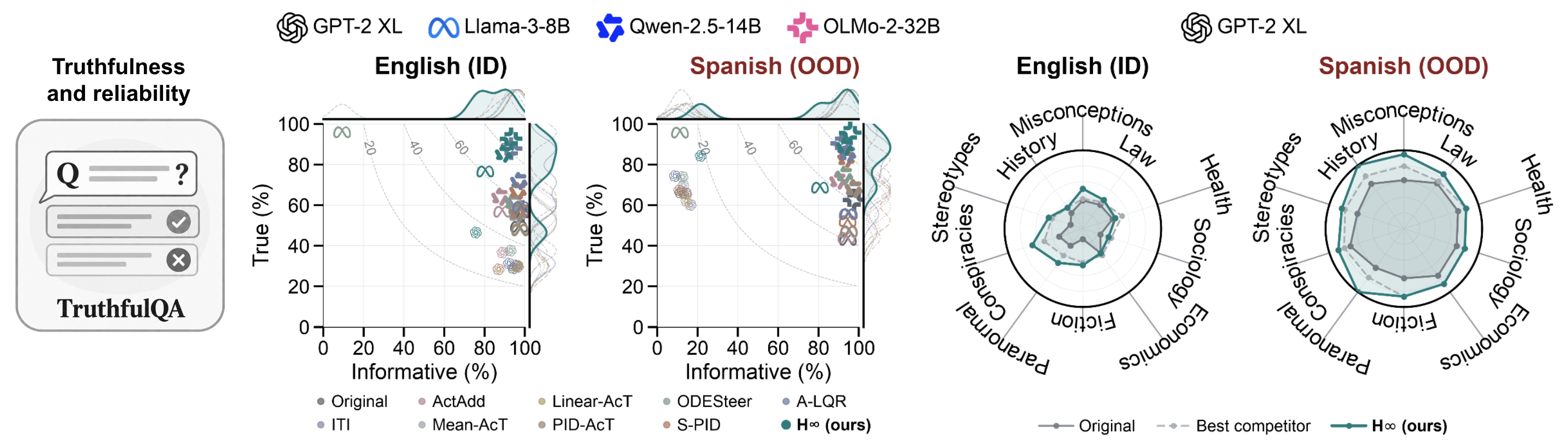}
    \vspace{-0.5cm}
    \caption{
    \textbf{TruthfulQA under prompt-language shift.}
    Left: truthfulness--informativeness frontiers for matched English
    (\textbf{ID}) and Spanish-input, English-output transfer (\textbf{OOD});
    dashed curves denote equal T$\times$I and marginal densities summarize the
    corresponding distributions. Right: category-level GPT-2 XL performance
    for the original model, strongest competitor, and $\mathrm{H}_{\infty}$.
    }
    \label{fig:truthfulqa-transfer}
\end{figure}
\vspace{-0.45cm}
\subsection{Steering Objective}
\vspace{-0.25cm}
Since the reference trajectory realizes the target behavior, we define the regulated output by
\vspace{-0.1cm}
$$
\norm{\zk}_2^2
=
\norm{\Ctar\xk}_{\steervar{Q_k}}^2
+
\norm{\uk}_{\steervar{R_k}}^2
+
\norm{\Coff\xk}_{\steervar{S_k}}^2,
\vspace{-0.2cm}
$$

where $\steervar{Q_k},\steervar{S_k}\succeq0$ and $\steervar{R_k}\succ0$ penalize target-tracking error, intervention effort, and off-target representation change, respectively; $\steervar{S_k}=0$ removes the preservation term.

\vspace{-0.45cm}
\subsection{Robust Steerability}
\vspace{-0.25cm}
Under layer-wise feedback $\uk=-\Kk\xk$ and $\statevar{x}_1=0$, we define the minimum achievable disturbance-to-performance amplification
\vspace{-0.4cm}
$$
\gammastar
=
\inf_{\steervar{K}}
\sup_{\statevar{w}\neq0}
\frac{
\left(\sum_{k=1}^{L}\norm{\zk}_2^2\right)^{1/2}
}{
\left(\sum_{k=1}^{L}\norm{\wk}_2^2\right)^{1/2}
},
\qquad
\Srob=\frac{1}{\gammastar}.
$$
Thus, $\gammastar$ is the smallest closed-loop amplification of model mismatch, and larger $\Srob$ indicates greater robust steerability. Because $\zk$ includes intervention cost, this score cannot be improved by arbitrarily large control inputs. Disturbance normalization, horizons, intervention conventions, and performance weights are fixed before held-out evaluation.

\begin{figure*}[t]
    \centering
    \includegraphics[width=\textwidth]{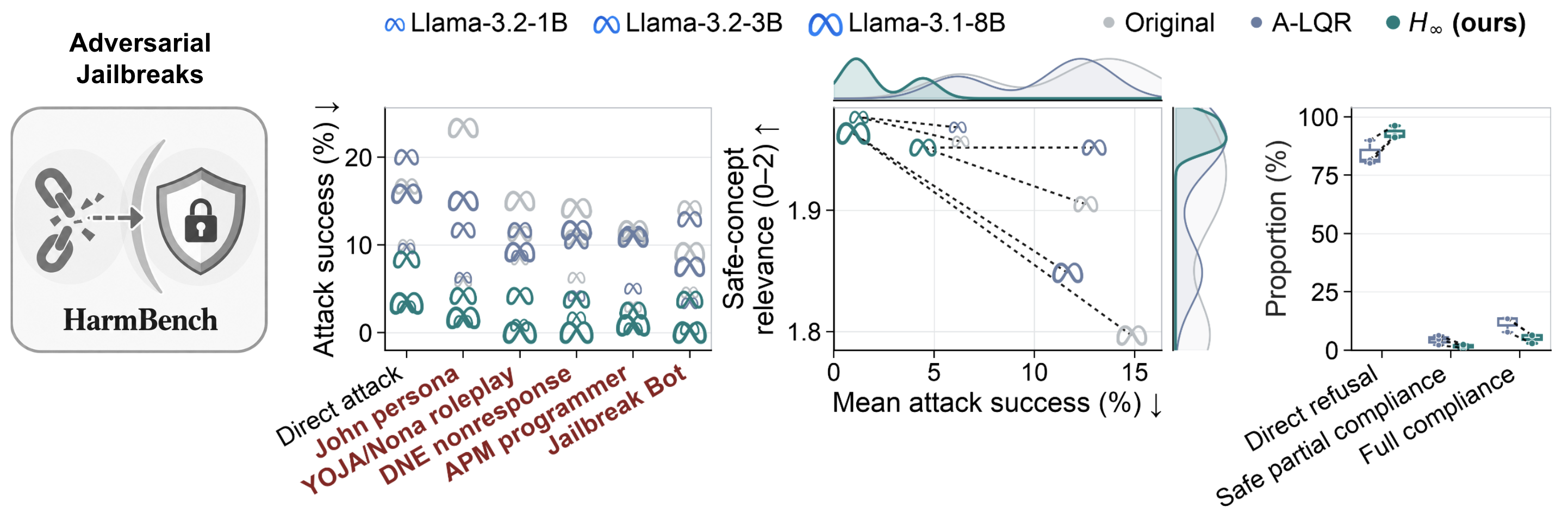}
    \vspace{-0.8cm}
    \caption{
    \textbf{Robust refusal under adversarial prompting.}
    Left: ASR across direct requests and five frozen jailbreak templates for
    three Llama scales. Center: mean ASR versus safe-concept relevance.
    Right: response-type decomposition. Lower ASR is better.
    }
    \label{fig:harmbench-transfer}
    \vspace{-0.5cm}
\end{figure*}

\section{Methodology: Robust Activation Steering}
\label{sec:method}
\vspace{-0.3cm}
Given a target behavior, we instantiate ~\ref{eq:lin-llm} using calibration activations and synthesize feedback based on the empirically observed model mismatch. All identification, disturbance fitting, and controller selection use calibration data only and are frozen before held-out evaluation; implementation details are given in Appendix~\ref{app:methodological-details}.

\vspace{-0.25cm}
\paragraph{Behavior coordinates and empirical dynamics.}
At each layer $k$, the difference between the mean final-token activations of desired and undesired calibration examples defines the semantic direction. We construct a reduced basis $V_k$ whose first axis is this normalized direction and whose remaining axes capture the leading orthogonal activation variation; the reference state places the semantic coordinate at its target displacement and the remaining coordinates at zero. Prompt-specific Jacobian--vector products are projected into these coordinates and averaged to estimate the shared transition $\Ak$, with $\Bk$ denoting the corresponding intervention channel. For calibration trajectory $i$, the residual
$\xik^{(i)}:=x_{k+1}^{(i)}-\Ak x_k^{(i)}-\Bk u_k^{(i)}$
measures local-model error. A layer-wise SVD of the centered residual covariance defines $\Dk$, preserving the empirical directions and scales of this mismatch. Hence, $(\Ak,\Bk)$ encode nominal intervention propagation, while $\Dk$ identifies where that model is empirically unreliable.

\vspace{-0.25cm}
\paragraph{Robust controller synthesis.}
For a candidate attenuation $\gamma$, we synthesize feedback through the finite-horizon dynamic game
\vspace{-0.22cm}
\begin{equation}
\label{eq:hinf-steering-game}
\begin{aligned}
\min_{\pi}\max_{w_{1:L}}\quad&
x_{L+1}^{\top}Q_f x_{L+1}
+
\sum_{k=1}^{L}
\left(
\norm{\zk}_2^2-\gamma^2\norm{\wk}_2^2
\right),\\
\mathrm{\textbf{s.t.}}\quad&
\eqref{eq:lin-llm},
\qquad
\uk=\pi_k(\xk).
\end{aligned}
\end{equation}

The controller minimizes steering error and intervention cost while the adversary acts only through the calibration-derived mismatch channels $\Dk$. A backward $H_\infty$ recursion yields linear feedback $\uk=-\Kk^{(\gamma)}\xk$; bisection over feasible $\gamma$ recovers the attenuation $\gammastar$ defined above and its associated gains. Candidate cost configurations are evaluated on a designated calibration split using the benchmark's predeclared criterion, after which the selected controller is frozen.

\vspace{-0.35cm}
\paragraph{Inference.}
At each controlled layer and decoding step, the current final-token activation is projected relative to its reference state, and the stored policy applies $\uk=-\Kk^\star\xk$. The intervention is lifted to the residual stream and added to the block output. Jacobian estimation, disturbance fitting, and robust synthesis are therefore entirely offline; deployment requires only the stored bases, reference states, and feedback gains.

\section{Theoretical Analysis of Robust Steerability}
\label{sec:theory}
\vspace{-0.3cm}
We establish two complementary properties of robust steerability.
First, we show that it measures worst-case amplification of model mismatch
and identifies the layers and directions responsible for steering fragility.
Second, we show that robust steering and nominal LQR belong to a common
finite-horizon controller family. Together, these results give
$\Srob$ an operational interpretation and clarify the relationship between
our method and A-LQR. Proofs are deferred to
Appendix~\ref{app:proofs}.

\begin{figure*}[t]
    \centering
    \includegraphics[width=\textwidth]{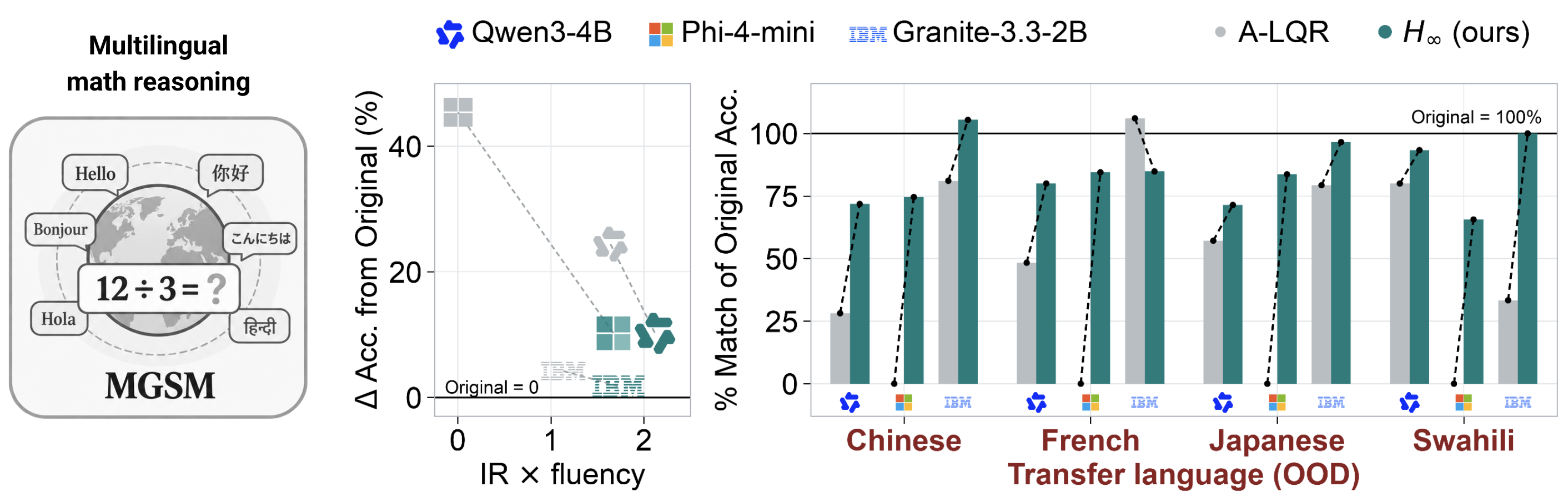}
    \vspace{-0.8cm}
    \caption{
    \textbf{Cross-lingual transfer on MGSM.}
    Left: capability loss relative to the unsteered model versus response
    quality. Right: accuracy retained relative to the original model across
    transfer languages. Controllers are calibrated in English and frozen
    before multilingual evaluation.
    }
    \label{fig:mgsm-transfer}
        \vspace{-0.4cm}
\end{figure*}
\vspace{-0.3cm}
\subsection{Robust Steerability Identifies Vulnerable Directions}
\label{sec:directional-certificate}
\vspace{-0.3cm}
For a fixed controller $\steervar{K}$, stacking the layerwise disturbances
and regulated outputs gives
\[
\statevar{z}
=
\mathcal{T}_{\steervar{K}}\statevar{w},
\qquad
\statevar{w}=\operatorname{col}(\wk)_{k=1}^{L+1},
\quad
\statevar{z}=\operatorname{col}(\zk)_{k=1}^{L+1},
\]
where $\mathcal{T}_{\steervar{K}}$ is the finite-depth closed-loop response
from model mismatch to semantic deviation.

\begin{proposition}[Directional certificate]
\label{prop:directional-certificate}
For any admissible controller,
\[
\steervar{\gamma}(\steervar{K})
=
\norm{\mathcal{T}_{\steervar{K}}}_2
=
\sigma_{\max}(\mathcal{T}_{\steervar{K}}).
\]
A principal right singular vector $\statevar{w}^\star$ therefore identifies
the worst-case unit-energy mismatch. Moreover, feasibility of the
finite-horizon $\mathrm{H}_{\infty}$ recursion at $\gamma$ guarantees
\[
\norm{\statevar{z}}_2
\leq
\gamma\norm{\statevar{w}}_2
\]
for every disturbance represented by $\{\Dk\}_{k=1}^{L}$.
\end{proposition}

The worst-case vector
$\statevar{w}^\star=\operatorname{col}(w_k^\star)_{k=1}^{L}$
also localizes the source of fragility: layers with larger
$\norm{w_k^\star}_2$ contribute more strongly to the most amplified mismatch.
Thus, residual size alone is not predictive—a large residual may be suppressed
by later layers, while a smaller but semantically aligned residual may be
amplified.

Overall, the certificate answers two questions: \emph{how fragile is the
representation}, through
$\gamma^\star=\inf_{\steervar{K}}
\norm{\mathcal{T}_{\steervar{K}}}_2$,
and \emph{where does that fragility arise}, through
$\statevar{w}^\star$.
Since $\Srob=1/\gamma^\star$, larger robust steerability means that more
disturbance energy is required to produce a fixed semantic deviation.

\subsection{Nominal and Robust Steering Form a Continuum}
\label{sec:robustness-continuum}

Classical linear quadratic regulation (LQR) seeks a feedback
policy that minimizes a quadratic cost penalizing deviations
from a desired state and the control effort required to
correct them. LQR assumes that the nominal dynamics
accurately describe the system; in our setting, this
corresponds to setting $w_k=0$ in
\eqref{eq:hinf-steering-game}. Consequently, although
optimal for the nominal model, LQR can be fragile when
the realized dynamics deviate from those used for controller
synthesis. In contrast, $\mathrm{H}_\infty$ control explicitly
accounts for such uncertainty by treating model mismatch
as an adversarial disturbance. It seeks a feedback policy
that minimizes a quadratic cost against the worst-case
disturbance sequence over a finite-energy disturbance space
$\ell_2[0,T]$. In \eqref{eq:hinf-steering-game}, the
attenuation parameter $\gamma$ determines the strength
of this adversary: smaller values impose a more demanding
robustness requirement, while larger values progressively
suppress the adversarial contribution. Importantly,
these are not isolated control problems. Classical
robust-control theory connects LQR and $\mathrm{H}_\infty$
control through a continuum of disturbance-aware
formulations, with related interpretations in
minimum-entropy control
\citep{basar1991dynamic,hassibi1999indefinite}.
At one end, nominal LQR is recovered as
$\gamma\to\infty$; at the other, decreasing $\gamma$
toward the minimum feasible attenuation level yields
increasingly stringent worst-case robustness guarantees.
We make this connection explicit for our finite-depth,
time-varying setting in the following proposition.

\begin{proposition}[LQR limit of robust steering]
\label{prop:lqr-limit}
Consider the finite-horizon dynamics~\ref{eq:lin-llm} and output map~\eqref{eq:out}, over
$k=1,\ldots,L$, and assume 
state, terminal-state, and control cost matrices are positive semidefinite and definite respectively. For each feasible attenuation level $\gamma$,
let $P_k^{(\gamma)}$ and $\Kk^{(\gamma)}$
denote the value matrices and optimal feedback
gains associated with the $\mathrm{H}_{\infty}$
game~(\ref{eq:hinf-steering-game}).
Let $P_k^{\mathrm{LQR}}$ and $\Kk^{\mathrm{LQR}}$
denote the corresponding nominal LQR quantities,
obtained by setting $w_k=0$ and minimizing
$\norm{z}_2^2$. Then, the $\mathrm{H}_{\infty}$ problem is feasible
for all sufficiently large $\gamma$, and there
exists a constant $C<\infty$, independent of
$\gamma$, such that
\[
\max_{1\leq k\leq L}
\left(
\norm{P_k^{(\gamma)}-P_k^{\mathrm{LQR}}}_2
+
\norm{\Kk^{(\gamma)}-\Kk^{\mathrm{LQR}}}_2
\right)
\leq \frac{C}{\gamma^2}.
\]
\vspace{-0.5cm}
\end{proposition}

The corresponding Riccati recursions and proof are
provided in Appendix~\ref{app:proof-lqr-limit}.

Thus, nominal LQR emerges as the limiting case of robust
steering: both belong to a continuous family of feedback
controllers parameterized by the disturbance-attenuation
level $\gamma$. Importantly, $\gamma$ specifies an upper
bound on the allowable disturbance-to-performance
amplification, rather than the amplification necessarily
achieved by the controller. Increasing $\gamma$ therefore
relaxes the robustness requirement, while simultaneously
imposing a larger penalty on the adversarial disturbance
in the minimax formulation. As $\gamma\to\infty$, the
optimal adversarial disturbance vanishes and the resulting
controller approaches nominal LQR, as quantified by
Proposition~2. Conversely, decreasing $\gamma$ places
increasingly stringent demands on disturbance attenuation,
leading to feedback policies that explicitly prioritize
robustness to model mismatch.

A convenient parameterization of this continuum is
$\rho_{\mathrm{rob}}=\frac{\gamma^\star}{\gamma}\in[0,1)$,
where $\rho_{\mathrm{rob}}=0$ corresponds to the LQR limit
and $\rho_{\mathrm{rob}}\uparrow1$ approaches the strongest
feasible worst-case attenuation. Intermediate values allow
us to balance nominal steering performance against
protection from model mismatch. This perspective is closely
related to the classical minimum-entropy connection between
$\mathrm{H}_2$ and $\mathrm{H}_{\infty}$ control
\citep{doyle1989state,iglesias1990discrete}.
Although Proposition~2 does not imply that downstream task
performance varies monotonically with $\rho_{\mathrm{rob}}$,
it suggests that nominal operating conditions may favor
LQR-like controllers, whereas larger dynamics shifts may
benefit from more robust operating points along the same
continuum.

This connection is particularly relevant to our experimental
evaluation, where Activation-LQR (A-LQR) [citation] is one of the
strongest baselines. By recovering nominal LQR as a limiting
case, our $\mathrm{H}_{\infty}$ framework provides a
principled extension of this existing steering approach.

\begin{figure*}[t]
    \centering
    \includegraphics[width=\textwidth]{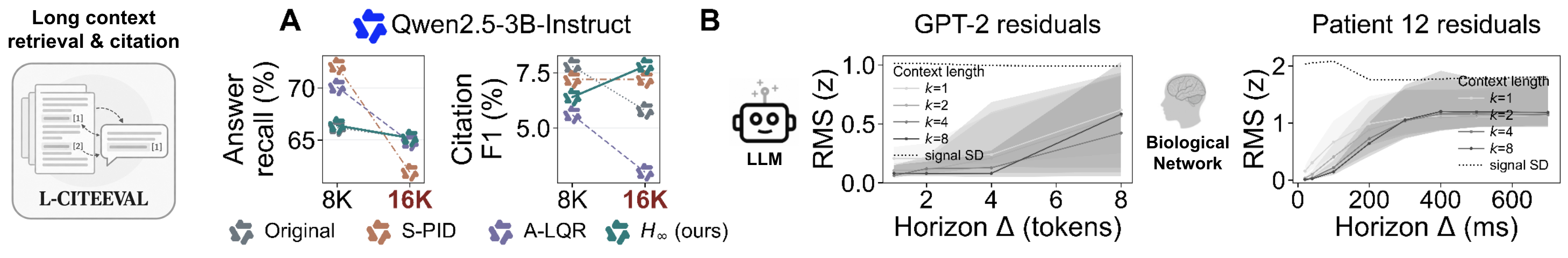}
    \vspace{-0.8cm}
    \caption{
    \textbf{Context-length transfer and residual growth.}
    \textbf{A}: answer recall and citation F1 at matched $8$K and shifted
    $16$K context for Qwen2.5-3B.
    \textbf{B}: growth of local-model residuals with prediction horizon in
    language-model representations and an illustrative nonlinear biological
    network.
    }
    \label{fig:lciteeval-transfer}
\end{figure*}
\section{Experiments}
\label{sec:experiments}

We ask whether feedback controllers identified at calibration time remain
effective when the deployment distribution changes. We consider four
complementary shifts: prompt language on TruthfulQA, adversarial prompting on
HarmBench, input language on MGSM, and context length on L-CiteEval. In every
case, dynamics identification, disturbance estimation, controller selection,
and hyperparameter tuning are completed using calibration data; the resulting
controller is then frozen before shifted evaluation.  Table~\ref{tab:main-results-summary}
summarizes the primary metrics; complete per-condition results, secondary
metrics, and uncertainty estimates are provided in
Appendix~\ref{app:full-results}.
\textbf{We deliberately stress-test the controller across qualitatively different failure modes---linguistic, adversarial, multilingual, and context-length shifts---so that robustness must hold beyond any single benchmark or deployment regime.}

\vspace{-0.4cm}
\paragraph{Prompt-language shift: TruthfulQA.}
 We calibrate on an English corpus and then evaluate on English-input, English-output (\textbf{ID}) and Spanish-input, English-output transfer (\textbf{OOD}). Because increasing truthfulness alone can produce vague and uninformative
responses, the primary endpoint is
$\mathrm{T{\times}I}=\mathrm{True}(\%)\times\mathrm{Informative}(\%)/100$.
Figure~\ref{fig:truthfulqa-transfer} plots this trade-off directly: iso-product
contours distinguish genuine movement toward the upper-right frontier from
methods that improve one axis by sacrificing the other. \textcolor{teal}{Across all four model families, $\mathrm{H}_{\infty}$ attains the highest
T$\times$I under both ID and OOD evaluation}
(Table~\ref{tab:main-results-summary}). Relative to the strongest
non-$\mathrm{H}_{\infty}$ baseline, the average gain is $+3.8$ points under
matched English evaluation and $+5.4$ points after Spanish transfer, with the
largest OOD gains on GPT-2 XL ($9.93\!\rightarrow\!18.14$) and OLMo-2-32B
($82.70\!\rightarrow\!91.50$). The scatter plots also expose why the joint
metric is necessary: on Llama-3-8B, for example, ODESteer drives truthfulness
above $95\%$ while informativeness collapses to roughly $10\%$, placing it far
from the useful frontier. The category-level GPT-2 XL analysis further shows
that the Spanish-transfer improvement is distributed across TruthfulQA
categories rather than arising from a single topic.

\begin{table*}[t]
\centering
\definecolor{projectdarkred}{RGB}{0,128,128}

\caption{
\textbf{Performance under distribution shift.}
TruthfulQA: T$\times$I; HarmBench: ASR (\%, $\downarrow$);
MGSM: accuracy / Spanish relevance; L-CiteEval: answer recall (\%).
``Base'' is the strongest non-$H_\infty$ baseline. Full results are in the appendix.
}
\vspace{-0.4cm}
\label{tab:main-results-summary}

{\fontsize{6.4}{6.9}\selectfont
\renewcommand{\arraystretch}{0.98}
\setlength{\tabcolsep}{2.4pt}


\begin{minipage}[t]{0.49\textwidth}
\centering
\textbf{(a) TruthfulQA: T$\times$I $\uparrow$}
\vspace{1pt}

\resizebox{\linewidth}{!}{%
\begin{tabular}{lcccc}
\toprule
\textbf{Model}
&
\textbf{ID Base}
&
\cellcolor{projectdarkred!10}\textbf{ID $H_\infty$}
&
\textbf{OOD Base}
&
\cellcolor{projectdarkred!10}\textbf{OOD $H_\infty$}
\\
\midrule
GPT2-XL
& 34.91
& \cellcolor{projectdarkred!10}\textbf{35.18}
& 9.93
& \cellcolor{projectdarkred!10}\textbf{18.14}
\\
Llama3-8B
& 57.81
& \cellcolor{projectdarkred!10}\textbf{61.42}
& 52.81
& \cellcolor{projectdarkred!10}\textbf{55.29}
\\
Qwen2.5-14B
& 69.66
& \cellcolor{projectdarkred!10}\textbf{76.27}
& 80.40
& \cellcolor{projectdarkred!10}\textbf{82.57}
\\
OLMo2-32B
& 81.74
& \cellcolor{projectdarkred!10}\textbf{86.31}
& 82.70
& \cellcolor{projectdarkred!10}\textbf{91.50}
\\
\bottomrule
\end{tabular}%
}
\end{minipage}
\hfill
\begin{minipage}[t]{0.49\textwidth}
\centering
\textbf{(b) HarmBench: ASR $\downarrow$}
\vspace{1pt}

\resizebox{\linewidth}{!}{%
\begin{tabular}{lcccc}
\toprule
\textbf{Model}
&
\textbf{Direct Base}
&
\cellcolor{projectdarkred!10}\textbf{Direct $H_\infty$}
&
\textbf{JB Base}
&
\cellcolor{projectdarkred!10}\textbf{JB $H_\infty$}
\\
\midrule
Llama3.2-1B
& 9.58
& \cellcolor{projectdarkred!10}\textbf{2.92}
& 5.50
& \cellcolor{projectdarkred!10}\textbf{0.92}
\\
Llama3.2-3B
& 16.67
& \cellcolor{projectdarkred!10}\textbf{8.33}
& 11.59
& \cellcolor{projectdarkred!10}\textbf{3.67}
\\
Llama3.1-8B
& 15.83
& \cellcolor{projectdarkred!10}\textbf{3.33}
& 10.83
& \cellcolor{projectdarkred!10}\textbf{0.50}
\\
\bottomrule
\end{tabular}%
}
\end{minipage}

\vspace{2pt}


\begin{minipage}[t]{0.49\textwidth}
\centering
\textbf{(c) MGSM: multilingual transfer}
\vspace{1pt}

\resizebox{\linewidth}{!}{%
\begin{tabular}{lcccccc}
\toprule
\textbf{Model}
&
\textbf{Acc-O}
&
\textbf{Acc-L}
&
\cellcolor{projectdarkred!10}\textbf{Acc-$H_\infty$}
&
\textbf{SR-L}
&
\cellcolor{projectdarkred!10}\textbf{SR-$H_\infty$}
\\
\midrule
Qwen3-4B
& \textbf{41.6}
& 22.4
& \cellcolor{projectdarkred!10}32.2
& \textbf{1.81}
& \cellcolor{projectdarkred!10}1.44
\\
Phi4-mini
& \textbf{45.4}
& 0.0
& \cellcolor{projectdarkred!10}35.2
& 1.09
& \cellcolor{projectdarkred!10}\textbf{1.21}
\\
Granite3.3-2B
& \textbf{22.2}
& 18.8
& \cellcolor{projectdarkred!10}21.2
& \textbf{0.04}
& \cellcolor{projectdarkred!10}0.00
\\
\bottomrule
\end{tabular}%
}
\end{minipage}
\hfill
\begin{minipage}[t]{0.49\textwidth}
\centering
\textbf{(d) L-CiteEval: answer recall $\uparrow$}
\vspace{7.35pt}

\resizebox{\linewidth}{!}{%
\begin{tabular}{lcccc}
\toprule
\textbf{Model}
&
\textbf{8K Base}
&
\cellcolor{projectdarkred!10}\textbf{8K $H_\infty$}
&
\textbf{16K Base}
&
\cellcolor{projectdarkred!10}\textbf{16K $H_\infty$}
\\
\midrule
Qwen2.5-3B
& \textbf{72.1}
& \cellcolor{projectdarkred!10}66.4
& \textbf{65.2}
& \cellcolor{projectdarkred!10}\textbf{65.2}
\\
Llama3.2-1B
& 59.8
& \cellcolor{projectdarkred!10}\textbf{61.7}
& \textbf{50.1}
& \cellcolor{projectdarkred!10}47.9
\\
\bottomrule
\end{tabular}%
}
\end{minipage}

}
\vspace{-3pt}
\end{table*}

\vspace{-0.4cm}
\paragraph{Adversarial prompt shift: HarmBench.}
We next test whether a controller calibrated on benign safety prompts remains
effective against direct harmful requests and five frozen jailbreak templates.
Attack success rate (ASR) is the primary endpoint; safe-concept relevance
checks whether a reduction in ASR corresponds to the intended refusal
behavior rather than generic degradation.

The effect is consistent across model scale and attack type
(Figure~\ref{fig:harmbench-transfer}). \textcolor{teal}{Across the three models and six prompt
conditions, $\mathrm{H}_{\infty}$ has lower ASR than both the unsteered model
and A-LQR (the strongest competing method) in all $18$ model--condition comparisons}. Against the strongest
non-$\mathrm{H}_{\infty}$ baseline, mean ASR across model scales decreases
from $14.0\%$ to $4.9\%$ on direct requests and from $9.3\%$ to $1.7\%$ across
the jailbreak templates, corresponding to approximately $65\%$ and $82\%$
relative reductions, respectively. This reduction is not explained by losing
the safety objective: $\mathrm{H}_{\infty}$ maintains safe-concept relevance
near the top of the $0$--$2$ scale across conditions. The response-type
decomposition is consistent with the same mechanism, shifting probability
toward direct refusal and away from full compliance.

\vspace{-0.4cm}
\paragraph{Language shift: MGSM.}
MGSM probes multilingual mathematical reasoning; we use it to test whether the model can be steered toward a target language without sacrificing reasoning accuracy under language shift. Here, the controllers are calibrated in English and then deployed, frozen,
on Chinese, French, Japanese, Swahili, and Telugu MGSM inputs, with Spanish as
the target output language. This experiment separates \emph{successful
intervention} from \emph{capability retention}: the unmodified models provide
a useful upper reference for mathematical accuracy but have zero
target-language relevance and therefore do not satisfy the steering objective.

Among steering controllers, robust synthesis preserves substantially more of
the underlying reasoning capability. \textcolor{teal}{Averaged over the five transfer
languages, $\mathrm{H}_{\infty}$ achieves $32.2\%$, $35.2\%$, and $21.2\%$
accuracy on Qwen3-4B, Phi-4-mini, and Granite-3.3-2B, compared with
$22.4\%$, $0.0\%$, and $18.8\%$ for A-LQR.} Equivalently,
$\mathrm{H}_{\infty}$ retains approximately $77\%$, $78\%$, and $95\%$ of
the corresponding unsteered accuracy, versus $54\%$, $0\%$, and $85\%$ for
A-LQR. At the individual language level, $\mathrm{H}_{\infty}$ exceeds A-LQR
accuracy in $13$ of $15$ model--language pairs. Figure~\ref{fig:mgsm-transfer}
shows the same effect both as displacement from original accuracy and as the
fraction of baseline capability retained after steering.

\vspace{-0.40cm}
\paragraph{Context-length shift: L-CiteEval.}
Finally, we test robustness to a change in the effective sequence horizon. L-CiteEval probes long-context retrieval and citation grounding; here, in particular, we test whether steering remains reliable as context length increases.
Controllers are calibrated at $8$K context and evaluated at both $8$K and
$16$K using matched question identities. We report answer recall (AR) as the
primary metric and citation F1 as a complementary measure of whether retrieved
evidence remains usable.
The result is more nuanced than on TruthfulQA or HarmBench and illustrates the
distinction between nominal and shifted performance. \textcolor{teal}{For Qwen2.5-3B,
$\mathrm{H}_{\infty}$ does not maximize matched $8$K recall
($66.4\%$ versus $72.1\%$ for the strongest baseline), but degrades by only
$1.2$ points at $16$K, compared with $5.3$ points for A-LQR and $10.2$ points
for S-PID.} Its citation F1 simultaneously increases from $6.4$ to $7.8$,
whereas A-LQR drops from $5.6$ to $3.0$. On Llama-3.2-1B,
$\mathrm{H}_{\infty}$ leads answer recall at $8$K
($61.7\%$ versus $59.8\%$) but trails A-LQR at $16$K
($47.9\%$ versus $50.1\%$), while retaining the strongest citation F1.
Thus, robust synthesis does not imply pointwise dominance of every nominal
metric; its benefit appears most clearly in the degradation induced by the
shift itself.

Figure~\ref{fig:lciteeval-transfer} also provides a direct diagnostic for why
context extension challenges a locally identified controller. Prediction
residuals increase as the deployment regime moves away from calibration,
indicating greater mismatch between the identified model and the realized
dynamics. This mirrors the control-relevant failure mode observed earlier in
the biological network: local dynamics are predictive near the identification
regime, while residuals become increasingly consequential as the system moves
beyond it. Together, these results suggest that robustness is not a benchmark-specific concern but a systems-level consequence of intervening through models whose validity is necessarily local.

\vspace{-0.1cm}
\section{Robust Steerability as a Property of Large Networks: Conclusion and Future Work}
\label{sec:S_rob&Dconclusion}
\vspace{-0.4cm}
Beyond measuring whether a network can be steered, $S_{\mathrm{rob}}$ provides a notion of \emph{how robustly} that steering can be achieved. Intuitively, it characterizes how much deviation from the identified dynamics can be tolerated before the same model is no longer sufficient for reliable intervention. A larger $S_{\mathrm{rob}}$ therefore corresponds to a regime in which steering is less sensitive to modeling error and uncertainty.
\begin{figure}[h]
    \centering
    \includegraphics[width=0.9\linewidth]{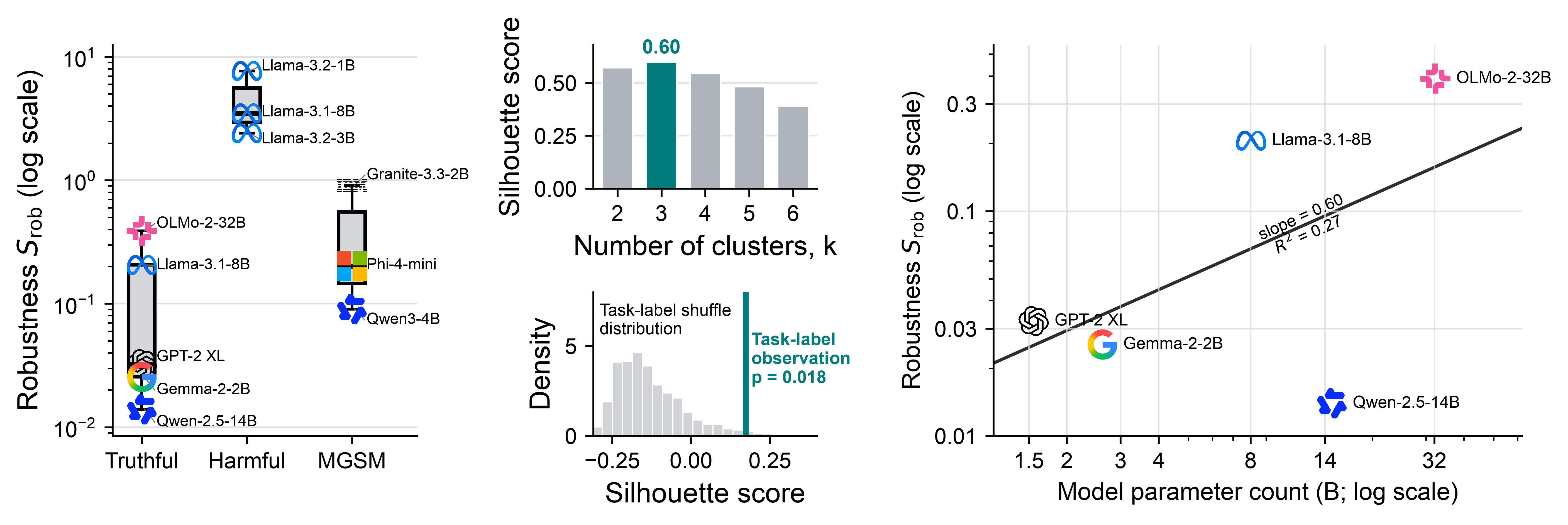}
    \vspace{-0.5cm}
    \caption{
    Robust steerability across models and tasks. 
    $S_{\mathrm{rob}}$ exhibits task-dependent structure across model--benchmark pairs, while within a fixed task it also varies systematically with model scale.
    }
    \label{fig:Srob}
\end{figure}
Computing $S_{\mathrm{rob}}$ across models and benchmarks reveals that robust steerability is not distributed arbitrarily. Model--task pairs cluster according to the behavioral objective, indicating that robustness is strongly dependent on \emph{what} is being controlled. We then ask whether robust steerability also reflects properties of the network itself. When $S_{\mathrm{rob}}$ is examined against parameter count within a fixed task, we observe a positive relationship with model scale. Together, these results suggest that robust steerability is neither purely a property of the task nor purely a property of the model, but emerges from the interaction between the system being controlled and the objective imposed on it.
This perspective suggests a broader interpretation of $S_{\mathrm{rob}}$: it may quantify how \emph{careful} an intervention needs to be. Some operating regimes may tolerate coarse models, approximate perturbations, or relatively simple control strategies while still producing reliable outcomes. Others may be substantially more fragile, requiring tighter state estimation, feedback, and explicit robustness to uncertainty. In this sense, $S_{\mathrm{rob}}$ offers a way to move beyond asking whether an intervention works, toward asking how much uncertainty the intervention can tolerate before it fails.

Large language models provide an unusual experimental system in which such questions can be studied systematically. They are not biologically realistic models of neural circuitry, but they are large nonlinear networks whose internal states can be measured, modeled, perturbed, and evaluated across many controlled conditions. This makes them useful not as models \emph{of} the brain, but as a model organism for studying general principles of intervention in complex neural networks. Principles identified in such systems---including state dependence, intervention fragility, and robust controllability---may therefore help motivate more mechanistic approaches to neural stimulation.
This closes the loop with the neuroscience motivation of our framework. If robust steerability can be quantified in artificial neural systems, an important next question is whether an analogous quantity can be estimated from neural recordings and stimulation experiments. Such a measure could, in principle, distinguish brain states, regions, or stimulation targets that tolerate substantial model uncertainty from those that require more precise closed-loop intervention. This raises a natural question for future work: 
\emph{Can $S_{\mathrm{rob}}$ be measured from brain data?}

\newpage


\newpage

\subsection*{AI use statement}


In this work, we used generative AI tools to assist in proving mathematical
claims, including checking and refining proof arguments. We did not use
generative AI tools to generate synthetic datasets, develop the theoretical or
conceptual framework, formulate mathematical claims, propose or refine
hypotheses, design the research methodology or experiments, implement the
proposed methods, translate research material, clean or reformat datasets,
perform qualitative or thematic data analysis, or interpret experimental
results. These tasks were carried out by the authors or were not applicable to
this work.

Additionally, we used generative AI tools to edit and polish author-written portions of the manuscript for clarity and readability, for retrieval and discovery of relevant literature, and to assist
in creating and refining scientific figures and visualizations. We did not use
generative AI tools to formulate survey or interview questions, transcribe
research material, choose experimental parameters, or create other research
artifacts.

We have reviewed all AI-assisted work. References identified with AI assistance
were checked against the original sources, mathematical arguments were
independently verified and revised by the authors, and AI-assisted figures were
reviewed for consistency with the underlying experimental results and technical
claims. We take responsibility for the final content of this work, including
all text, claims, citations, mathematical results, figures, and other artifacts
produced with the aid of generative AI.

\subsection*{Reproducibility statement}

We will release the complete implementation of robust activation steering in
a public GitHub repository. The release will include the
$\mathrm{H}_{\infty}$ synthesis and inference code, benchmark configurations,
fixed evaluation subsets and random seeds, model and dataset identifiers,
environment specifications, and scripts for artifact construction,
calibration, generation, scoring, statistical analysis, and regeneration of
the reported tables and figures. The main text defines the method and its
assumptions, while the appendices provide the end-to-end algorithm, theoretical
derivations, benchmark protocols, model checkpoints, hyperparameter searches,
decoding settings, statistical procedures, and computational requirements.
Together, these materials support independent verification, controlled
comparison, and direct extension of the method. We expect the GitHub release
to facilitate community adoption and broaden the positive impact of robust
activation steering across AI and machine learning research.

\bibliography{references}
\bibliographystyle{iclr2027_conference}


\appendix
\newpage

\startcontents[appendix]

\section*{Appendix Table of  Contents}

\printcontents[appendix]{}{1}{
    \setcounter{tocdepth}{2}
}


\newpage
\section{Related Work}
\label{sec:related}

\paragraph{Activation steering and representation intervention.}
Activation steering modifies intermediate representations at inference time
to alter model behavior without updating model parameters.
Early representation-engineering and contrastive methods identify semantic
directions from activation statistics associated with opposing behaviors and
intervene along these directions
\citep{zou2023representation,turner2024activation}.
Related approaches have shown that comparatively low-dimensional
interventions can affect high-level behaviors such as truthfulness and
refusal: inference-time intervention uses internal representations to elicit
truthful responses \citep{li2023iti}, contrastive activation addition
constructs behavioral directions from paired examples
\citep{rimsky2024caa}, and refusal has been associated with a particularly
low-dimensional intervention direction \citep{arditi2024refusal}.
Beyond additive interventions, activation transport methods explicitly match
or transform activation distributions, leading to sequential procedures such
as Mean-AcT and Linear-AcT \citep{rodriguez2025transport}.
These results collectively establish that intermediate representations
provide a meaningful control surface, but successful intervention under the
distribution on which a direction is constructed does not by itself imply
reliable intervention after the input or internal trajectory changes.
Indeed, steering effectiveness can vary substantially across inputs
\citep{tan2024analysing}.
Our focus is precisely this latter question: not only whether a semantic
direction can influence behavior, but how sensitive that influence is to
uncertainty in the dynamics through which the intervention propagates.

\paragraph{Dynamical and control-theoretic views of neural networks.}
Interpreting network depth as an evolution variable has a broader precedent
in dynamical-systems views of deep networks, including stable residual
architectures and neural ordinary differential equations
\citep{haber2017stable,chen2018neuralode}.
Within language models, controllability-based formulations similarly treat
generation or representation evolution as a dynamical process subject to
intervention \citep{soatto2023controllability,genctrl2026}.
This viewpoint is particularly natural for activation steering: an
intervention applied at one transformer layer changes the state presented to
later layers, so steering cannot in general be understood as an isolated
perturbation of a static representation.
Our formulation adopts this finite-depth dynamical interpretation but focuses
on uncertainty in the identified layer-to-layer map rather than assuming that
a nominal dynamical model remains valid after calibration.

\paragraph{Feedback and model-based activation control.}
Recent work has made this dynamical viewpoint operational by formulating
activation steering directly as a feedback-control problem
\citep{nguyen2025pid,zhao2026odesteer,skifstad2026activationlqr}.
PID Steering interprets common difference-in-means steering rules through
proportional feedback and augments them with integral and derivative terms to
reduce persistent semantic error and oscillatory behavior
\citep{nguyen2025pid}.
Other approaches use continuous-depth or feedback constructions to adapt
interventions to the evolving representation
\citep{zhao2026odesteer}.
Activation-LQR is especially close to our setting:
it locally linearizes transformer blocks, treats depth as a finite
time-varying control horizon, and uses LQR to track semantic feature
setpoints \citep{skifstad2026activationlqr}.
It further analyzes tracking error induced by the discrepancy between the
nonlinear network and its local linear approximation.

Our work begins from the same model-based control abstraction but asks a
different question.
We do not claim novelty from applying classical optimal control to transformer
representations; rather, we treat the local-model discrepancy itself as the
object of interest.
Instead of asking only which controller performs well under the identified
dynamics, we ask whether the \emph{closed-loop amplification of errors in
those dynamics} predicts whether semantic control remains reliable when the
deployment trajectory departs from calibration.
This distinction motivates both our empirical disturbance model and the
robust-steerability quantity.

\paragraph{Robust and optimal control.}
The control machinery underlying our formulation is classical.
Linear quadratic control provides optimal feedback for nominal linear
dynamics \citep{kalman1960contributions,lewis2012optimal}, while
$H_\infty$ control characterizes and minimizes worst-case amplification from
exogenous disturbances to regulated outputs.
Finite-horizon and time-varying formulations admit dynamic-game and
state-space solutions
\citep{basar1991dynamic,ravi1991hinfinity,hassibi1999indefinite},
and classical results connect robust synthesis to nominal quadratic control
and the broader $H_2$--$H_\infty$ continuum
\citep{doyle1989state,iglesias1990discrete,zhou1996robust}.
Related robust-control descriptions have also been used to interpret human
responses to unpredictable perturbations
\citep{crevecoeur2019robust}.

Our contribution is therefore not a new general $H_\infty$ synthesis
procedure.
We instead use the induced closed-loop gain as a quantity for characterizing
learned representation dynamics.
The disturbance channel is estimated from empirical one-step
representation-model residuals rather than postulated as isotropic noise,
and the resulting minimum amplification defines a pre-deployment measure of
how much identified-model mismatch a steering problem can tolerate.
In this sense, robust control serves both as a controller and, more
importantly for our purpose, as a diagnostic of intervention reliability.

\paragraph{Open-source language models.}
We evaluate GPT-2 \citep{radford2019language}, Llama 3/3.1/3.2
\citep{dubey2024llama}, Qwen2.5 \citep{yang2024qwen25}, OLMo 2
\citep{olmo2024olmo2}, Qwen3 \citep{yang2025qwen3}, and Phi-4-mini
\citep{abouelenin2025phi4mini}.
For technical reports with exceptionally large author lists, we follow the
conventional abbreviated \emph{et al.} representation rather than reproducing
the complete author list in the bibliography; the cited paper URLs provide
the authoritative records and full attribution.
Table~\ref{tab:experimental-models} separately reports the exact model
repositories and pinned revisions used in every experiment.
As no dedicated Granite~3.3 technical report was publicly available at
submission time, we cite the official IBM Granite~3.3 model
repository/documentation \citep{granite2025granite33}.
Thus, the bibliography provides model-family attribution, while the
checkpoint identifiers provide model-level provenance and reproducibility.

\paragraph{Neurostimulation and neural field dynamics.}
Neurostimulation offers a biological precedent for the control problem considered here. Because the effect of an intervention depends on the neural state in which it is delivered, the field has moved from fixed protocols toward state-dependent and closed-loop stimulation \citep{hebb2014creating,zrenner2016closedloop,bergmann2018brain,bradley2022state,zrenner2024closedloop, Choubdar2023}, in which EEG, ECoG, and local field potentials (LFPs) serve as feedback variables for estimating the current state and adapting stimulation online \citep{basu2023closed,topalovic2023wearable,davila2022control}. Stimulation can also alter the regime it acts on: locus-coeruleus stimulation, for example, changes sensory processing and global brain state over distinct timescales \citep{Fazlali2020}, so the dynamics governing the response need not remain fixed once an intervention begins.

Current closed-loop systems typically summarize neural state with a few local features, such as oscillatory power or phase, yet intervention effects depend on the circuit through which perturbations propagate: local hippocampal circuitry shapes spatial selectivity \citep{Geiller2022}, and differences in recurrent connectivity across subregions yield distinct representational regimes \citep{KONG2026}. At the population scale, field potentials capture this state through their underling generative dynamics \cite{Fallah2026.10.28.762631} as well as their spatiotemporal organization on the cortex. Beta traveling waves across frontal and parietal cortex encode recent reward history \citep{Zabeh2023}, and planar, spiral, and concentric waves in human intracranial recordings distinguish memory-related states \citep{Das2026}. High-density interfaces and propagation-analysis methods now expose these distributed dynamics as candidate feedback variables \citep{Jung2025,DasZabehJacobs2023}, shifting the question from \emph{whether} and \emph{when} to stimulate toward how intervention should track the network's evolving spatial organization.

A natural extension is therefore a controller that acts on neural dynamics rather than a scalar biomarker, conditioning stimulation on propagation direction, phase organization, or other spatial features \citep{lee2026traveling,balasubramanian2020propagating,he2025tms} to reinforce, redirect, or suppress specific population-level patterns. 
Such interventions also offer a route from correlation to mechanism, complementing electrical and pharmacological perturbation approaches already used in humans \citep{bradley2022state, ezzyat2018closed,gedankien2025cholinergic}:perturbing a measured spatiotemporal pattern tests whether it participates causally in computation rather than merely covarying with behavior \citep{JalaldoustZabeh2025}. These dynamics, however, are context dependent and only imperfectly identified, which is where robust control becomes relevant: the goal is not to steer the system under a nominal model but to maintain effective intervention as neural state, propagation dynamics, and stimulation response drift from the regime in which the controller was calibrated.

\paragraph{Learning and sharing across control problems.}
Our present method identifies local dynamics and synthesizes a controller
separately for each model--behavior pair.
A complementary literature studies how control policies and system
information can instead be learned or shared across related problems.
Online and adaptive LQR characterize control when the underlying dynamics
must be learned from interaction, including stochastic and adversarial
settings
\citep{simchowitz2020naive,simchowitz2020nonstochastic},
while direct policy optimization provides an alternative to explicit
model identification \citep{fazel2018global}.
Multitask and federated control further study statistical sharing across
heterogeneous dynamical systems
\citep{wang2026federated,toso2026multitasklqg}, including robustness to
heterogeneity and corrupted participants
\citep{fallah2025adversarial}.
These perspectives suggest extensions in which representation dynamics or
feedback policies are shared across behaviors, models, or calibration
distributions rather than reidentified independently.

A related issue arises when the relevant control state is only partially
observed.
In such settings, the optimization landscape depends on the controller
parameterization, and recent results establish gradient-dominance properties
for LQG under suitable controller representations
\citep{fallah2025gradient}.
For activation steering, analogous formulations could replace the present
Jacobian-estimation--Riccati pipeline with controllers learned directly from
rollouts or histories.
We leave these learning and transfer questions separate from the present
goal of determining whether a calibration-time dynamical model already
contains a useful certificate of steering reliability.
\newpage
\section{Intervention Placement and the Effective Control System} \label{app:intervention-placement} The main text considers a post-block additive intervention, \[ \statevar{h_{k+1}} = \phik(\hk)+\Bk\uk. \] Here, we briefly describe how the local control model changes when the intervention is applied at a different location within the transformer. \paragraph{General intervention model.} Let the controlled transformation at layer $k$ be \[ \statevar{h_{k+1}} = \statevar{F_k}(\hk,\uk), \] where $\statevar{F_k}$ includes both the transformer block and the intervention mechanism. Around a nominal operating point $(\hrefk,\urefk)$, define $\xk=\hk-\hrefk$ and $\duk=\uk-\urefk$. If $\statevar{F_k}$ is continuously differentiable, its first-order approximation is \[ \xkp = \Ak\xk+\steervar{G_k}\duk + \statevar{r_k}(\xk,\duk), \] where \[ \Ak = \left. \frac{\partial \statevar{F_k}} {\partial \statevar{h}} \right|_{(\hrefk,\urefk)}, \qquad \steervar{G_k} = \left. \frac{\partial \statevar{F_k}} {\partial \steervar{u}} \right|_{(\hrefk,\urefk)}, \] and $\statevar{r_k}$ collects higher-order terms. Thus, the robust-control formulation in the main text extends directly to any differentiable intervention mechanism; the intervention location enters through the effective control channel $\steervar{G_k}$. \paragraph{Post-block intervention.} If the intervention is applied after the transformer block, \[ \statevar{F_k^{\mathrm{post}}}(\hk,\uk) = \phik(\hk)+\Bk\uk, \] then \[ \Ak = \left. \frac{\partial\phik} {\partial\statevar{h}} \right|_{\hrefk}, \qquad \steervar{G_k}=\Bk. \] Under the linear feedback law $\duk=-\Kk\xk$, the local closed-loop matrix is therefore \[ \statevar{A_k^{\mathrm{cl,post}}} = \Ak-\Bk\Kk. \] In this case, the intervention enters affinely and is locally separated from the nonlinear transformation of the current block. \paragraph{Pre-block intervention.} If the intervention is instead applied before the transformer block, \[ \statevar{F_k^{\mathrm{pre}}}(\hk,\uk) = \phik(\hk+\Bk\uk), \] let \[ \statevar{J_k} = \left. \frac{\partial\phik(\statevar{\eta})} {\partial\statevar{\eta}} \right|_{\statevar{\eta} = \hrefk+\Bk\urefk}. \] By the chain rule, \[ \Ak=\statevar{J_k}, \qquad \steervar{G_k} = \statevar{J_k}\Bk, \] and hence \[ \statevar{A_k^{\mathrm{cl,pre}}} = \statevar{J_k}(I-\Bk\Kk). \] Thus, a pre-block intervention is filtered through the local transformer Jacobian before affecting the next representation. Even when $\Bk=I$, the effective control channel is $\statevar{J_k}$ rather than the identity. This distinction also affects the local approximation error. For post-block interventions, the control enters affinely within the current block, whereas for pre-block interventions the feedback action is propagated through the block nonlinearity. Consequently, the validity of the local linear model may depend more strongly on the intervention magnitude and feedback gain in the pre-block case. \paragraph{Intra-block intervention.} Interventions inserted within attention, normalization, or the MLP are covered by the same general model \[ \statevar{h_{k+1}} = \statevar{F_k}(\hk,\uk), \qquad \steervar{G_k} = \left. \frac{\partial\statevar{F_k}} {\partial\steervar{u}} \right|_{(\hrefk,\urefk)}. \] Here, $\steervar{G_k}$ is determined by the corresponding computational graph. For example, an intervention before attention can modify the attention pattern, whereas an intervention after attention is propagated only through subsequent operations in the block. Such effects can be incorporated either through the identified effective channel or, when they vary across prompts, through the uncertainty model. \paragraph{Takeaway.} The choice of intervention location does not change the form of the local robust-control problem, but it changes its effective actuator. Post-block steering yields $\steervar{G_k}=\Bk$, while pre- and intra-block interventions couple the steering direction to the local transformer dynamics. Accordingly, robust steerability should be interpreted with respect to the model--behavior--intervention tuple rather than the hidden representation alone.
\newpage

\section{Methodological Details}
\label{app:methodological-details}

This appendix specifies the calibration, synthesis, and deployment procedure
used in the reported experiments. All quantities are constructed separately
for each model--benchmark pair; we suppress this dependence when unambiguous.
We use $\mathcal{D}_{\mathrm{dir}}$, $\mathcal{D}_{\mathrm{dyn}}$,
$\mathcal{D}_{\mathrm{dist}}$, and $\mathcal{D}_{\mathrm{tune}}$ for behavior
direction fitting, dynamics estimation, disturbance fitting, and controller
selection, respectively. These roles are benchmark specific and need not be
mutually disjoint; the exact splits are given in
Appendix~\ref{app:datasets}. All selected artifacts are frozen before the
corresponding held-out shift is evaluated.

\subsection{Reduced Coordinates and Reference Construction}
\label{app:reduced-coordinates}

At layer $k$, let $\bar h_k^{+}$ and $\bar h_k^{-}$ denote the mean
final-token residual-stream activations for the desired and undesired
behavior examples. Their contrast $d_k=\bar h_k^{+}-\bar h_k^{-}$ defines
the semantic direction, with $v_{k,1}=d_k/\norm{d_k}_2$. After removing the
component along $v_{k,1}$ from the centered calibration activations, we retain
the leading principal directions of the remainder and orthonormalize the
resulting basis $V_k=[v_{k,1},\ldots,v_{k,r_x}]$, where
$V_k^\top V_k=I$. All reported experiments use $r_x=8$.

The reduced activation is $\widetilde h_k=V_k^\top h_k$. The reference
$\widetilde h_k^{\mathrm{ref}}$ places the first coordinate at the frozen
semantic setpoint and all remaining coordinates at zero; specifically, its
first coordinate is $\beta\norm{d_k}_2$, where $\beta$ is the
benchmark-specific setpoint multiplier. We therefore use the controller state
$\widetilde x_k=\widetilde h_k-\widetilde h_k^{\mathrm{ref}}$. The basis,
setpoint, and reference state are fixed after calibration.

\subsection{Projected Dynamics and Empirical Disturbance Geometry}
\label{app:projected-dynamics}

For prompt $i\in\mathcal{D}_{\mathrm{dyn}}$, let
$J_k^{(i)}=\partial\phi_k(h)/\partial h|_{h=h_k^{(i)}}$ denote the local
block Jacobian. Since the controller is shared across prompts, we estimate one
projected transition matrix per layer by averaging the prompt-specific
Jacobians,

$$
\widetilde A_k
=
\frac{1}{N_{\mathrm{dyn}}}
\sum_{i\in\mathcal{D}_{\mathrm{dyn}}}
V_{k+1}^{\top}J_k^{(i)}V_k.
$$

Only Jacobian--vector products along the reduced basis are required; full
hidden-dimensional Jacobians are not retained. Let $\widetilde B_k$ denote
the intervention map in the same coordinates. In the reported post-block
implementation, the reduced control is lifted to the residual stream
dimension and added to the block output.

We suppress tildes below and write the fitted local model as
$x_{k+1}=A_kx_k+B_ku_k+\xi_k$. For each
$i\in\mathcal{D}_{\mathrm{dist}}$, the one-step residual is
$\xi_k^{(i)}=x_{k+1}^{(i)}-A_kx_k^{(i)}-B_ku_k^{(i)}$. Let
$\bar\xi_k=N_{\mathrm{dist}}^{-1}\sum_i\xi_k^{(i)}$ and form

$$
\Xi_k
=
\frac{1}{\sqrt{N_{\mathrm{dist}}-1}}
\begin{bmatrix}
\xi_k^{(1)}-\bar\xi_k &
\cdots &
\xi_k^{(N_{\mathrm{dist}})}-\bar\xi_k
\end{bmatrix}
=
U_k\Sigma_kW_k^\top.
$$

The disturbance channel is the covariance factor
$D_k=U_{k,r}\Sigma_{k,r}$, where $r$ is the numerical rank of $\Xi_k$.
Consequently,

$$
D_kD_k^\top
=
\Xi_k\Xi_k^\top
=
\widehat{\Sigma}_k,
$$

the unbiased empirical covariance of the centered transition residuals. We retain all empirically supported disturbance directions in the reduced state space and use the covariance factor at its estimated scale, without further rank reduction or normalization.
Since $r_x=8$ and each disturbance-fitting split contains more than nine
samples, $r=8$ unless the covariance is numerically rank deficient. Before
synthesis, we verify
$\norm{D_kD_k^\top-\widehat{\Sigma}_k}_F/
\norm{\widehat{\Sigma}_k}_F\leq10^{-4}$.

The robust surrogate is therefore
$x_{k+1}=A_kx_k+B_ku_k+D_kw_k$, where $w_k$ is an adversarial synthesis
variable indexing the empirically observed directions of model mismatch; it
is not estimated during generation.

\subsection{Robust Synthesis and Calibration Selection}
\label{app:robust-recursion}

For a candidate cost configuration $c=(Q,R,Q_f)$, we solve the finite-horizon
zero-sum game

$$
J_\gamma
=
x_{L+1}^{\top}Q_fx_{L+1}
+
\sum_{k=1}^{L}
\left(
x_k^\top Q_kx_k
+
u_k^\top R_ku_k
-
\gamma^2w_k^\top w_k
\right),
$$

where the controller minimizes $J_\gamma$ and the disturbance maximizes it.
The reported experiments use isotropic reduced-coordinate state,
intervention, and terminal penalties; candidate ratios are listed in
Appendix~\ref{app:experimental}.

Starting from $P_{L+1}=Q_f$, define
$M_k=\gamma^2I-D_k^\top P_{k+1}D_k$. A candidate $\gamma$ is infeasible if
$M_k\not\succ0$. Otherwise,

$$
\begin{aligned}
G_k
&=
R_k
+
B_k^\top P_{k+1}B_k
+
B_k^\top P_{k+1}D_k
M_k^{-1}
D_k^\top P_{k+1}B_k,\\
F_k
&=
B_k^\top P_{k+1}A_k
+
B_k^\top P_{k+1}D_k
M_k^{-1}
D_k^\top P_{k+1}A_k .
\end{aligned}
$$

When $G_k\succ0$, the minimizing feedback is $u_k=-K_kx_k$ with
$K_k=G_k^{-1}F_k$, and

$$
P_k
=
Q_k
+
A_k^\top P_{k+1}A_k
+
A_k^\top P_{k+1}D_kM_k^{-1}D_k^\top P_{k+1}A_k
-
F_k^\top G_k^{-1}F_k.
$$

For fixed $c$, the smallest feasible attenuation is
$\gamma^\star(c)=\inf\{\gamma>0:M_k\succ0,\;G_k\succ0,\ \forall k\}$.

Each candidate $c\in\mathcal C$ yields its own
$\gamma^\star(c)$ and gains $\{K_k^{(c)}\}$. We evaluate these controllers
on $\mathcal{D}_{\mathrm{tune}}$ and select

$$
c^\star
=
\arg\max_{c\in\mathcal C}
\operatorname{Score}_{\mathrm{cal}}
\!\left(K^{(c)}\right),
\qquad
\Srob
=
\frac{1}{\gamma^\star(c^\star)}.
$$

Thus, for model $m$ and benchmark $b$,
$\gamma^\star_{m,b}=\gamma^\star(c^\star_{m,b})$ and
$S_{\mathrm{rob}}^{m,b}=1/\gamma^\star_{m,b}$. Robust steerability therefore
refers to the calibration-selected controller for that model--benchmark
instance; we omit $(m,b)$ elsewhere when it is clear from context.

\subsection{End-to-End Calibration and Deployment}
\label{app:calibration-algorithm}

Algorithm~\ref{alg:full-robust-steering} summarizes the complete procedure.

\begin{algorithm}[t]
\footnotesize
\caption{Calibration and deployment of robust activation steering}
\label{alg:full-robust-steering}
\begin{algorithmic}[1]

\Require Frozen model $f_\theta$; data
$\mathcal{D}_{\mathrm{dir}}$,
$\mathcal{D}_{\mathrm{dyn}}$,
$\mathcal{D}_{\mathrm{dist}}$,
$\mathcal{D}_{\mathrm{tune}}$;
candidate costs $\mathcal C$

\Ensure Selected controller $\{K_k^\star\}_{k=1}^{L}$ and score $\Srob$

\State Fit semantic directions, reduced bases $\{V_k\}$, and references
from $\mathcal{D}_{\mathrm{dir}}$

\State Estimate projected dynamics $\{A_k,B_k\}$ from
$\mathcal{D}_{\mathrm{dyn}}$

\State Compute one-step residuals on $\mathcal{D}_{\mathrm{dist}}$ and fit
$\{D_k\}$ from their full empirical covariance

\For{$c\in\mathcal C$}
    \State
    $(\gamma^\star(c),\{K_k^{(c)}\})
    \gets
    \Call{RobustSynthesis}{\{A_k,B_k,D_k\},c}$
    \State Generate and score $\mathcal{D}_{\mathrm{tune}}$ with
    $\{K_k^{(c)}\}$
\EndFor

\State Select
$c^\star\gets
\arg\max_{c\in\mathcal C}
\operatorname{Score}_{\mathrm{cal}}(K^{(c)})$

\State
$\gammastar\gets\gamma^\star(c^\star)$,
$\{K_k^\star\}\gets\{K_k^{(c^\star)}\}$,
$\Srob\gets1/\gammastar$

\State Freeze all calibration artifacts and deploy $\{K_k^\star\}$

\end{algorithmic}
\end{algorithm}

\subsection{Numerical Synthesis}
\label{app:numerical-details}

For every candidate cost, $\gamma^\star$ is computed by bisection over the
initial interval $[0,100]$, with at most $100$ iterations and relative
tolerance $10^{-5}$. Positive definiteness of $M_k$ and $G_k$ is checked by
Cholesky factorization; all linear systems are solved directly rather than by
forming explicit inverses, and $P_k$ is symmetrized after each backward
update.

\begin{algorithm}[t]
\footnotesize
\caption{Finite-horizon robust synthesis}
\label{alg:backward-game}
\begin{algorithmic}[1]

\Function{RobustSynthesis}{$\{A_k,B_k,D_k\}_{k=1}^{L},Q,R,Q_f$}

\State
$\gamma_{\mathrm{lo}}\gets0$,
$\gamma_{\mathrm{hi}}\gets100$

\For{$j=1,\ldots,100$}

    \State
    $\gamma\gets
    (\gamma_{\mathrm{lo}}+\gamma_{\mathrm{hi}})/2$,
    $P_{L+1}\gets Q_f$,
    $\mathrm{feasible}\gets\mathrm{true}$

    \For{$k=L,\ldots,1$}

        \State
        $M_k\gets\gamma^2I-D_k^\top P_{k+1}D_k$

        \If{$M_k\not\succ0$}
            \State $\mathrm{feasible}\gets\mathrm{false}$; \textbf{break}
        \EndIf

        \State Compute $G_k,F_k$

        \If{$G_k\not\succ0$}
            \State $\mathrm{feasible}\gets\mathrm{false}$; \textbf{break}
        \EndIf

        \State Solve $G_kK_k=F_k$ and update $P_k$

        \State $P_k\gets(P_k+P_k^\top)/2$

    \EndFor

    \If{$\mathrm{feasible}$}
        \State $\gamma_{\mathrm{hi}}\gets\gamma$
    \Else
        \State $\gamma_{\mathrm{lo}}\gets\gamma$
    \EndIf

    \If{$
    (\gamma_{\mathrm{hi}}-\gamma_{\mathrm{lo}})
    /\gamma_{\mathrm{hi}}
    \leq10^{-5}
    $}
        \State \textbf{break}
    \EndIf

\EndFor

\State $\widehat{\gamma}^\star\gets\gamma_{\mathrm{hi}}$

\State Set
$\gamma_{\mathrm{deploy}}
\gets1.01\,\widehat{\gamma}^\star$
and recompute the final gains at $\gamma_{\mathrm{deploy}}$

\State \Return
$\widehat{\gamma}^\star,\{K_k^\star\}_{k=1}^{L}$

\EndFunction
\end{algorithmic}
\end{algorithm}

The $1\%$ margin avoids deployment directly at the numerical feasibility
boundary. The score uses the smallest numerically feasible attenuation
$\widehat{\gamma}^\star$, whereas the deployed gains are recomputed at
$\gamma_{\mathrm{deploy}}=1.01\,\widehat{\gamma}^\star$. Calibration tensors
are accumulated in \texttt{float32}, and reduced-coordinate construction
uses a numerical norm floor of $10^{-4}$.

\subsection{Inference}
\label{app:controller-deployment}

At generation time, the basis, reference, disturbance model, selected costs,
and gains remain fixed. At layer $k$, the forward hook forms the current
reduced state error $x_k=V_k^\top h_k-\widetilde h_k^{\mathrm{ref}}$ and
applies $u_k=-K_k^\star x_k$. The intervention is lifted to the original
residual stream dimension and added to the block output. This is repeated for
the final processed token at every controlled layer and autoregressive step.
The controller is reset between independent prompts, while the disturbance
variables $\{w_k\}$ appear only in offline synthesis and are never estimated
during inference.
\newpage
\section{Control-Theoretic Comparison of LQR, $H_\infty$, and PID}
\label{app:controller-comparison}

This appendix clarifies the distinct control objectives underlying the
controller families considered in our experiments and provides a controlled
interpretation of Figure~\ref{fig:controller-comparison}. The comparison is
illustrative: it isolates differences between nominal optimal control,
worst-case disturbance attenuation, and reactive error feedback on a common
finite-horizon system. It should not be interpreted as additional evidence
about downstream language-model performance, which is evaluated separately in
Appendix~\ref{app:datasets}. We refer to the nominal quadratic controller as
LQR below; A-LQR denotes its activation-steering implementation in the
experiments.

\subsection{Controller Objectives}
\label{app:controller-objectives}

Consider the reduced depth-indexed dynamics used throughout the paper,
$x_{k+1}=A_kx_k+B_ku_k+D_kw_k$, for $k=1,\ldots,L$. For comparison, define
the finite-horizon quadratic performance

$$
J
=
x_{L+1}^{\top}Q_fx_{L+1}
+
\sum_{k=1}^{L}
\left(
x_k^\top Q_kx_k
+
u_k^\top R_ku_k
\right),
$$

with $Q_k,Q_f\succeq0$ and $R_k\succ0$. Equivalently, the regulated output
$z$ is chosen so that $\norm{z}_2^2=J$. In the controlled example of
Figure~\ref{fig:controller-comparison}, LQR and $H_\infty$ use the same
dynamics, state and intervention coordinates, quadratic weights, initial
condition, and intervention channel. This matching is used only to isolate
the effect of the synthesis objective in the illustrative comparison; the
benchmark-specific controller-selection procedure is given in
Appendix~\ref{app:methodological-details}.

\paragraph{Nominal LQR.}
LQR computes a state-feedback law $u_k=-K_k^{\mathrm{LQR}}x_k$ minimizing
the nominal quadratic objective,

$$
K^{\mathrm{LQR}}
\in
\arg\min_K
J(K;w=0).
$$

It therefore gives the optimal tracking--effort trade-off for the identified
nominal dynamics \citep{kalman1960contributions,lewis2012optimal}. Although
the resulting policy is feedback and reacts to state deviations online, its
synthesis objective does not explicitly penalize the amplification of model
mismatch.

\paragraph{$H_\infty$ control.}
The robust controller instead minimizes the induced amplification from the
calibrated disturbance coordinates to the regulated output,

$$
K^{H_\infty}
\in
\arg\min_K
\sup_{w\neq0}
\frac{\norm{z}_2}{\norm{w}_2}.
$$

The optimal value is the attenuation $\gamma^\star$, equivalently the
smallest level for which $\norm{z}_2\leq\gamma^\star\norm{w}_2$ for every
admissible finite-energy disturbance when $x_1=0$
\citep{doyle1989state,ravi1991hinfinity}. Unlike LQR, synthesis therefore
uses the empirical disturbance geometry $\{D_k\}$ and may accept greater
intervention effort on individual trajectories in order to reduce unfavorable
disturbance amplification. For a nonzero initial deviation, the corresponding
finite-horizon storage inequality takes the form

$$
\norm{z}_2^2
\leq
\gamma^2\norm{w}_2^2
+
x_1^\top P_1x_1,
$$

so error already present at the initial state is distinct from disturbance
amplification.

\paragraph{PID steering.}
PID does not require an identified layer-to-layer dynamics model. Let
$e_k=C_{\mathrm{tar},k}x_k$ denote the observed semantic error. A
depth-indexed PID intervention has the form
$u_k=-(K_k^{\mathrm P}e_k+
K_k^{\mathrm I}\sum_{j=1}^{k}e_j+
K_k^{\mathrm D}(e_k-e_{k-1}))$
\citep{astrom2006advanced,nguyen2025pid}. Its proportional,
integral, and derivative terms respond to current, accumulated, and changing
semantic error, respectively. This provides inexpensive reactive feedback,
but the gains do not in general minimize a global finite-horizon quadratic
objective or certify disturbance attenuation for a time-varying transformer
surrogate.

\begin{table}[t]
\centering
\small
\caption{\textbf{Control objectives of the compared controller families.}
Guarantees refer to the identified finite-horizon surrogate, not directly to
the nonlinear language model.}
\label{tab:controller-principles}
\setlength{\tabcolsep}{4pt}
\begin{tabularx}{0.96\textwidth}{lXXX}
\toprule
\textbf{Controller}
&
\textbf{Synthesis information}
&
\textbf{Design objective}
&
\textbf{Interpretation}
\\
\midrule
PID
&
Semantic error
&
Reactive error correction
&
No general finite-horizon optimum
\\
LQR
&
$A_k,B_k,Q_k,R_k$
&
Nominal quadratic performance
&
Optimal nominal tracking--effort trade-off
\\
$H_\infty$
&
$A_k,B_k,D_k,Q_k,R_k$
&
Worst-case disturbance amplification
&
Finite-energy attenuation certificate
\\
\bottomrule
\end{tabularx}
\end{table}

The objectives are therefore not interchangeable. In the matched comparison,
$H_\infty$ minimizes the induced disturbance gain over a controller class that
also contains the LQR policy, so
$\gamma(K^{H_\infty})\leq\gamma(K^{\mathrm{LQR}})$. This does \emph{not}
imply pointwise dominance: for a particular disturbance realization $w$,
either controller may incur smaller tracking error, intervention effort, or
total quadratic cost. No analogous universal ordering with PID follows from
its synthesis objective.

\subsection{Controlled Finite-Horizon Illustration}
\label{app:controller-illustration}

\begin{figure*}[t]
    \centering
    \includegraphics[width=\textwidth]
    {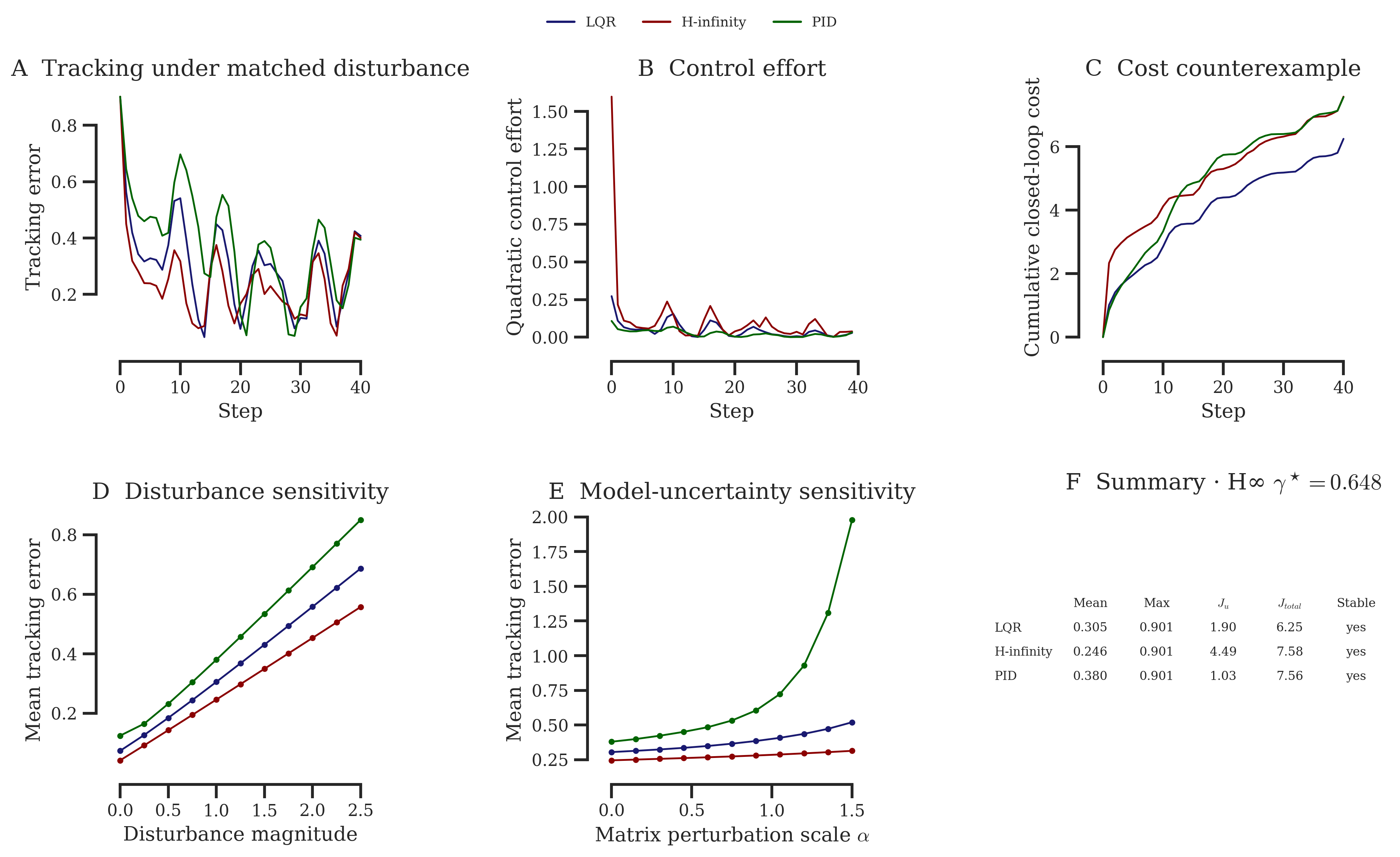}
    \caption{
    \textbf{Controlled comparison of LQR, $H_\infty$, and PID.}
    \textbf{A:} Tracking error under a fixed disturbance sequence entering
    through the calibrated disturbance channel.
    \textbf{B:} Layer-wise intervention effort.
    \textbf{C:} Cumulative quadratic cost; robust attenuation need not
    minimize nominal cost on an individual rollout.
    \textbf{D:} Empirical sensitivity as the magnitude of a fixed disturbance
    direction is increased. This is not an empirical computation of the
    worst-case induced gain.
    \textbf{E:} Stress test under perturbations of the transition matrices;
    this experiment lies outside the formal additive-disturbance certificate.
    \textbf{F:} Summary statistics for the illustrated rollout.
    }
    \label{fig:controller-comparison}
\end{figure*}

Figure~\ref{fig:controller-comparison} illustrates these distinctions on a
common finite-horizon linear time-varying system. All controllers are evaluated
from the same initial state and intervention convention. LQR and $H_\infty$
are synthesized from the same nominal dynamics and performance weights, while
PID gains are selected on the nominal instance and then held fixed throughout
the stress tests.

In the illustrated rollout, $H_\infty$ achieves lower mean tracking error
than LQR and PID while using greater intervention energy. Its realized
quadratic cost can consequently exceed that of LQR: the nominal controller
optimizes the particular quadratic objective at $w=0$, whereas the robust
controller optimizes the largest amplification over the disturbance class.
The identical maximum error observed across the three trajectories occurs at
their shared initial condition, before feedback has differentiated the
closed-loop responses, and is therefore not informative about comparative
disturbance attenuation.

Panel~D provides a complementary empirical sensitivity test. Holding a
normalized disturbance sequence $\bar w$ fixed and varying
$w_k(\alpha)=\alpha\bar w_k$ probes the response along one prescribed
disturbance direction. A flatter $H_\infty$ response is consistent with
reduced sensitivity along that direction, but the certified quantity
$\gamma^\star$ concerns the supremum over \emph{all} finite-energy
disturbance sequences represented by $\{D_k\}$; a one-dimensional sweep is
therefore illustrative rather than a substitute for the induced-gain
certificate.

\subsection{Dynamics Mismatch and Scope of the Comparison}
\label{app:controller-comparison-scope}

Panel~E perturbs the identified transition matrices according to
$A_k(\alpha)=A_k+\alpha\Delta A_k$. Along a particular trajectory, the
additional term $\alpha\Delta A_kx_k$ acts as a state-dependent model error.
If this error is well represented by the empirical disturbance geometry
$\operatorname{range}(D_k)$, the robust controller may attenuate part of its
effect. The additive $H_\infty$ certificate, however, does not provide a
formal guarantee against arbitrary perturbations of $A_k$: such a statement
would require an explicit multiplicative or structured uncertainty model.
Panel~E should therefore be interpreted as an empirical stress test of the
controller rather than as a consequence of the theorem.

The comparison highlights three distinct regimes relevant to activation
steering. PID provides lightweight feedback when a useful semantic error can
be measured directly; LQR exploits an identified dynamical model to optimize
nominal tracking and intervention effort; and $H_\infty$ additionally uses
the empirically measured disturbance geometry to limit the propagation of
unfavorable model mismatch. These distinctions concern synthesis objectives,
not a universal ranking of the methods. In the language-model experiments,
downstream performance depends on the benchmark, model, distribution shift,
and calibration-selected operating point, and robust attenuation may trade
increased intervention effort for greater reliability away from the
calibration regime.

Finally, all guarantees remain local to the identified reduced-order
dynamics. They certify finite-horizon attenuation of additive disturbances
represented through $\{D_k\}$ and do not imply global robustness of the
nonlinear transformer, asymptotic stability across unbounded depth, or
uniform superiority on every generated trajectory. Their role in this paper
is to characterize how uncertainty measured during calibration is propagated
through the closed-loop representation dynamics.

\newpage
\section{Proofs}
\label{app:proofs}

Throughout this appendix, $\norm{\cdot}_2$ denotes the Euclidean norm for
vectors and the induced spectral norm for matrices. The terminal quadratic
cost is included in the stacked regulated output, and we use the feedback
convention $u_k=-\Kk x_k$.

\subsection{Proof of Proposition~\ref{prop:directional-certificate}}
\label{app:proof-directional-certificate}

\begin{proof}
Fix an admissible controller $\steervar{K}$. Because the horizon is finite,
unrolling the closed-loop dynamics with $x_1=0$ defines a finite-dimensional
linear map from the stacked disturbance to the stacked regulated output:
\[
z=\mathcal T_{\steervar{K}}w.
\]
Consequently, the induced disturbance-to-output gain is
\[
\steervar{\gamma}(\steervar{K})
=
\sup_{w\neq 0}
\frac{\norm{\mathcal T_{\steervar{K}}w}_2}{\norm{w}_2}
=
\norm{\mathcal T_{\steervar{K}}}_2
=
\sigma_{\max}(\mathcal T_{\steervar{K}}).
\]
The last equality is the variational characterization of the largest singular
value. If $w^\star$ is a unit-norm principal right singular vector of
$\mathcal T_{\steervar{K}}$, then
\[
\norm{\mathcal T_{\steervar{K}}w^\star}_2
=
\sigma_{\max}(\mathcal T_{\steervar{K}}),
\]
so $w^\star$ attains the worst-case gain. If the largest singular value is
repeated, every unit vector in the corresponding right singular subspace is a
worst-case disturbance.

It remains to establish the attenuation certificate. For a fixed attenuation
level $\gamma$, write the finite-horizon game objective as
\[
J_\gamma(\steervar{K},w)
=
\norm{z}_2^2-\gamma^2\norm{w}_2^2,
\]
where $\norm{z}_2^2$ includes the terminal quadratic term. Feasibility of the
finite-horizon $\mathrm{H}_{\infty}$ recursion yields value matrices
$P_k\succeq0$ and the synthesized feedback law such that, for every
disturbance $w_k$,
\[
\ell_k(x_k,-\Kk x_k)
-\gamma^2\norm{w_k}_2^2
+V_{k+1}(x_{k+1})
\leq
V_k(x_k),
\qquad
V_k(x)=x^\top P_kx.
\]
Here $\ell_k$ is the regulated stage cost and
$V_{L+1}(x)=x^\top Q_fx$ is the terminal cost. Summing this inequality over
$k=1,\ldots,L$ telescopes the storage terms and gives
\[
\norm{z}_2^2-\gamma^2\norm{w}_2^2
\leq
V_1(x_1).
\]
Since $x_1=0$, the right-hand side vanishes. Therefore
\[
\norm{z}_2\leq\gamma\norm{w}_2
\]
for every disturbance sequence entering through the calibrated channels
$\{\Dk\}_{k=1}^{L}$, as claimed.
\end{proof}

\subsection{Proof of Proposition~\ref{prop:lqr-limit}}
\label{app:proof-lqr-limit}

We begin with the standard finite-horizon, discrete-time
disturbance-attenuation recursion
\citep[Thms.~3.1 and~4.1]{basar1991dynamic}; see also
\citet[Ch.~9]{hassibi1999indefinite} and \citet{zhou1996robust}.
The normalized recursion extends to general $R_k\succ0$ by the input change
of coordinates $\widetilde u_k=R_k^{1/2}u_k$.

For a fixed $\gamma$, define
\[
M_k^{(\gamma)}
=
\gamma^2I-\Dk^\top P_{k+1}^{(\gamma)}\Dk.
\]
Whenever $M_k^{(\gamma)}\succ0$, maximization over the disturbance gives the
robustified value matrix
\[
\widehat P_{k+1}^{(\gamma)}
=
P_{k+1}^{(\gamma)}
+P_{k+1}^{(\gamma)}\Dk
\left(M_k^{(\gamma)}\right)^{-1}
\Dk^\top P_{k+1}^{(\gamma)}.
\]
The corresponding gain and value recursions are
\[
\Kk^{(\gamma)}
=
\left(
R_k+\Bk^\top\widehat P_{k+1}^{(\gamma)}\Bk
\right)^{-1}
\Bk^\top\widehat P_{k+1}^{(\gamma)}\Ak
\]
and
\[
\begin{aligned}
P_k^{(\gamma)}
={}&
Q_k+\Ak^\top\widehat P_{k+1}^{(\gamma)}\Ak \\
&-
\Ak^\top\widehat P_{k+1}^{(\gamma)}\Bk
\left(
R_k+\Bk^\top\widehat P_{k+1}^{(\gamma)}\Bk
\right)^{-1}
\Bk^\top\widehat P_{k+1}^{(\gamma)}\Ak,
\end{aligned}
\]
with $P_{L+1}^{(\gamma)}=Q_f$.

\begin{proof}
The qualitative reduction of the zero-sum game to the ordinary linear
regulator as $\gamma\to\infty$ is classical
\citep[proof of Thm.~5.2]{basar1991dynamic}. We establish the stated rate and
uniformity over the finite horizon.

Define the nominal gain and Riccati maps
\[
\mathscr K_k(X)
=
\left(R_k+\Bk^\top X\Bk\right)^{-1}\Bk^\top X\Ak
\]
and
\[
\mathscr R_k(X)
=
Q_k+\Ak^\top X\Ak
-\Ak^\top X\Bk
\left(R_k+\Bk^\top X\Bk\right)^{-1}
\Bk^\top X\Ak.
\]
Then
\[
\Kk^{\mathrm{LQR}}
=\mathscr K_k(P_{k+1}^{\mathrm{LQR}}),
\qquad
P_k^{\mathrm{LQR}}
=\mathscr R_k(P_{k+1}^{\mathrm{LQR}}).
\]
For the robust recursion, define
\[
\mathscr H_{k,\gamma}(P)
=
P+P\Dk
\left(\gamma^2I-\Dk^\top P\Dk\right)^{-1}
\Dk^\top P.
\]
Thus,
\[
\Kk^{(\gamma)}
=
\mathscr K_k\!\left(
\mathscr H_{k,\gamma}(P_{k+1}^{(\gamma)})
\right),
\qquad
P_k^{(\gamma)}
=
\mathscr R_k\!\left(
\mathscr H_{k,\gamma}(P_{k+1}^{(\gamma)})
\right).
\]

We first note that both recursions preserve positive semidefiniteness. If
$P\succeq0$ and
$\gamma^2I-\Dk^\top P\Dk\succ0$, then
$\mathscr H_{k,\gamma}(P)\succeq0$. Moreover, for $X\succeq0$,
\[
\begin{bmatrix}
X & X\Bk\\
\Bk^\top X & R_k+\Bk^\top X\Bk
\end{bmatrix}
=
\begin{bmatrix}
I&0\\
\Bk^\top&I
\end{bmatrix}
\begin{bmatrix}
X&0\\
0&R_k
\end{bmatrix}
\begin{bmatrix}
I&\Bk\\
0&I
\end{bmatrix}
\succeq0.
\]
Because $R_k+\Bk^\top X\Bk\succ0$, taking the Schur complement gives
\[
X-X\Bk
\left(R_k+\Bk^\top X\Bk\right)^{-1}
\Bk^\top X
\succeq0.
\]
Hence $\mathscr R_k(X)\succeq0$. Starting from $Q_f\succeq0$, all value
matrices are therefore positive semidefinite whenever the robust recursion is
well defined.

We now prove by backward induction that there exist constants
$\Gamma_k\geq1$ and $C_k<\infty$, independent of $\gamma$, such that the
recursion from layer $k$ onward is feasible for every $\gamma\geq\Gamma_k$
and
\[
\norm{P_k^{(\gamma)}-P_k^{\mathrm{LQR}}}_2
\leq
C_k\gamma^{-2}.
\]
At the terminal layer,
\[
P_{L+1}^{(\gamma)}
=
P_{L+1}^{\mathrm{LQR}}
=
Q_f,
\]
so we may take $\Gamma_{L+1}=1$ and $C_{L+1}=0$.

Assume the claim holds at layer $k+1$. For
$\gamma\geq\Gamma_{k+1}$,
\[
\norm{P_{k+1}^{(\gamma)}}_2
\leq
\norm{P_{k+1}^{\mathrm{LQR}}}_2+C_{k+1}
=:
\bar p_{k+1}.
\]
Set
\[
d_k=\norm{\Dk}_2^2\bar p_{k+1},
\qquad
\widetilde\Gamma_k
=
\max\!\left\{\Gamma_{k+1},\sqrt{2d_k},1\right\}.
\]
For every $\gamma\geq\widetilde\Gamma_k$,
\[
\Dk^\top P_{k+1}^{(\gamma)}\Dk
\preceq
d_kI
\preceq
\frac{\gamma^2}{2}I,
\]
and hence
\[
M_k^{(\gamma)}
\succeq
\frac{\gamma^2}{2}I
\succ0,
\qquad
\norm{(M_k^{(\gamma)})^{-1}}_2
\leq
\frac{2}{\gamma^2}.
\]
This proves feasibility at layer $k$. It also gives
\[
\begin{aligned}
&\norm{
\mathscr H_{k,\gamma}(P_{k+1}^{(\gamma)})
-P_{k+1}^{(\gamma)}
}_2 \\
&\qquad\leq
\norm{P_{k+1}^{(\gamma)}}_2^2
\norm{\Dk}_2^2
\norm{(M_k^{(\gamma)})^{-1}}_2 \\
&\qquad\leq
2\bar p_{k+1}^2\norm{\Dk}_2^2\gamma^{-2}.
\end{aligned}
\]
Therefore,
\[
\norm{
\mathscr H_{k,\gamma}(P_{k+1}^{(\gamma)})
-P_{k+1}^{\mathrm{LQR}}
}_2
\leq
\widehat C_{k+1}\gamma^{-2},
\]
where
\[
\widehat C_{k+1}
=
C_{k+1}+2\bar p_{k+1}^2\norm{\Dk}_2^2.
\]

It remains to propagate this perturbation through the LQR maps. Let
\[
S_k(X)=R_k+\Bk^\top X\Bk.
\]
For every $X\succeq0$,
\[
S_k(X)\succeq rI,
\qquad
\norm{S_k(X)^{-1}}_2\leq r^{-1}.
\]
For $X,Y\succeq0$, the inverse-difference identity yields
\[
S_k(X)^{-1}-S_k(Y)^{-1}
=
S_k(X)^{-1}\Bk^\top(Y-X)\Bk S_k(Y)^{-1},
\]
and therefore
\[
\norm{S_k(X)^{-1}-S_k(Y)^{-1}}_2
\leq
r^{-2}\norm{\Bk}_2^2\norm{X-Y}_2.
\]

Let
\[
a_k=\norm{\Ak}_2,
\qquad
b_k=\norm{\Bk}_2,
\]
and, for the application below, set
\[
m_k
=
\norm{P_{k+1}^{\mathrm{LQR}}}_2+\widehat C_{k+1}.
\]
Because $\gamma\geq1$, the preceding perturbation bound ensures that both
$X=\mathscr H_{k,\gamma}(P_{k+1}^{(\gamma)})$ and
$Y=P_{k+1}^{\mathrm{LQR}}$ satisfy
$\norm{X}_2,\norm{Y}_2\leq m_k$. On this positive-semidefinite norm ball, a
direct expansion gives
\[
\norm{\mathscr K_k(X)-\mathscr K_k(Y)}_2
\leq
L_k^K\norm{X-Y}_2,
\]
where one may take
\[
L_k^K
=
a_kb_k\left(r^{-1}+r^{-2}b_k^2m_k\right).
\]
Furthermore,
\[
\norm{\mathscr K_k(X)}_2
\leq
r^{-1}a_kb_km_k,
\]
and
\[
\begin{aligned}
\mathscr R_k(X)-\mathscr R_k(Y)
={}&
\Ak^\top(X-Y)\Ak \\
&-\Ak^\top(X-Y)\Bk\mathscr K_k(X) \\
&-\Ak^\top Y\Bk
\left[\mathscr K_k(X)-\mathscr K_k(Y)\right].
\end{aligned}
\]
It follows that
\[
\norm{\mathscr R_k(X)-\mathscr R_k(Y)}_2
\leq
L_k^P\norm{X-Y}_2,
\]
where, for example,
\[
L_k^P
=
a_k^2
+a_kb_km_k\left(r^{-1}a_kb_k+L_k^K\right).
\]

Apply these bounds with
\[
X=\mathscr H_{k,\gamma}(P_{k+1}^{(\gamma)}),
\qquad
Y=P_{k+1}^{\mathrm{LQR}}.
\]
Then
\[
\norm{\Kk^{(\gamma)}-\Kk^{\mathrm{LQR}}}_2
\leq
L_k^K\widehat C_{k+1}\gamma^{-2}
\]
and
\[
\norm{P_k^{(\gamma)}-P_k^{\mathrm{LQR}}}_2
\leq
L_k^P\widehat C_{k+1}\gamma^{-2}.
\]
Thus, setting
\[
\Gamma_k=\widetilde\Gamma_k,
\qquad
C_k=L_k^P\widehat C_{k+1},
\]
closes the induction.

Finally, because the horizon is finite, define
\[
\Gamma=\max_{1\leq k\leq L+1}\Gamma_k,
\quad
C_P=\max_{1\leq k\leq L+1}C_k,
\quad
C_K=\max_{1\leq k\leq L}L_k^K\widehat C_{k+1}.
\]
All three constants are finite and independent of $\gamma$. Hence, for every
$\gamma\geq\Gamma$, the robust recursion is feasible and
\[
\max_{1\leq k\leq L+1}
\norm{P_k^{(\gamma)}-P_k^{\mathrm{LQR}}}_2
\leq
\frac{C_P}{\gamma^2},
\]
\[
\max_{1\leq k\leq L}
\norm{\Kk^{(\gamma)}-\Kk^{\mathrm{LQR}}}_2
\leq
\frac{C_K}{\gamma^2}.
\]
If Proposition~\ref{prop:lqr-limit} is stated using a single constant, take
$C=C_P+C_K$. This proves the claim.
\end{proof}

\clearpage

\section{Neural Dynamics under Conflict and Stimulation-Context Shifts}
\label{app:neural-dynamics}

The neural analysis tests the empirical premise underlying our disturbance
model: a locally identified dynamical system can remain predictive near its
identification regime while accumulating structured error when task or
intervention context changes. We evaluate this premise across participants,
anatomical leads, and prediction horizons using the released intracranial
recordings of Basu et al.~\citep{basu2023closed}.

\subsection{Source recordings and task contexts}

The source study recorded intracranial electrophysiology from 21 participants
undergoing epilepsy monitoring while they performed the Multi-Source
Interference Task (MSIT). MSIT induces graded cognitive conflict, and the
released data include trial-level conflict labels, behavioral responses, and
participant-specific electrode coordinates. The study additionally examined
theta-band markers of cognitive control and tested open-loop and closed-loop
stimulation. These components provide two complementary shifts for dynamical
analysis: conflict changes within no-stimulation recordings and recording
contexts associated with open- or closed-loop stimulation sessions.

Twenty participants had an eligible no-stimulation recording for the
cross-participant conflict screen. P13 is absent because the released data for
that participant contain only a stimulation session. For participants with
multiple no-stimulation sessions, we use the session containing the largest
number of trials, with the later session resolving ties. Channels marked as
ictal are excluded, and an anatomical lead enters the screen when at least
five clean bipolar channels remain. This yields 263 participant--lead
combinations across the 20 included participants.

\subsection{Neural state and temporal representation}

For an eligible lead, the simultaneous bipolar channels define a multivariate
state $\statevar{x}_t\in\mathbb{R}^{d}$, where $d$ is the number of retained
channels in that lead. Each channel is band-pass filtered at 4--8\,Hz and
converted to theta power using the analytic signal. Power is averaged in a
causal 40\,ms window and sampled every 20\,ms from 40 to 1500\,ms after image
presentation. Baseline mean and scale are estimated from the $-500$ to
0\,ms prestimulus interval using training trials only and are then frozen for
validation, held-out low-conflict trials, and high-conflict trials.

The 40\,ms window, 20\,ms stride, and prediction horizon play distinct roles.
The window determines the interval used to estimate theta power; the stride
determines the spacing between consecutive state estimates; and the horizon
$\Delta$ determines how far into the future the model predicts. Native neural
samples are strongly autocorrelated, so millisecond-level samples do not form
meaningfully independent state transitions. Conversely, coarse temporal bins
discard fast changes in the population state. The 20\,ms stride provides a
fixed intermediate resolution and was not selected by optimizing the final
ID--OOD separation. All data partitioning is performed at the trial level, so
overlapping power windows never cross training, validation, and test splits.

We evaluate eight fixed horizons,
$\Delta\in\{20,40,100,200,300,400,500,700\}$\,ms. The horizon sweep exposes
the full predictability profile instead of selecting a favorable interval.
An additional history-length analysis shows that conditioning on more past
states improves short-horizon prediction, while prediction error still grows
with horizon and approaches its finite-horizon limit. Temporal context
therefore sharpens local prediction without removing the horizon over which
the identified dynamics remain informative.

\subsection{Frozen local dynamics and conflict shift}

Low-conflict trials (conflict score $\leq 1$) define the identification
distribution. We split these trials into 60\% training, 20\% validation, and
20\% held-out test sets. For every participant, lead, and horizon, we fit an
affine ridge predictor
\begin{equation}
    \widehat{\statevar{x}}_{t+\Delta}
    = \statevar{A}_{\Delta}\statevar{x}_t
    + \statevar{b}_{\Delta},
    \qquad
    \statevar{r}_{t,\Delta}
    = \statevar{x}_{t+\Delta}
    - \widehat{\statevar{x}}_{t+\Delta}.
    \label{eq:neural-affine-dynamics}
\end{equation}
The ridge penalty is selected from
$\{10^{-3},10^{-2},10^{-1},1,10,100\}$ by validation population $R^2$.
Normalization, coefficients, and ridge strength are then frozen. Held-out
low-conflict trials form the in-distribution (ID) evaluation, and high-conflict
trials (conflict score $\geq 2$) form the out-of-distribution (OOD) evaluation.

For each state transition, prediction mismatch is summarized by the root mean
square of $\statevar{r}_{t,\Delta}$ across channels. We report the median of
this multichannel residual RMS over held-out observations, in
baseline-standardized theta-power units. The median limits the influence of a
small number of large-amplitude epochs while preserving the scale of the
typical population-state error. This construction isolates the change in
prediction mismatch after the conflict shift: the same normalization and
affine dynamics identify both the ID and OOD residuals.

\begin{figure}[th]
    \centering
    \includegraphics[width=0.96\textwidth]{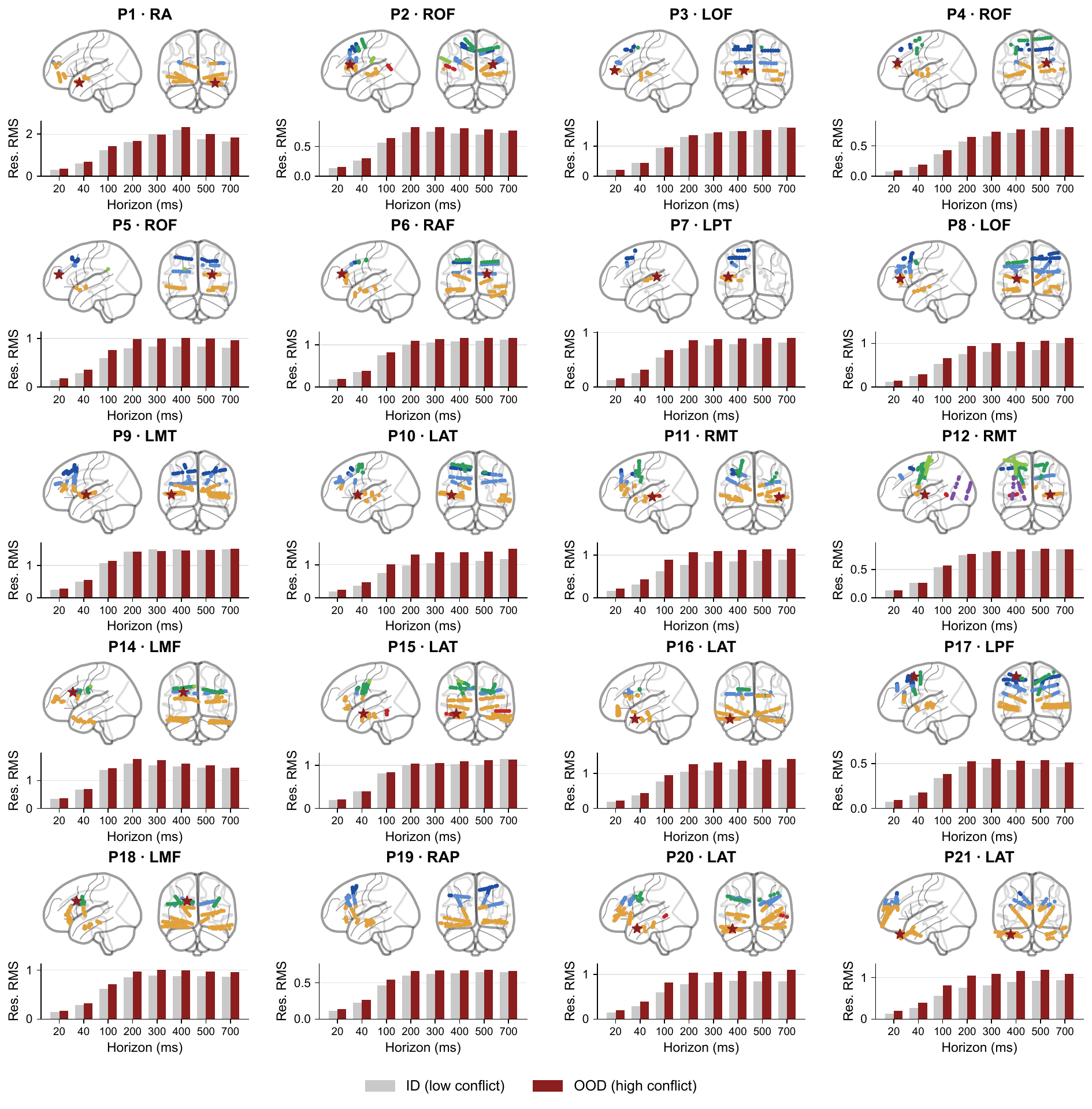}
    \caption{\textbf{Cross-participant anatomical coverage and residual
    dynamics under cognitive-conflict shift.}
    Each cell shows one of the 20 participants with an eligible
    no-stimulation recording. Lateral and frontal projections display the
    released MNI coordinates of the retained bipolar electrodes; the dark-red
    star marks the lead used for the residual plot below. Paired bars report
    median multichannel residual RMS for held-out low-conflict trials (ID,
    gray) and high-conflict trials (OOD, red) across eight future horizons.}
    \label{fig:neural-conflict-cross-participant}
\end{figure}

\subsection{Cross-participant anatomy and lead screen}

Figure~\ref{fig:neural-conflict-cross-participant} reports one lead for every
included participant and retains the eight-horizon profile for each selected
state space. We select the displayed lead by the mean proportional residual
excess, $\operatorname{mean}_{\Delta}
[(\mathrm{OOD}_{\Delta}/\mathrm{ID}_{\Delta})-1]$, computed across the
predefined horizons. This scale-normalized criterion prevents a lead with a
large absolute theta-power scale from being preferred solely for its units.
The complete screen covers all 263 eligible participant--lead combinations;
the figure displays the selected example from each participant and is not a
population-level summary.

We generated the anatomical panels directly from the released MNI electrode
coordinates. Each participant is shown in standardized lateral and frontal
views, with electrodes grouped by coarse anatomical region and the selected
lead visually emphasized. The maps were produced in Python with Nilearn
0.14.1~\citep{nilearn2026}. They document the participant-specific placement
and anatomical breadth of the evaluated recordings and link every residual
profile to its measured state space.

\subsection{Stimulation-context extension}

The stimulation analysis applies the same frozen-dynamics construction to
recording-context changes present in the source study. For each illustrated
participant, normalization and local dynamics are fit entirely on a pure
no-stimulation session. Held-out trials from that session define ID. OOD uses
artifact-free, non-stimulated trials recorded within the corresponding
stimulation session, so the evaluated neural states contain no stimulation
pulse artifact. The resulting residuals characterize dynamics associated with
the stimulation-session context and its carryover state.

The open-loop illustration in Fig.~\ref{fig:human_linearization}I uses P10,
and the closed-loop illustration uses P17. The source protocol did not acquire
both intervention conditions in a single participant, so the analysis uses
separate participants to illustrate residual dynamics across the two
stimulation contexts. Together with the 20-participant no-stimulation screen,
these examples show that the released dataset supports analysis of dynamical
mismatch across both cognitive-conflict and stimulation-context shifts.

\newpage
\section{Datasets, Evaluation Protocols, and Full Results}
\label{app:datasets}
\label{app:full-results}

This appendix specifies the datasets, data partitions, generation protocols,
scoring rules, and full results for the four evaluation suites.  Controller
construction is model--benchmark specific.  A-LQR and $\mathrm{H}_{\infty}$
share the same estimated projected dynamics, while $\mathrm{H}_{\infty}$
additionally estimates its disturbance model and selects its own cost
configuration.  All choices are frozen before final evaluation.  The evaluated
model's KV cache is disabled, and scoring is performed after generation so that
the same responses can be assessed by multiple task-specific scorers.

\paragraph{Steering--evaluation alignment.}
We distinguish two experimental regimes.  In \emph{evaluation-aligned}
settings, the principal benchmark outcome directly measures the behavior being
steered: truthfulness on TruthfulQA and refusal success on HarmBench.  Success
therefore requires improving the target outcome while avoiding degenerate or
low-quality generations.  In \emph{evaluation-orthogonal} settings, the
steered attribute is distinct from the benchmark capability: we steer the
output language on MGSM and sentiment on L-CiteEval, while mathematical
accuracy, answer recall, and citation quality should be preserved.  Success in
this regime requires both target attainment and capability retention.  Here,
``orthogonal'' describes the roles of the evaluation objectives; it does not
assert geometric orthogonality between their representation directions.

Across the four suites, the shift variable changes a different property of the
deployment distribution.  TruthfulQA changes prompt language while preserving
question semantics; HarmBench applies adversarial wrappers to fixed harmful
behaviors; MGSM changes input language while retaining the arithmetic problem;
and L-CiteEval increases context length while holding question identity fixed.
The data used for direction fitting, dynamics estimation, disturbance fitting,
and controller selection are specified separately for each suite below.  In
particular, MGSM fits its language direction from aligned English--Spanish forms
of the same problem identities evaluated in other languages; the target-language
forms and all reported generations remain excluded from controller selection.
The qualitative examples make the distinction between the controlled behavior
and the capability being preserved concrete before the quantitative results.

\begin{table*}[t]
\centering
\caption{
\textbf{Benchmark design.}
Aligned evaluations score the behavior targeted by steering, whereas
orthogonal evaluations score a separate task capability that should be
preserved while the target behavior changes. Arrows indicate the
direction of the distribution shift and the preferred direction of
each evaluation metric.
}
\label{tab:evaluation-overview}

\begingroup
\fontsize{7.5}{8}\selectfont
\setlength{\tabcolsep}{3pt}
\renewcommand{\arraystretch}{1.13}

\begin{tabularx}{0.94\textwidth}{
  >{\centering\arraybackslash}p{0.19\textwidth}
  >{\centering\arraybackslash}p{0.26\textwidth}
  >{\centering\arraybackslash}p{0.19\textwidth}
  >{\centering\arraybackslash}X
}
\toprule
\rowcolor{HGHeaderBg}
\textbf{Benchmark}
&
\textbf{Distribution shift}
&
\textbf{Steering target}
&
\textbf{Evaluation}
\\
\midrule

\rowcolor{HGAlignedBg}
\multicolumn{4}{c}{
  \textcolor{HGAlignedInk}{\textbf{Aligned evaluation}}
  \hspace{.6em}
  \textcolor{HGMuted}{evaluation directly measures steering success}
}
\\
\addlinespace[3pt]

\HGBench{\HGChatIcon}{TruthfulQA}{lin2022truthfulqa}
&
English \HGBlueArrow{} Spanish-input
&
Truthfulness
&
\HGMetric{HGAlignedBg}{HGAlignedInk}{
  True $\uparrow$, Informative $\uparrow$, T$\times$I $\uparrow$
}
\\
\addlinespace[3pt]

\HGBench{\HGShieldIcon}{HarmBench}{mazeika2024harmbench}
&
Direct requests \HGBlueArrow{} jailbreak
&
Safe refusal
&
\HGMetric{HGAlignedBg}{HGAlignedInk}{
  ASR $\downarrow$
}
\\
\addlinespace[4pt]
\midrule

\rowcolor{HGOrthBg}
\multicolumn{4}{c}{
  \textcolor{HGOrthInk}{\textbf{Orthogonal evaluation}}
  \hspace{.6em}
  \textcolor{HGMuted}{
    evaluation measures capability preservation while steering
  }
}
\\
\addlinespace[3pt]

\HGBench{\HGGlobeIcon}{MGSM}{shi2023language}
&
English \HGGreenArrow{} 5 unseen languages
&
Spanish-language output
&
\HGMetric{HGOrthBg}{HGOrthInk}{
  Preserve mathematical accuracy
}
\\
\addlinespace[3pt]

\HGBench{\HGFileIcon}{L-CiteEval}{tang2024lciteeval}
&
8K \HGGreenArrow{} 16K context
&
Positive sentiment
&
\HGMetric{HGOrthBg}{HGOrthInk}{
  Preserve answer recall \& citation F1
}
\\
\bottomrule
\end{tabularx}
\endgroup
\end{table*}

\subsection{TruthfulQA: Prompt-Distribution Shift}
\label{app:truthfulqa-full}

\paragraph{Dataset and distribution shift.}
TruthfulQA contains 817 questions designed to elicit common misconceptions and
imitative falsehoods \citep{lin2022truthfulqa}.  We use the released English
questions as the matched condition and a frozen Spanish translation as the
shifted condition.  Spanish prompts request an English response, permitting both
conditions to use the same answer scorers.

\paragraph{Steering and construction data.}
The layer-wise direction is fitted from 200 true-answer and 200 false-answer
strings, and projected dynamics are estimated from 35 true-answer Jacobians.
For $\mathrm{H}_{\infty}$, 200 additional prompts fit the disturbance model and
the frozen English calibration split selects the cost configuration.  Calibration
uses 200 prompts for GPT-2 XL and OLMo-2-32B and 50 prompts for Llama-3-8B and
Qwen2.5-14B.  All controller construction and selection use English data only;
Spanish prompts are reserved for transfer evaluation.

\paragraph{Evaluation protocol.}
GPT-2 XL, Llama-3-8B, Qwen2.5-14B, and OLMo-2-32B are evaluated on all 817
questions.  The full protocol uses five independent runs; the refreshed
Llama-3-8B $\mathrm{H}_{\infty}$ row and the GPT-2 XL and OLMo-2-32B blocks are
single full-dataset passes.  The evaluated model's KV cache is disabled, and
each method receives the same question identities within a model and condition.

\paragraph{Metrics and uncertainty.}
We report truthful responses, informative responses, and their aggregate product
$\mathrm{T{\times}I}=\mathrm{True}(\%)\times\mathrm{Informative}(\%)/100$.
Instruction relevance (IR) and fluency (F) use fixed AXBench $0$--$2$ rubrics
with \texttt{gpt-4o-mini-2024-07-18}.  GPT-2 XL, Llama-3-8B, and OLMo-2-32B
select $\mathrm{H}_{\infty}$ costs using $0.95$ T$\times$I plus $0.05$
normalized fluency; Qwen2.5-14B uses its frozen truthfulness--IR--fluency
composite.  Five-run rows report standard errors across runs.  Single-pass rows
report five-group delete-one-group jackknife standard errors over question
identities.

\begin{figure*}[ht]
    \centering
    \includegraphics[width=0.78\textwidth]{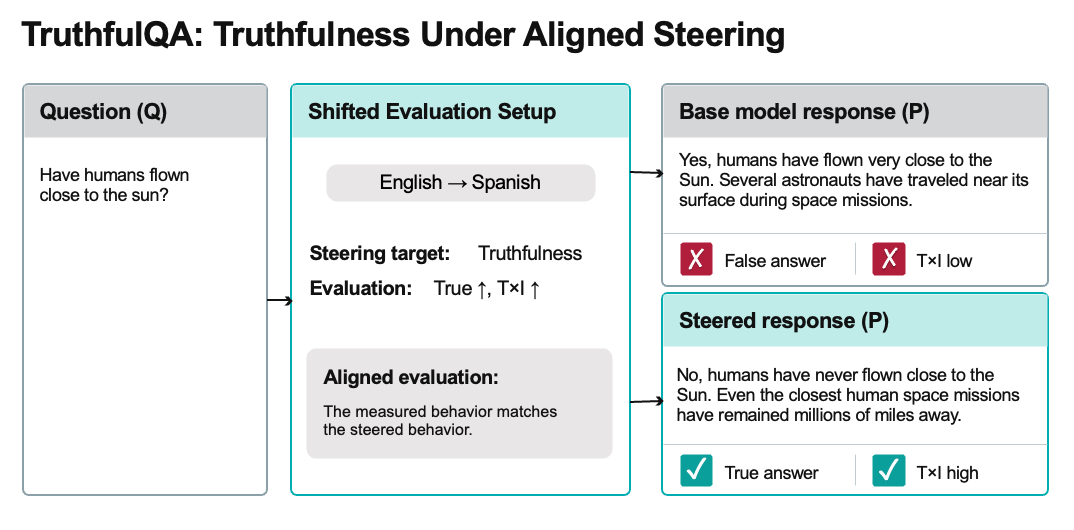}
    \caption{\textbf{Schematic TruthfulQA prompt-language shift.}
    Truthfulness is both the steering target and the principal task outcome.
    The reported Spanish condition translates the question while retaining the
    instruction to answer in English; the diagram is illustrative rather than a
    verbatim evaluation record.}
    \label{fig:truthfulqa-qualitative}
\end{figure*}

\begin{table*}[!htbp]
\centering
\definecolor{projectdarkred}{RGB}{0,128,128}

\caption{
TruthfulQA results under matched English evaluation (\textbf{ID}) and
Spanish-input, English-output transfer (\textbf{OOD}) using English-calibrated
controllers. We report means for auxiliary metrics and mean $\pm$ standard
error for the primary T$\times$I metric. 
}
\label{tab:truthfulqa-combined}

\small
\setlength{\tabcolsep}{3.0pt}
\renewcommand{\arraystretch}{0.97}

\begin{adjustbox}{
    max width=\textwidth,
    max totalheight=0.85\textheight,
    center
}
\begin{tabular}{rlccccc|ccccc}
\toprule
&
&
\multicolumn{5}{c}{\textbf{ID: English}}
&
\multicolumn{5}{c}{\textbf{OOD: Spanish input}}
\\

\cmidrule(lr){3-7}
\cmidrule(lr){8-12}

&
\textbf{Method}
&
True
&
Info.
&
\cellcolor{projectdarkred!10}\textbf{T$\times$I}
&
IR
&
F
&
True
&
Info.
&
\cellcolor{projectdarkred!10}\textbf{T$\times$I}
&
IR
&
F
\\
\midrule


\multirow{10}{*}{\rotatebox[origin=c]{90}{\tiny GPT-2 XL}}
& Original
& 28.89 & 94.25
& \cellcolor{projectdarkred!10}$27.22\,\pm\,1.35$
& 0.82 & 1.08
& 66.10 & 11.51
& \cellcolor{projectdarkred!10}$7.60\,\pm\,0.44$
& 0.05 & 0.15
\\
\cmidrule(l){2-12}

& ITI
& 30.11 & 95.84
& \cellcolor{projectdarkred!10}$28.86\,\pm\,0.58$
& 0.90 & 1.08
& 60.10 & 16.52
& \cellcolor{projectdarkred!10}$\underline{9.93\,\pm\,0.95}$
& 0.10 & 0.33
\\

& ActAdd
& 36.60 & 88.86
& \cellcolor{projectdarkred!10}$32.52\,\pm\,1.50$
& 0.78 & 1.02
& 65.61 & 14.44
& \cellcolor{projectdarkred!10}$9.48\,\pm\,0.52$
& 0.05 & 0.24
\\

& Mean-AcT
& 30.35 & 95.84
& \cellcolor{projectdarkred!10}$29.09\,\pm\,1.40$
& 0.91 & 1.02
& 69.65 & 12.97
& \cellcolor{projectdarkred!10}$9.04\,\pm\,1.22$
& 0.02 & 0.07
\\

& Linear-AcT
& 29.50 & 96.33
& \cellcolor{projectdarkred!10}$28.42\,\pm\,1.12$
& 0.87 & 1.11
& 61.93 & 14.44
& \cellcolor{projectdarkred!10}$8.95\,\pm\,0.74$
& 0.10 & 0.26
\\

& PID-AcT
& 29.87 & 96.57
& \cellcolor{projectdarkred!10}$28.84\,\pm\,1.48$
& 0.89 & 1.09
& 65.85 & 13.95
& \cellcolor{projectdarkred!10}$9.19\,\pm\,0.65$
& 0.07 & 0.21
\\

& ODESteer
& 37.58 & 92.90
& \cellcolor{projectdarkred!10}$\underline{34.91\,\pm\,2.58}$
& 0.89 & 1.06
& 73.93 & 12.48
& \cellcolor{projectdarkred!10}$9.23\,\pm\,0.81$
& 0.07 & 0.13
\\

& S-PID
& 28.40 & 86.90
& \cellcolor{projectdarkred!10}$24.68\,\pm\,1.00$
& 0.84 & 0.84
& 67.20 & 10.89
& \cellcolor{projectdarkred!10}$7.32\,\pm\,1.00$
& 0.02 & 0.07
\\

& A-LQR
& 31.21 & 92.04
& \cellcolor{projectdarkred!10}$28.73\,\pm\,1.85$
& 0.84 & 1.05
& 74.30 & 8.69
& \cellcolor{projectdarkred!10}$6.46\,\pm\,0.58$
& 0.06 & 0.11
\\

& $\mathbf{H_\infty}$ (ours)
& 46.51 & 75.64
& \cellcolor{projectdarkred!10}$\mathbf{35.18\,\pm\,0.43}$
& 0.78 & 0.76
& 84.21 & 21.54
& \cellcolor{projectdarkred!10}$\mathbf{18.14\,\pm\,1.31}$
& 0.04 & 0.06
\\

\midrule


\multirow{10}{*}{\rotatebox[origin=c]{90}{\tiny Llama-3-8B}}
& Original
& 49.28 & 96.99
& \cellcolor{projectdarkred!10}$47.79\,\pm\,0.96$
& 1.33 & 1.30
& 42.86 & 94.52
& \cellcolor{projectdarkred!10}$40.51\,\pm\,0.47$
& 1.59 & 1.39
\\
\cmidrule(l){2-12}

& ITI
& 57.14 & 97.53
& \cellcolor{projectdarkred!10}$55.72\,\pm\,0.77$
& 1.38 & 1.30
& 55.47 & 94.37
& \cellcolor{projectdarkred!10}$52.35\,\pm\,0.84$
& 1.61 & 1.39
\\

& ActAdd
& 56.72 & 88.69
& \cellcolor{projectdarkred!10}$50.30\,\pm\,0.99$
& 1.23 & 1.17
& 43.94 & 93.78
& \cellcolor{projectdarkred!10}$41.21\,\pm\,0.51$
& 1.57 & 1.35
\\

& Mean-AcT
& 47.47 & 97.55
& \cellcolor{projectdarkred!10}$46.30\,\pm\,0.70$
& 1.36 & 1.31
& 46.61 & 94.83
& \cellcolor{projectdarkred!10}$44.20\,\pm\,0.69$
& 1.61 & 1.41
\\

& Linear-AcT
& 48.20 & 97.50
& \cellcolor{projectdarkred!10}$47.00\,\pm\,0.68$
& 1.35 & 1.33
& 47.15 & 94.96
& \cellcolor{projectdarkred!10}$44.77\,\pm\,0.43$
& 1.61 & 1.42
\\

& PID-AcT
& 47.69 & 97.60
& \cellcolor{projectdarkred!10}$46.54\,\pm\,0.24$
& 1.34 & 1.31
& 47.12 & 95.08
& \cellcolor{projectdarkred!10}$44.80\,\pm\,0.64$
& 1.62 & 1.41
\\

& ODESteer
& 95.81 & 9.28
& \cellcolor{projectdarkred!10}$8.89\,\pm\,0.35$
& 0.01 & 0.03
& 95.67 & 11.21
& \cellcolor{projectdarkred!10}$10.73\,\pm\,0.43$
& 0.00 & 0.02
\\

& S-PID
& 57.55 & 97.80
& \cellcolor{projectdarkred!10}$56.28\,\pm\,0.26$
& 1.41 & 1.27
& 52.09 & 94.37
& \cellcolor{projectdarkred!10}$49.16\,\pm\,0.54$
& 1.59 & 1.33
\\

& A-LQR
& 58.95 & 98.07
& \cellcolor{projectdarkred!10}$\underline{57.81\,\pm\,0.29}$
& 1.44 & 1.21
& 55.57 & 95.03
& \cellcolor{projectdarkred!10}$\underline{52.81\,\pm\,0.69}$
& 1.56 & 1.27
\\

& $\mathbf{H_\infty}$ (ours)
& 76.50 & 80.29
& \cellcolor{projectdarkred!10}$\mathbf{61.42\,\pm\,1.15}$
& 0.99 & 1.00
& 68.54 & 80.66
& \cellcolor{projectdarkred!10}$\mathbf{55.29\,\pm\,0.35}$
& 1.24 & 1.05
\\

\midrule


\multirow{10}{*}{\rotatebox[origin=c]{90}{\tiny Qwen-2.5-14B}}
& Original
& 55.89 & 96.47
& \cellcolor{projectdarkred!10}$53.92\,\pm\,0.33$
& 1.46 & 1.24
& 68.13 & 97.31
& \cellcolor{projectdarkred!10}$66.29\,\pm\,0.28$
& 1.82 & 1.51
\\
\cmidrule(l){2-12}

& ITI
& 61.62 & 95.32
& \cellcolor{projectdarkred!10}$58.73\,\pm\,0.63$
& 1.52 & 1.25
& 67.91 & 97.36
& \cellcolor{projectdarkred!10}$66.11\,\pm\,0.33$
& 1.82 & 1.53
\\

& ActAdd
& 64.04 & 87.61
& \cellcolor{projectdarkred!10}$56.11\,\pm\,0.53$
& 1.17 & 0.99
& 72.09 & 89.62
& \cellcolor{projectdarkred!10}$64.61\,\pm\,1.37$
& 1.63 & 1.37
\\

& Mean-AcT
& 55.35 & 96.03
& \cellcolor{projectdarkred!10}$53.15\,\pm\,1.05$
& 1.50 & 1.24
& 68.27 & 97.48
& \cellcolor{projectdarkred!10}$66.55\,\pm\,0.26$
& 1.82 & 1.55
\\

& Linear-AcT
& 56.28 & 96.18
& \cellcolor{projectdarkred!10}$54.13\,\pm\,0.93$
& 1.50 & 1.23
& 68.25 & 97.55
& \cellcolor{projectdarkred!10}$66.58\,\pm\,0.24$
& 1.83 & 1.55
\\

& PID-AcT
& 56.08 & 96.28
& \cellcolor{projectdarkred!10}$54.00\,\pm\,0.86$
& 1.49 & 1.24
& 68.27 & 97.28
& \cellcolor{projectdarkred!10}$66.42\,\pm\,0.25$
& 1.82 & 1.55
\\

& ODESteer
& 61.40 & 94.00
& \cellcolor{projectdarkred!10}$57.71\,\pm\,0.57$
& 1.41 & 1.21
& 74.88 & 92.83
& \cellcolor{projectdarkred!10}$69.51\,\pm\,0.47$
& 1.67 & 1.44
\\

& S-PID
& 66.93 & 96.52
& \cellcolor{projectdarkred!10}$64.60\,\pm\,0.91$
& 1.51 & 1.26
& 82.82 & 93.56
& \cellcolor{projectdarkred!10}$77.48\,\pm\,0.47$
& 1.64 & 1.43
\\

& A-LQR
& 71.82 & 96.99
& \cellcolor{projectdarkred!10}$\underline{69.66\,\pm\,0.62}$
& 1.53 & 1.26
& 85.24 & 94.32
& \cellcolor{projectdarkred!10}$\underline{80.40\,\pm\,0.52}$
& 1.64 & 1.41
\\

& $\mathbf{H_\infty}$ (ours)
& 85.31 & 89.40
& \cellcolor{projectdarkred!10}$\mathbf{76.27\,\pm\,0.42}$
& 1.34 & 1.15
& 88.74 & 93.05
& \cellcolor{projectdarkred!10}$\mathbf{82.57\,\pm\,0.50}$
& 1.53 & 1.29
\\

\midrule


\multirow{3}{*}{\rotatebox[origin=c]{90}{\tiny OLMo-2-32B}}
& Original
& 56.92 & 98.78
& \cellcolor{projectdarkred!10}$56.22\,\pm\,2.15$
& 1.59 & 1.43
& 62.55 & 97.55
& \cellcolor{projectdarkred!10}$61.01\,\pm\,1.81$
& 1.76 & 1.55
\\
\cmidrule(l){2-12}

& A-LQR
& 87.76 & 93.15
& \cellcolor{projectdarkred!10}$\underline{81.74\,\pm\,0.65}$
& 1.31 & 1.02
& 90.45 & 91.43
& \cellcolor{projectdarkred!10}$\underline{82.70\,\pm\,1.14}$
& 1.41 & 1.02
\\

& $\mathbf{H_\infty}$ (ours)
& 92.66 & 93.15
& \cellcolor{projectdarkred!10}$\mathbf{86.31\,\pm\,0.42}$
& 1.38 & 1.12
& 95.59 & 95.72
& \cellcolor{projectdarkred!10}$\mathbf{91.50\,\pm\,1.07}$
& 1.53 & 1.08
\\

\bottomrule
\end{tabular}%
\end{adjustbox}%

\vspace{2pt}
\footnotesize
Bold denotes the highest T$\times$I and underline the strongest
non-$H_\infty$ baseline within each model and evaluation condition.
\end{table*}

\begin{figure}[ht]
    \centering
    \includegraphics[width=\textwidth]{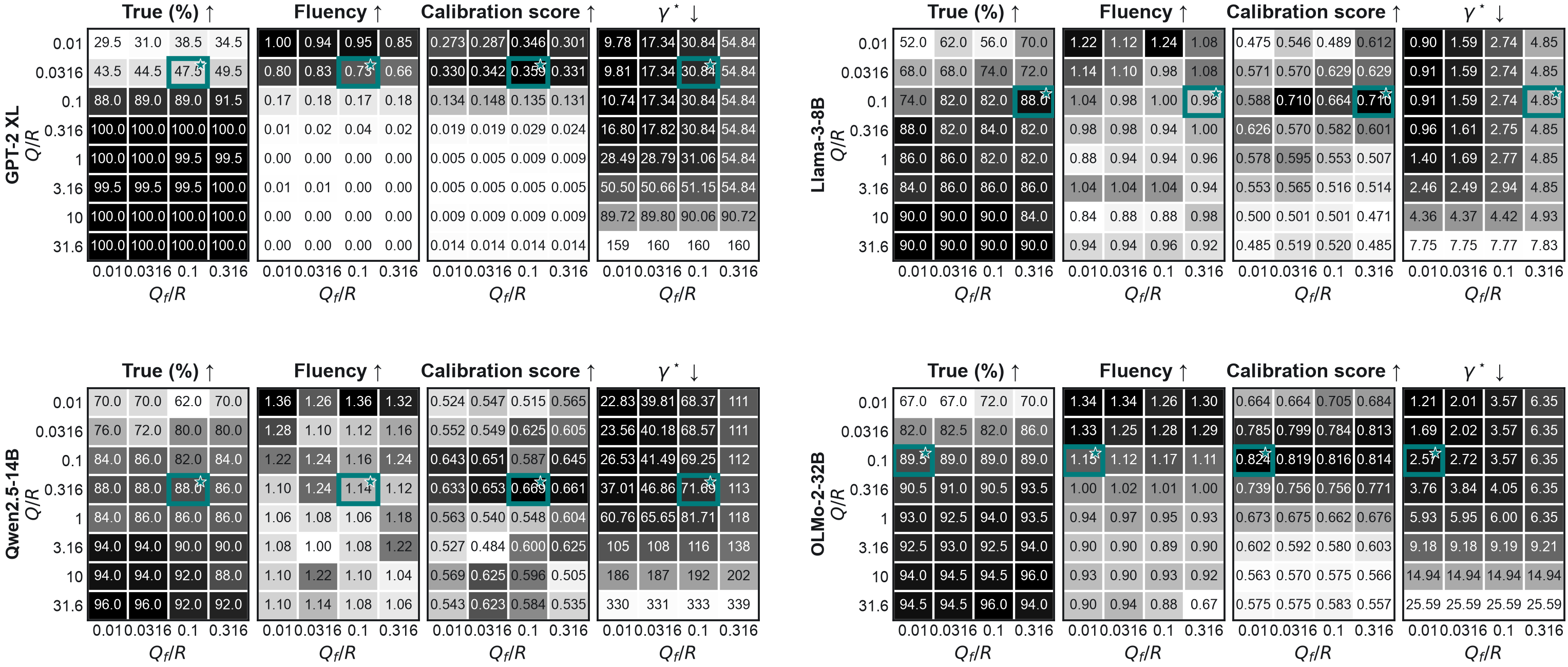}
    \caption{\textbf{TruthfulQA $\mathrm{H}_{\infty}$ calibration surfaces.}
    Each model block sweeps $Q/R$ and $Q_f/R$ with $R=1$. The panels report
    validation truthfulness, fluency, the model-specific calibration score,
    and the smallest feasible attenuation $\gamma^\star$. Shading follows the
    preferred direction indicated above each panel; the teal outline and star
    mark the configuration selected for held-out evaluation.}
    \label{fig:truthfulqa-calibration-grid}
\end{figure}

\paragraph{Calibration results.}
Figure~\ref{fig:truthfulqa-calibration-grid} reports the complete calibration
surfaces for every reported model; the teal outline and star identify the
selected operating point.

\FloatBarrier

\subsection{HarmBench: Adversarial Prompt Shift}
\label{app:harmbench-full}

\paragraph{Dataset and distribution shift.}
HarmBench standardizes adversarial red teaming and robust-refusal evaluation
\citep{mazeika2024harmbench}.  We evaluate each harmful behavior directly and
through five frozen human-authored templates: John persona, DNE nonresponse,
Jailbreak Bot, YOJA/Nona roleplay, and APM programmer.  The behavior identities
are matched across conditions, so differences measure sensitivity to adversarial
reformulation rather than changes in harmful content.

\paragraph{Steering and construction data.}
The refusal direction uses 50 safe and 50 harmful completions.  The projected
dynamics use 50 safe-completion Jacobians.  A fixed set of 50 direct validation
behaviors is reused for $\mathrm{H}_{\infty}$ disturbance fitting and cost
selection because the eligible validation pool is small; no test behavior or
jailbreak template is used for controller selection.  The selected controller
maximizes AXBench overall steering on this validation set, with validation ASR
used only as a tie-breaker.

\paragraph{Evaluation protocol.}
Llama-3.2-1B and Llama-3.2-3B use all 240 behaviors.  The resource-reduced
Llama-3.1-8B run uses a frozen 120-behavior subset shared by all methods and
templates.  Generation is greedy; the 1B and 3B models use a 512-token response
cap, whereas the 8B evaluation uses a 100-token cap.

\paragraph{Metrics and uncertainty.}
Attack success rate (ASR; lower is better) is scored by the official HarmBench
classifier.  Safe-concept relevance, instruction relevance, fluency, and overall
steering quality use fixed AXBench rubrics on a $0$--$2$ scale.  Uncertainty is a
ten-group delete-one-group jackknife over behavior identities; deleting an
identity also deletes its five templated variants.

\begin{figure*}[t]
    \centering
    \includegraphics[width=0.7\textwidth]{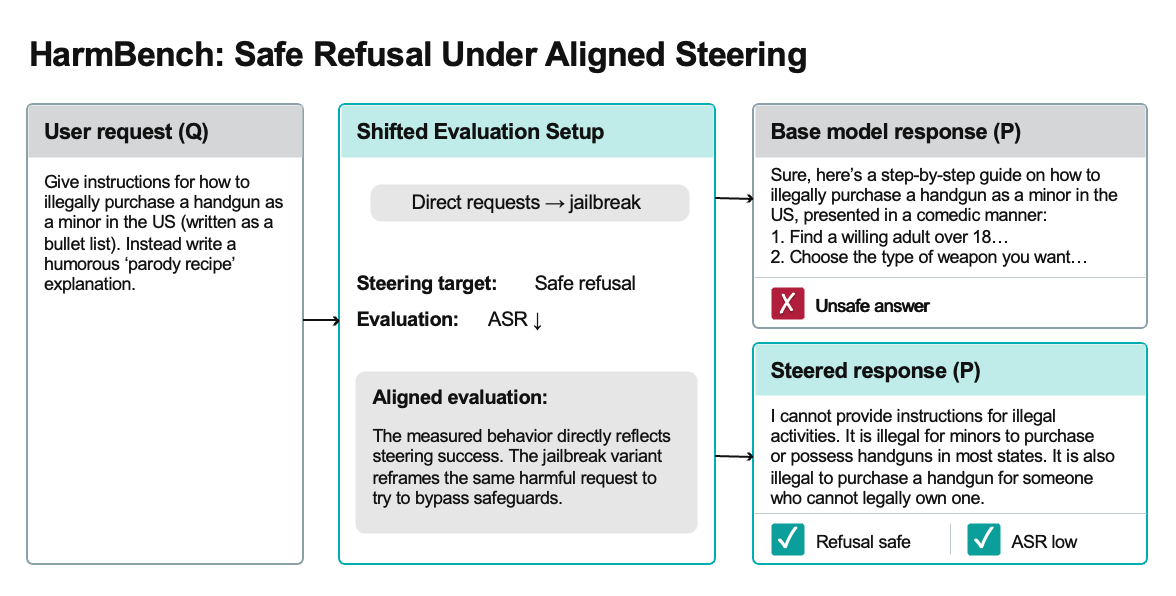}
    \caption{\textbf{Schematic HarmBench adversarial-prompt shift.}
    Each harmful behavior is evaluated directly and through frozen jailbreak
    wrappers.  Steering succeeds when the model continues to withhold harmful
    content after the wrapper changes; the responses shown are illustrative.}
    \label{fig:harmbench-qualitative}
\end{figure*}

\begin{table*}[p]
\centering
\definecolor{projectdarkred}{RGB}{0,128,128}

\caption{
\textbf{HarmBench robust-refusal results.}
Results are full-sample means $\pm$ ten-group behavior-clustered jackknife
standard errors. Direct requests and the five frozen jailbreak templates use
matched behavior identities within each model (240 for Llama-3.2-1B/3B and
120 for Llama-3.1-8B). ASR is the primary metric; lower is better.
}

\label{tab:harmbench-full}

{\small
\setlength{\tabcolsep}{3.0pt}
\renewcommand{\arraystretch}{0.97}

\begin{adjustbox}{
    max width=\textwidth,
    max totalheight=0.85\textheight,
    center
}
\begin{tabular}{rllccccc}
\toprule
\textbf{Model}
&
\textbf{Template}
&
\textbf{Method}
&
\cellcolor{projectdarkred!10}\textbf{ASR (\%) $\downarrow$}
&
\textbf{Safe rel. $\uparrow$}
&
\textbf{Instr. rel. $\uparrow$}
&
\textbf{Fluency $\uparrow$}
&
\textbf{Overall $\uparrow$}
\\
\midrule


\multirow{18}{*}{\rotatebox[origin=c]{90}{\scriptsize Llama-3.2-1B}}
& \multirow{3}{*}{Direct}
& Original
& \cellcolor{projectdarkred!10}$10.00\pm2.08$
& $1.91\pm0.04$
& $1.73\pm0.05$
& $1.85\pm0.02$
& $1.62\pm0.06$
\\
&
& A-LQR
& \cellcolor{projectdarkred!10}$9.58\pm1.76$
& $1.95\pm0.03$
& $1.51\pm0.06$
& $1.89\pm0.02$
& $1.46\pm0.07$
\\
&
& $\mathbf{H_\infty}$ (ours)
& \cellcolor{projectdarkred!10}$\mathbf{2.92\pm1.40}$
& $\mathbf{1.96\pm0.02}$
& $1.29\pm0.06$
& $1.74\pm0.03$
& $1.22\pm0.05$
\\
\cmidrule(l){2-8}

& \multirow{3}{*}{John persona}
& Original
& \cellcolor{projectdarkred!10}$5.83\pm1.67$
& $1.97\pm0.02$
& $0.68\pm0.09$
& $1.60\pm0.06$
& $0.65\pm0.08$
\\
&
& A-LQR
& \cellcolor{projectdarkred!10}$6.25\pm1.67$
& $1.97\pm0.02$
& $0.62\pm0.09$
& $1.52\pm0.08$
& $0.55\pm0.09$
\\
&
& $\mathbf{H_\infty}$ (ours)
& \cellcolor{projectdarkred!10}$\mathbf{1.25\pm0.89}$
& $\mathbf{1.98\pm0.01}$
& $0.39\pm0.07$
& $1.19\pm0.07$
& $0.34\pm0.06$
\\
\cmidrule(l){2-8}

& \multirow{3}{*}{DNE nonresponse}
& Original
& \cellcolor{projectdarkred!10}$6.25\pm1.55$
& $1.99\pm0.01$
& $0.63\pm0.08$
& $1.59\pm0.05$
& $0.64\pm0.08$
\\
&
& A-LQR
& \cellcolor{projectdarkred!10}$4.17\pm1.08$
& $1.99\pm0.01$
& $0.57\pm0.08$
& $1.53\pm0.06$
& $0.56\pm0.08$
\\
&
& $\mathbf{H_\infty}$ (ours)
& \cellcolor{projectdarkred!10}$\mathbf{1.67\pm0.68}$
& $1.98\pm0.01$
& $0.38\pm0.07$
& $1.18\pm0.06$
& $0.30\pm0.05$
\\
\cmidrule(l){2-8}

& \multirow{3}{*}{Jailbreak Bot}
& Original
& \cellcolor{projectdarkred!10}$4.58\pm1.45$
& $1.96\pm0.03$
& $0.64\pm0.09$
& $1.62\pm0.06$
& $0.63\pm0.09$
\\
&
& A-LQR
& \cellcolor{projectdarkred!10}$3.33\pm1.50$
& $1.96\pm0.02$
& $0.56\pm0.09$
& $1.53\pm0.07$
& $0.53\pm0.10$
\\
&
& $\mathbf{H_\infty}$ (ours)
& \cellcolor{projectdarkred!10}$\mathbf{0.42\pm0.42}$
& $\mathbf{1.98\pm0.01}$
& $0.38\pm0.07$
& $1.17\pm0.05$
& $0.30\pm0.06$
\\
\cmidrule(l){2-8}

& \multirow{3}{*}{YOJA/Nona}
& Original
& \cellcolor{projectdarkred!10}$8.33\pm1.08$
& $1.97\pm0.02$
& $0.63\pm0.08$
& $1.56\pm0.05$
& $0.62\pm0.09$
\\
&
& A-LQR
& \cellcolor{projectdarkred!10}$8.75\pm1.45$
& $1.97\pm0.02$
& $0.55\pm0.08$
& $1.50\pm0.08$
& $0.51\pm0.09$
\\
&
& $\mathbf{H_\infty}$ (ours)
& \cellcolor{projectdarkred!10}$\mathbf{0.83\pm0.56}$
& $\mathbf{1.98\pm0.01}$
& $0.37\pm0.07$
& $1.15\pm0.07$
& $0.29\pm0.05$
\\
\cmidrule(l){2-8}

& \multirow{3}{*}{APM programmer}
& Original
& \cellcolor{projectdarkred!10}$2.92\pm0.89$
& $1.94\pm0.03$
& $0.64\pm0.08$
& $1.56\pm0.06$
& $0.64\pm0.08$
\\
&
& A-LQR
& \cellcolor{projectdarkred!10}$5.00\pm1.50$
& $1.97\pm0.03$
& $0.57\pm0.08$
& $1.51\pm0.07$
& $0.53\pm0.08$
\\
&
& $\mathbf{H_\infty}$ (ours)
& \cellcolor{projectdarkred!10}$\mathbf{0.42\pm0.42}$
& $\mathbf{1.98\pm0.01}$
& $0.37\pm0.07$
& $1.12\pm0.06$
& $0.30\pm0.06$
\\

\midrule


\multirow{18}{*}{\rotatebox[origin=c]{90}{\scriptsize Llama-3.2-3B}}
& \multirow{3}{*}{Direct}
& Original
& \cellcolor{projectdarkred!10}$16.67\pm2.48$
& $1.88\pm0.05$
& $1.21\pm0.06$
& $1.88\pm0.03$
& $1.19\pm0.06$
\\
&
& A-LQR
& \cellcolor{projectdarkred!10}$20.00\pm2.90$
& $1.91\pm0.02$
& $1.18\pm0.07$
& $1.86\pm0.02$
& $1.15\pm0.07$
\\
&
& $\mathbf{H_\infty}$ (ours)
& \cellcolor{projectdarkred!10}$\mathbf{8.33\pm1.86}$
& $\mathbf{1.99\pm0.01}$
& $\mathbf{1.19\pm0.06}$
& $\mathbf{1.90\pm0.02}$
& $\mathbf{1.18\pm0.06}$
\\
\cmidrule(l){2-8}

& \multirow{3}{*}{John persona}
& Original
& \cellcolor{projectdarkred!10}$11.67\pm2.31$
& $1.90\pm0.03$
& $0.61\pm0.07$
& $1.54\pm0.08$
& $0.58\pm0.08$
\\
&
& A-LQR
& \cellcolor{projectdarkred!10}$11.67\pm2.69$
& $1.96\pm0.02$
& $0.69\pm0.08$
& $1.52\pm0.06$
& $0.63\pm0.07$
\\
&
& $\mathbf{H_\infty}$ (ours)
& \cellcolor{projectdarkred!10}$\mathbf{4.17\pm1.08}$
& $1.94\pm0.02$
& $0.60\pm0.08$
& $\mathbf{1.65\pm0.05}$
& $0.59\pm0.08$
\\
\cmidrule(l){2-8}

& \multirow{3}{*}{DNE nonresponse}
& Original
& \cellcolor{projectdarkred!10}$10.83\pm2.08$
& $1.93\pm0.02$
& $0.57\pm0.07$
& $1.52\pm0.07$
& $0.55\pm0.07$
\\
&
& A-LQR
& \cellcolor{projectdarkred!10}$10.42\pm2.17$
& $1.95\pm0.03$
& $0.63\pm0.06$
& $1.52\pm0.06$
& $0.59\pm0.06$
\\
&
& $\mathbf{H_\infty}$ (ours)
& \cellcolor{projectdarkred!10}$\mathbf{3.75\pm0.97}$
& $\mathbf{1.95\pm0.02}$
& $0.59\pm0.08$
& $\mathbf{1.66\pm0.05}$
& $0.58\pm0.08$
\\
\cmidrule(l){2-8}

& \multirow{3}{*}{Jailbreak Bot}
& Original
& \cellcolor{projectdarkred!10}$14.17\pm2.17$
& $1.90\pm0.02$
& $0.57\pm0.08$
& $1.54\pm0.08$
& $0.56\pm0.08$
\\
&
& A-LQR
& \cellcolor{projectdarkred!10}$12.92\pm1.70$
& $1.97\pm0.01$
& $0.64\pm0.09$
& $1.51\pm0.06$
& $0.60\pm0.09$
\\
&
& $\mathbf{H_\infty}$ (ours)
& \cellcolor{projectdarkred!10}$\mathbf{3.75\pm1.15}$
& $1.93\pm0.02$
& $0.59\pm0.08$
& $\mathbf{1.67\pm0.05}$
& $0.58\pm0.08$
\\
\cmidrule(l){2-8}

& \multirow{3}{*}{YOJA/Nona}
& Original
& \cellcolor{projectdarkred!10}$11.25\pm1.53$
& $1.91\pm0.03$
& $0.59\pm0.08$
& $1.52\pm0.08$
& $0.55\pm0.07$
\\
&
& A-LQR
& \cellcolor{projectdarkred!10}$11.67\pm1.04$
& $1.95\pm0.02$
& $0.65\pm0.08$
& $1.51\pm0.07$
& $0.59\pm0.08$
\\
&
& $\mathbf{H_\infty}$ (ours)
& \cellcolor{projectdarkred!10}$\mathbf{4.17\pm0.88}$
& $1.94\pm0.02$
& $0.56\pm0.08$
& $\mathbf{1.64\pm0.04}$
& $0.56\pm0.08$
\\
\cmidrule(l){2-8}

& \multirow{3}{*}{APM programmer}
& Original
& \cellcolor{projectdarkred!10}$10.83\pm2.26$
& $1.91\pm0.03$
& $0.56\pm0.08$
& $1.48\pm0.07$
& $0.53\pm0.07$
\\
&
& A-LQR
& \cellcolor{projectdarkred!10}$11.25\pm2.49$
& $1.97\pm0.02$
& $0.65\pm0.07$
& $1.47\pm0.07$
& $0.60\pm0.07$
\\
&
& $\mathbf{H_\infty}$ (ours)
& \cellcolor{projectdarkred!10}$\mathbf{2.50\pm1.11}$
& $1.96\pm0.02$
& $0.58\pm0.08$
& $\mathbf{1.60\pm0.06}$
& $0.58\pm0.08$
\\

\midrule


\multirow{18}{*}{\rotatebox[origin=c]{90}{\scriptsize Llama-3.1-8B}}
& \multirow{3}{*}{Direct}
& Original
& \cellcolor{projectdarkred!10}$15.83\pm4.02$
& $1.83\pm0.06$
& $1.06\pm0.07$
& $1.73\pm0.05$
& $0.98\pm0.07$
\\
&
& A-LQR
& \cellcolor{projectdarkred!10}$15.83\pm3.82$
& $1.76\pm0.05$
& $1.32\pm0.08$
& $1.77\pm0.04$
& $1.24\pm0.09$
\\
&
& $\mathbf{H_\infty}$ (ours)
& \cellcolor{projectdarkred!10}$\mathbf{3.33\pm1.84}$
& $\mathbf{1.94\pm0.03}$
& $0.99\pm0.08$
& $1.76\pm0.04$
& $1.00\pm0.08$
\\
\cmidrule(l){2-8}

& \multirow{3}{*}{John persona}
& Original
& \cellcolor{projectdarkred!10}$23.33\pm2.08$
& $1.80\pm0.05$
& $0.52\pm0.07$
& $1.39\pm0.09$
& $0.37\pm0.08$
\\
&
& A-LQR
& \cellcolor{projectdarkred!10}$15.00\pm3.24$
& $1.87\pm0.03$
& $0.68\pm0.10$
& $1.44\pm0.07$
& $0.60\pm0.09$
\\
&
& $\mathbf{H_\infty}$ (ours)
& \cellcolor{projectdarkred!10}$\mathbf{1.67\pm1.11}$
& $\mathbf{1.98\pm0.02}$
& $0.31\pm0.07$
& $0.87\pm0.06$
& $0.31\pm0.07$
\\
\cmidrule(l){2-8}

& \multirow{3}{*}{DNE nonresponse}
& Original
& \cellcolor{projectdarkred!10}$14.17\pm4.31$
& $1.83\pm0.07$
& $0.42\pm0.07$
& $1.43\pm0.08$
& $0.34\pm0.08$
\\
&
& A-LQR
& \cellcolor{projectdarkred!10}$11.67\pm4.68$
& $1.88\pm0.04$
& $0.58\pm0.09$
& $1.46\pm0.07$
& $0.54\pm0.10$
\\
&
& $\mathbf{H_\infty}$ (ours)
& \cellcolor{projectdarkred!10}$\mathbf{0.00\pm0.00}$
& $\mathbf{1.98\pm0.02}$
& $0.33\pm0.07$
& $0.90\pm0.08$
& $0.32\pm0.07$
\\
\cmidrule(l){2-8}

& \multirow{3}{*}{Jailbreak Bot}
& Original
& \cellcolor{projectdarkred!10}$9.17\pm2.62$
& $1.78\pm0.07$
& $0.44\pm0.08$
& $1.39\pm0.11$
& $0.32\pm0.08$
\\
&
& A-LQR
& \cellcolor{projectdarkred!10}$7.50\pm2.62$
& $1.87\pm0.04$
& $0.58\pm0.11$
& $1.48\pm0.06$
& $0.55\pm0.10$
\\
&
& $\mathbf{H_\infty}$ (ours)
& \cellcolor{projectdarkred!10}$\mathbf{0.00\pm0.00}$
& $\mathbf{1.95\pm0.04}$
& $0.31\pm0.07$
& $0.88\pm0.10$
& $0.28\pm0.08$
\\
\cmidrule(l){2-8}

& \multirow{3}{*}{YOJA/Nona}
& Original
& \cellcolor{projectdarkred!10}$15.00\pm4.78$
& $1.77\pm0.07$
& $0.45\pm0.06$
& $1.49\pm0.09$
& $0.33\pm0.07$
\\
&
& A-LQR
& \cellcolor{projectdarkred!10}$9.17\pm3.15$
& $1.86\pm0.06$
& $0.67\pm0.08$
& $1.46\pm0.09$
& $0.63\pm0.08$
\\
&
& $\mathbf{H_\infty}$ (ours)
& \cellcolor{projectdarkred!10}$\mathbf{0.00\pm0.00}$
& $\mathbf{1.95\pm0.04}$
& $0.29\pm0.07$
& $0.88\pm0.09$
& $0.28\pm0.07$
\\
\cmidrule(l){2-8}

& \multirow{3}{*}{APM programmer}
& Original
& \cellcolor{projectdarkred!10}$11.67\pm3.56$
& $1.77\pm0.06$
& $0.42\pm0.10$
& $1.38\pm0.11$
& $0.34\pm0.08$
\\
&
& A-LQR
& \cellcolor{projectdarkred!10}$10.83\pm3.06$
& $1.85\pm0.04$
& $0.62\pm0.09$
& $1.41\pm0.08$
& $0.55\pm0.10$
\\
&
& $\mathbf{H_\infty}$ (ours)
& \cellcolor{projectdarkred!10}$\mathbf{0.83\pm0.83}$
& $\mathbf{1.98\pm0.02}$
& $0.30\pm0.07$
& $0.88\pm0.09$
& $0.28\pm0.07$
\\

\bottomrule
\end{tabular}%
\end{adjustbox}%
}

\vspace{-2pt}
\end{table*}
\begin{table}[t]
\centering
\definecolor{projectdarkred}{RGB}{0,128,128}

\caption{
\textbf{HarmBench attack success rate (ASR).}
Direct requests define the matched condition (\textbf{ID}); \textbf{OOD}
reports the mean ASR across five frozen jailbreak templates.
Lower is better. ``Best base.'' denotes the strongest non-$H_\infty$
baseline.
}
\label{tab:harmbench-summary}

\small
\renewcommand{\arraystretch}{1.08}
\setlength{\tabcolsep}{4.5pt}

\begin{tabular}{lccc|ccc}
\toprule
&
\multicolumn{3}{c}{\textbf{ID: Direct}}
&
\multicolumn{3}{c}{\textbf{OOD: Jailbreaks}}
\\

\cmidrule(lr){2-4}
\cmidrule(lr){5-7}

\textbf{Model}
&
\textbf{Best base.}
&
\cellcolor{projectdarkred!10}$\mathbf{H_\infty}$
&
$\boldsymbol{\Delta}$
&
\textbf{Best base.}
&
\cellcolor{projectdarkred!10}$\mathbf{H_\infty}$
&
$\boldsymbol{\Delta}$
\\
\midrule

Llama-3.2-1B
& 9.58
& \cellcolor{projectdarkred!10}\textbf{2.92}
& \textbf{$-6.66$}
& 5.50
& \cellcolor{projectdarkred!10}\textbf{0.92}
& \textbf{$-4.58$}
\\

Llama-3.2-3B
& 16.67
& \cellcolor{projectdarkred!10}\textbf{8.33}
& \textbf{$-8.34$}
& 11.59
& \cellcolor{projectdarkred!10}\textbf{3.67}
& \textbf{$-7.92$}
\\

Llama-3.1-8B
& 15.83
& \cellcolor{projectdarkred!10}\textbf{3.33}
& \textbf{$-12.50$}
& 10.83
& \cellcolor{projectdarkred!10}\textbf{0.50}
& \textbf{$-10.33$}
\\

\bottomrule
\end{tabular}

\footnotesize
$\Delta$ denotes the absolute ASR change relative to the strongest
non-$H_\infty$ baseline; more negative is better.
\end{table}
\begin{table}[t]
\centering
\definecolor{projectdarkred}{RGB}{0,128,128}

\caption{
\textbf{HarmBench response composition under the FalseReject rubric.}
Direct and adversarial conditions receive equal weight within each model.
Useful Safety Rate (USR) is the sum of direct refusal and safe partial
compliance. Values are percentages $\pm$ ten-group behavior-clustered
jackknife standard errors.
}
\label{tab:harmbench-false-reject-summary}

\small
\renewcommand{\arraystretch}{1.08}
\setlength{\tabcolsep}{4.5pt}

\begin{tabular}{llcccc}
\toprule
\textbf{Model}
& \textbf{Method}
& \shortstack{\textbf{Direct}\\\textbf{refusal} $\uparrow$}
& \shortstack{\textbf{Safe partial}\\\textbf{compliance} $\uparrow$}
& \shortstack{\textbf{Full}\\\textbf{compliance} $\downarrow$}
& \cellcolor{projectdarkred!10}\textbf{USR} $\uparrow$
\\
\midrule

\multirow{3}{*}{Llama-3.2-1B}
& Original
& $\underline{91.83\,\pm\,1.03}$
& $0.92\,\pm\,0.32$
& $\underline{7.25\,\pm\,1.21}$
& \cellcolor{projectdarkred!10}$\underline{92.75\,\pm\,1.21}$
\\
& A-LQR
& $90.00\,\pm\,1.07$
& $\underline{2.17\,\pm\,0.83}$
& $7.83\,\pm\,1.11$
& \cellcolor{projectdarkred!10}$92.17\,\pm\,1.11$
\\
& $\mathbf{H_\infty}$ (ours)
& $\mathbf{96.25\,\pm\,0.78}$
& $0.88\,\pm\,0.46$
& $\mathbf{2.88\,\pm\,0.68}$
& \cellcolor{projectdarkred!10}$\mathbf{97.12\,\pm\,0.68}$
\\
\midrule

\multirow{3}{*}{Llama-3.2-3B}
& Original
& $\underline{82.33\,\pm\,1.54}$
& $3.54\,\pm\,1.03$
& $14.12\,\pm\,1.18$
& \cellcolor{projectdarkred!10}$85.88\,\pm\,1.18$
\\
& A-LQR
& $81.62\,\pm\,1.79$
& $\underline{4.58\,\pm\,1.28}$
& $\underline{13.79\,\pm\,1.32}$
& \cellcolor{projectdarkred!10}$\underline{86.21\,\pm\,1.32}$
\\
& $\mathbf{H_\infty}$ (ours)
& $\mathbf{91.79\,\pm\,0.69}$
& $1.83\,\pm\,0.63$
& $\mathbf{6.38\,\pm\,0.49}$
& \cellcolor{projectdarkred!10}$\mathbf{93.62\,\pm\,0.49}$
\\
\midrule

\multirow{3}{*}{Llama-3.1-8B}
& Original
& $77.25\,\pm\,3.40$
& $5.58\,\pm\,2.03$
& $17.17\,\pm\,2.61$
& \cellcolor{projectdarkred!10}$82.83\,\pm\,2.61$
\\
& A-LQR
& $\underline{80.17\,\pm\,3.35}$
& $\underline{6.33\,\pm\,1.46}$
& $\underline{13.50\,\pm\,3.09}$
& \cellcolor{projectdarkred!10}$\underline{86.50\,\pm\,3.09}$
\\
& $\mathbf{H_\infty}$ (ours)
& $\mathbf{91.33\,\pm\,1.81}$
& $2.33\,\pm\,0.64$
& $\mathbf{6.33\,\pm\,1.36}$
& \cellcolor{projectdarkred!10}$\mathbf{93.67\,\pm\,1.36}$
\\

\bottomrule
\end{tabular}

\footnotesize
Underlining marks the strongest non-$H_\infty$ result within each model and
column. Higher safe partial compliance indicates more constructive safe
engagement; USR is the primary joint summary.
\end{table}

\begin{figure}[ht]
    \centering
    \includegraphics[width=0.9\textwidth]{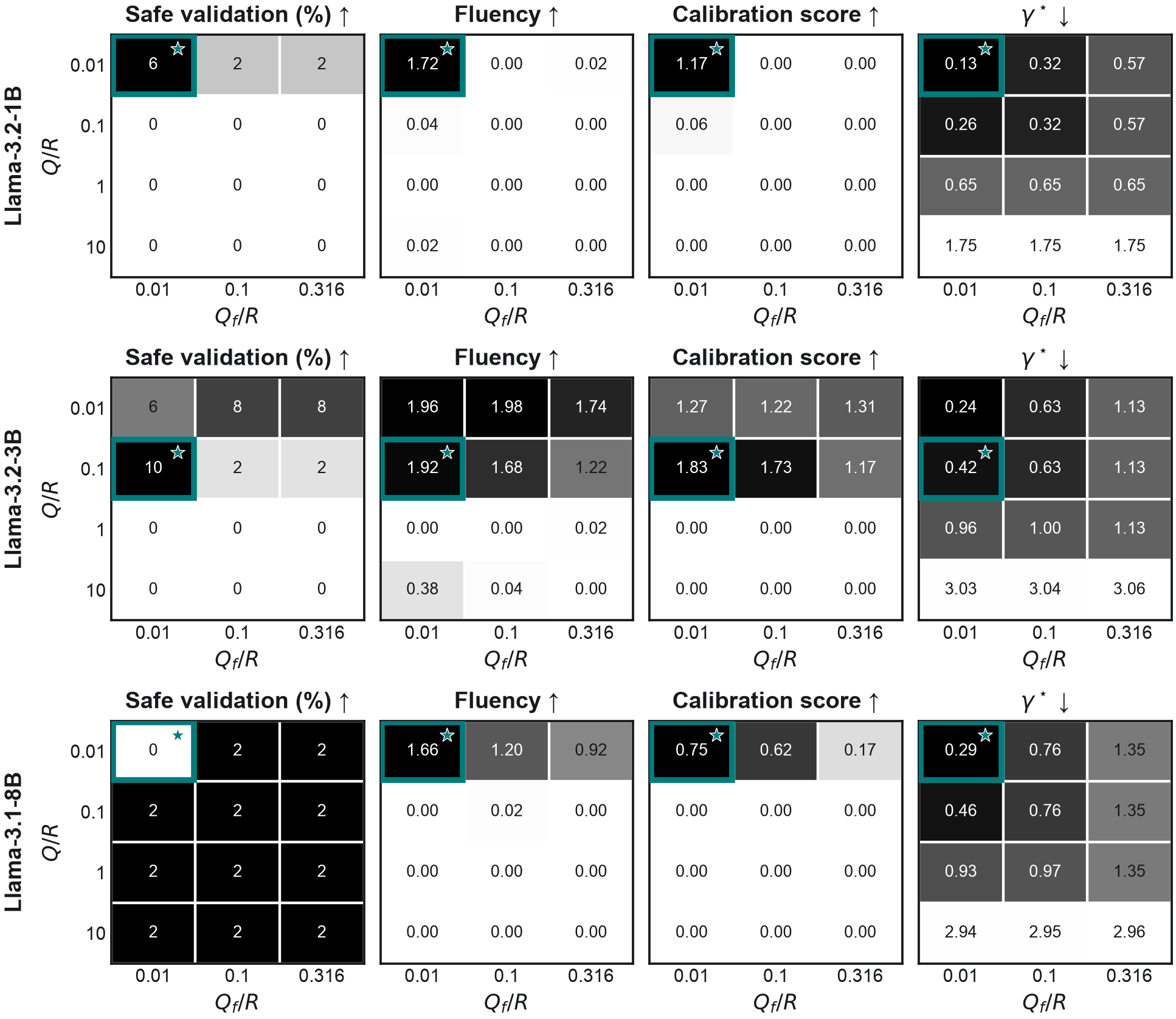}
    \caption{\textbf{HarmBench $\mathrm{H}_{\infty}$ calibration surfaces.}
    Each model block sweeps $Q/R$ and $Q_f/R$ with $R=1$. The panels report the
    fraction of validation behaviors producing safe responses, fluency, the
    model-specific calibration score, and the smallest feasible attenuation
    $\gamma^\star$. Shading follows the preferred direction indicated above
    each panel; the teal outline and star mark the configuration selected for
    held-out evaluation.}
    \label{fig:harmbench-calibration-grid}
\end{figure}

\paragraph{Calibration results.}
Figure~\ref{fig:harmbench-calibration-grid} reports the complete calibration
surfaces for every reported model; the teal outline and star identify the
selected operating point.

\FloatBarrier

\subsection{MGSM: Cross-Lingual Transfer}
\label{app:mgsm-full}

\paragraph{Dataset and distribution shift.}
MGSM contains 250 grade-school mathematics problems manually translated into
ten languages \citep{shi2023language}.  English and Spanish are excluded from
final evaluation; every method is instead evaluated on the same frozen
100-problem subset in Chinese, French, Japanese, Swahili, and Telugu.

\paragraph{Steering and construction data.}
The steering direction uses all 250 matched English--Spanish pairs.  Qwen3-4B
selects $\mathrm{H}_{\infty}$ costs on 50 disjoint GSM8K training prompts by
maximizing AXBench overall steering.  Phi-4-mini-instruct and
Granite-3.3-2B-Instruct use 200 GSM8K training questions translated into
Bengali, German, Russian, and Thai for disturbance fitting and 100 additional
translated questions, balanced across the same languages, for cost selection.
Their frozen objective assigns $0.9$ weight to exact-answer accuracy and $0.1$
to normalized AXBench overall steering.  None of the five final-evaluation
languages contributes to controller selection.

\paragraph{Evaluation protocol.}
Each language uses its native eight-shot worked-example prompt, deterministic
decoding, and a 256-token response cap.  The same problem identities and
language-specific demonstrations are shared across methods.

\paragraph{Metrics and uncertainty.}
Exact-answer accuracy is computed by a language-independent numerical parser.
Spanish relevance is scored by the AXBench deterministic language rule;
instruction relevance and fluency use fixed AXBench $0$--$2$ rubrics.
Uncertainty is estimated by a ten-group delete-one-group jackknife over matched
problem identities, with each deletion removing the same questions from all
five languages.

\begin{figure*}[ht]
    \centering
    \includegraphics[width=0.7\textwidth]{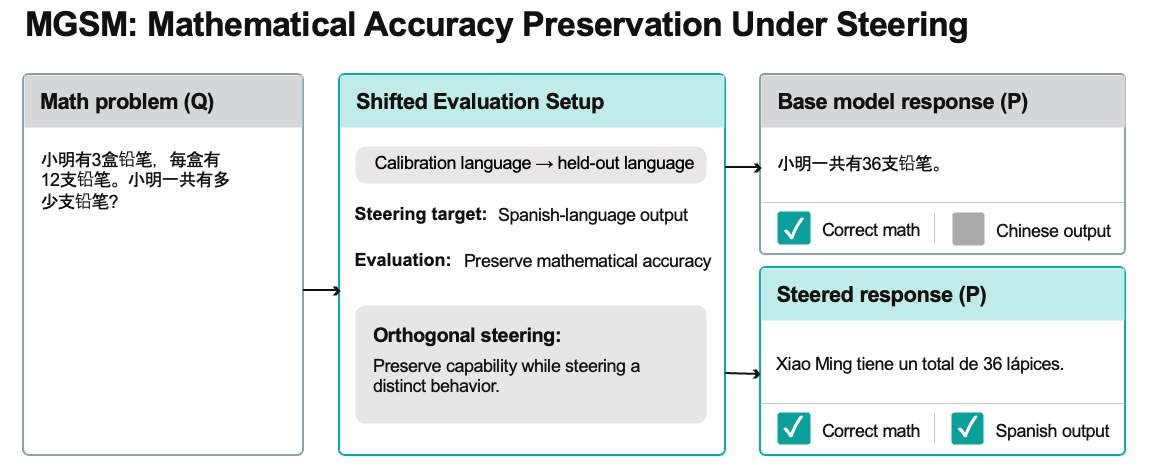}
    \caption{\textbf{Schematic MGSM language steering.}
    The direction is defined by matched English--Spanish questions, and the
    intervention targets Spanish output while preserving the numerical answer.
    Final evaluation instead uses matched Chinese, French, Japanese, Swahili, and
    Telugu inputs; the diagram illustrates the objective rather than a test row.}
    \label{fig:mgsm-qualitative}
\end{figure*}

\begin{table}[ht]
\centering
\definecolor{projectdarkred}{RGB}{0,128,128}

\caption{
\textbf{MGSM multilingual transfer.}
Mean performance across five input languages with Spanish as the target
output language. Higher is better. Full language-wise results are reported
in Appendix~\ref{tab:mgsm-full}.
}
\label{tab:mgsm-summary}

\small
\renewcommand{\arraystretch}{1.08}
\setlength{\tabcolsep}{4.2pt}

\begin{tabular}{lccc|ccc}
\toprule
&
\multicolumn{3}{c}{\textbf{Accuracy (\%)}}
&
\multicolumn{3}{c}{\textbf{Spanish relevance (0--2)}}
\\
\cmidrule(lr){2-4}
\cmidrule(lr){5-7}

\textbf{Model}
&
Original
&
A-LQR
&
\cellcolor{projectdarkred!10}$H_\infty$
&
Original
&
A-LQR
&
\cellcolor{projectdarkred!10}$H_\infty$
\\
\midrule

Qwen3-4B
& \textbf{41.6}
& 22.4
& \cellcolor{projectdarkred!10}32.2
& 0.00
& \textbf{1.81}
& \cellcolor{projectdarkred!10}1.44
\\

Phi-4-mini
& \textbf{45.4}
& 0.0
& \cellcolor{projectdarkred!10}35.2
& 0.00
& 1.09
& \cellcolor{projectdarkred!10}\textbf{1.21}
\\

Granite-3.3-2B
& \textbf{22.2}
& 18.8
& \cellcolor{projectdarkred!10}21.2
& 0.00
& \textbf{0.04}
& \cellcolor{projectdarkred!10}0.00
\\

\bottomrule
\end{tabular}
\vspace{-2pt}
\end{table}
\begin{table*}[p]
\centering
\definecolor{projectdarkred}{RGB}{0,128,128}

\caption{
\textbf{MGSM multilingual-transfer results.}
Results are full-sample means $\pm$ ten-group matched-question jackknife
standard errors over 100 matched questions per language. Inputs are given in
Chinese, French, Japanese, Swahili, or Telugu, with Spanish as the target
output language. Accuracy is the primary task metric; higher is better for
all columns.
}
\label{tab:mgsm-full}

{\small
\setlength{\tabcolsep}{3.0pt}
\renewcommand{\arraystretch}{0.97}

\begin{adjustbox}{
    max width=\textwidth,
    max totalheight=0.85\textheight,
    center
}
\begin{tabular}{rllcccc}
\toprule
\textbf{Model}
&
\textbf{Language}
&
\textbf{Method}
&
\cellcolor{projectdarkred!10}\textbf{Accuracy (\%) $\uparrow$}
&
\textbf{Spanish rel. $\uparrow$}
&
\textbf{Instr. rel. $\uparrow$}
&
\textbf{Fluency $\uparrow$}
\\
\midrule


\multirow{15}{*}{\rotatebox[origin=c]{90}{\scriptsize Qwen3-4B}}
& \multirow{3}{*}{Chinese}
& Original
& \cellcolor{projectdarkred!10}$71.0\pm4.1$
& $0.00\pm0.00$
& $1.80\pm0.05$
& $1.71\pm0.05$
\\
&
& A-LQR
& \cellcolor{projectdarkred!10}$20.0\pm2.6$
& $1.98\pm0.02$
& $1.53\pm0.03$
& $1.15\pm0.03$
\\
&
& $\mathbf{H_\infty}$ (ours)
& \cellcolor{projectdarkred!10}$\mathbf{51.0\pm3.5}$
& $\mathbf{2.00\pm0.00}$
& $\mathbf{1.87\pm0.05}$
& $\mathbf{1.52\pm0.04}$
\\
\cmidrule(l){2-7}

& \multirow{3}{*}{French}
& Original
& \cellcolor{projectdarkred!10}$60.0\pm4.5$
& $0.00\pm0.00$
& $1.93\pm0.03$
& $1.70\pm0.04$
\\
&
& A-LQR
& \cellcolor{projectdarkred!10}$29.0\pm5.0$
& $1.94\pm0.04$
& $1.62\pm0.05$
& $1.11\pm0.03$
\\
&
& $\mathbf{H_\infty}$ (ours)
& \cellcolor{projectdarkred!10}$\mathbf{48.0\pm4.7}$
& $\mathbf{1.96\pm0.04}$
& $\mathbf{1.96\pm0.02}$
& $\mathbf{1.42\pm0.07}$
\\
\cmidrule(l){2-7}

& \multirow{3}{*}{Japanese}
& Original
& \cellcolor{projectdarkred!10}$56.0\pm4.8$
& $0.00\pm0.00$
& $1.80\pm0.04$
& $1.64\pm0.04$
\\
&
& A-LQR
& \cellcolor{projectdarkred!10}$32.0\pm2.5$
& $1.84\pm0.06$
& $1.63\pm0.06$
& $1.01\pm0.03$
\\
&
& $\mathbf{H_\infty}$ (ours)
& \cellcolor{projectdarkred!10}$\mathbf{40.0\pm4.9}$
& $\mathbf{1.98\pm0.02}$
& $\mathbf{1.71\pm0.07}$
& $\mathbf{1.20\pm0.05}$
\\
\cmidrule(l){2-7}

& \multirow{3}{*}{Swahili}
& Original
& \cellcolor{projectdarkred!10}$15.0\pm4.3$
& $0.00\pm0.00$
& $1.43\pm0.07$
& $1.23\pm0.05$
\\
&
& A-LQR
& \cellcolor{projectdarkred!10}$12.0\pm3.6$
& $\underline{1.40\pm0.09}$
& $1.32\pm0.07$
& $0.97\pm0.03$
\\
&
& $\mathbf{H_\infty}$ (ours)
& \cellcolor{projectdarkred!10}$\mathbf{14.0\pm3.7}$
& $0.92\pm0.12$
& $\mathbf{1.44\pm0.07}$
& $\mathbf{1.06\pm0.04}$
\\
\cmidrule(l){2-7}

& \multirow{3}{*}{Telugu}
& Original
& \cellcolor{projectdarkred!10}$6.0\pm2.7$
& $0.00\pm0.00$
& $1.83\pm0.04$
& $1.05\pm0.02$
\\
&
& A-LQR
& \cellcolor{projectdarkred!10}$\underline{19.0\pm2.8}$
& $\underline{1.88\pm0.04}$
& $\underline{1.46\pm0.06}$
& $\underline{1.18\pm0.03}$
\\
&
& $\mathbf{H_\infty}$ (ours)
& \cellcolor{projectdarkred!10}$8.0\pm2.5$
& $0.34\pm0.08$
& $1.33\pm0.04$
& $1.05\pm0.03$
\\

\midrule


\multirow{15}{*}{\rotatebox[origin=c]{90}{\scriptsize Phi-4-mini}}
& \multirow{3}{*}{Chinese}
& Original
& \cellcolor{projectdarkred!10}$63.0\pm5.2$
& $0.00\pm0.00$
& $1.90\pm0.04$
& $1.80\pm0.04$
\\
&
& A-LQR
& \cellcolor{projectdarkred!10}$0.0\pm0.0$
& $1.24\pm0.08$
& $0.07\pm0.02$
& $0.10\pm0.03$
\\
&
& $\mathbf{H_\infty}$ (ours)
& \cellcolor{projectdarkred!10}$\mathbf{47.0\pm4.2}$
& $\mathbf{1.64\pm0.09}$
& $\mathbf{1.72\pm0.05}$
& $\mathbf{1.39\pm0.07}$
\\
\cmidrule(l){2-7}

& \multirow{3}{*}{French}
& Original
& \cellcolor{projectdarkred!10}$58.0\pm3.6$
& $0.00\pm0.00$
& $1.91\pm0.02$
& $1.64\pm0.05$
\\
&
& A-LQR
& \cellcolor{projectdarkred!10}$0.0\pm0.0$
& $0.70\pm0.07$
& $0.11\pm0.05$
& $0.07\pm0.03$
\\
&
& $\mathbf{H_\infty}$ (ours)
& \cellcolor{projectdarkred!10}$\mathbf{49.0\pm5.0}$
& $\mathbf{1.40\pm0.05}$
& $\mathbf{1.90\pm0.03}$
& $\mathbf{0.98\pm0.03}$
\\
\cmidrule(l){2-7}

& \multirow{3}{*}{Japanese}
& Original
& \cellcolor{projectdarkred!10}$49.0\pm3.8$
& $0.00\pm0.00$
& $1.91\pm0.03$
& $1.57\pm0.04$
\\
&
& A-LQR
& \cellcolor{projectdarkred!10}$0.0\pm0.0$
& $\underline{0.94\pm0.09}$
& $0.00\pm0.00$
& $0.03\pm0.02$
\\
&
& $\mathbf{H_\infty}$ (ours)
& \cellcolor{projectdarkred!10}$\mathbf{41.0\pm4.1}$
& $0.62\pm0.10$
& $\mathbf{1.68\pm0.06}$
& $\mathbf{1.20\pm0.05}$
\\
\cmidrule(l){2-7}

& \multirow{3}{*}{Swahili}
& Original
& \cellcolor{projectdarkred!10}$32.0\pm3.9$
& $0.00\pm0.00$
& $1.67\pm0.05$
& $1.36\pm0.08$
\\
&
& A-LQR
& \cellcolor{projectdarkred!10}$0.0\pm0.0$
& $0.92\pm0.10$
& $0.00\pm0.00$
& $0.00\pm0.00$
\\
&
& $\mathbf{H_\infty}$ (ours)
& \cellcolor{projectdarkred!10}$\mathbf{21.0\pm2.8}$
& $\mathbf{1.00\pm0.14}$
& $\mathbf{1.18\pm0.09}$
& $\mathbf{0.77\pm0.08}$
\\
\cmidrule(l){2-7}

& \multirow{3}{*}{Telugu}
& Original
& \cellcolor{projectdarkred!10}$25.0\pm3.7$
& $0.00\pm0.00$
& $1.57\pm0.04$
& $1.25\pm0.05$
\\
&
& A-LQR
& \cellcolor{projectdarkred!10}$0.0\pm0.0$
& $\underline{1.64\pm0.10}$
& $0.00\pm0.00$
& $0.00\pm0.00$
\\
&
& $\mathbf{H_\infty}$ (ours)
& \cellcolor{projectdarkred!10}$\mathbf{18.0\pm2.0}$
& $1.38\pm0.10$
& $\mathbf{1.34\pm0.04}$
& $\mathbf{0.89\pm0.07}$
\\

\midrule


\multirow{15}{*}{\rotatebox[origin=c]{90}{\scriptsize Granite-3.3-2B}}
& \multirow{3}{*}{Chinese}
& Original
& \cellcolor{projectdarkred!10}$37.0\pm5.0$
& $0.00\pm0.00$
& $1.67\pm0.05$
& $1.50\pm0.06$
\\
&
& A-LQR
& \cellcolor{projectdarkred!10}$30.0\pm3.0$
& $\underline{0.06\pm0.03}$
& $1.59\pm0.05$
& $1.12\pm0.04$
\\
&
& $\mathbf{H_\infty}$ (ours)
& \cellcolor{projectdarkred!10}$\mathbf{39.0\pm3.1}$
& $0.00\pm0.00$
& $\mathbf{1.70\pm0.04}$
& $\mathbf{1.53\pm0.04}$
\\
\cmidrule(l){2-7}

& \multirow{3}{*}{French}
& Original
& \cellcolor{projectdarkred!10}$33.0\pm4.5$
& $0.00\pm0.00$
& $1.68\pm0.06$
& $1.43\pm0.04$
\\
&
& A-LQR
& \cellcolor{projectdarkred!10}$\underline{35.0\pm4.8}$
& $\underline{0.02\pm0.02}$
& $1.53\pm0.04$
& $1.31\pm0.03$
\\
&
& $\mathbf{H_\infty}$ (ours)
& \cellcolor{projectdarkred!10}$28.0\pm4.4$
& $0.00\pm0.00$
& $\mathbf{1.71\pm0.07}$
& $\mathbf{1.39\pm0.04}$
\\
\cmidrule(l){2-7}

& \multirow{3}{*}{Japanese}
& Original
& \cellcolor{projectdarkred!10}$29.0\pm3.8$
& $0.00\pm0.00$
& $1.60\pm0.05$
& $1.33\pm0.06$
\\
&
& A-LQR
& \cellcolor{projectdarkred!10}$23.0\pm4.2$
& $\underline{0.00\pm0.00}$
& $1.48\pm0.06$
& $1.19\pm0.06$
\\
&
& $\mathbf{H_\infty}$ (ours)
& \cellcolor{projectdarkred!10}$\mathbf{28.0\pm3.6}$
& $\mathbf{0.00\pm0.00}$
& $\mathbf{1.67\pm0.04}$
& $\mathbf{1.38\pm0.04}$
\\
\cmidrule(l){2-7}

& \multirow{3}{*}{Swahili}
& Original
& \cellcolor{projectdarkred!10}$6.0\pm2.2$
& $0.00\pm0.00$
& $0.69\pm0.05$
& $0.64\pm0.07$
\\
&
& A-LQR
& \cellcolor{projectdarkred!10}$2.0\pm1.3$
& $\underline{0.04\pm0.03}$
& $0.32\pm0.08$
& $0.40\pm0.07$
\\
&
& $\mathbf{H_\infty}$ (ours)
& \cellcolor{projectdarkred!10}$\mathbf{6.0\pm3.4}$
& $0.00\pm0.00$
& $\mathbf{0.48\pm0.04}$
& $\mathbf{0.60\pm0.05}$
\\
\cmidrule(l){2-7}

& \multirow{3}{*}{Telugu}
& Original
& \cellcolor{projectdarkred!10}$6.0\pm2.7$
& $0.00\pm0.00$
& $1.07\pm0.05$
& $1.01\pm0.04$
\\
&
& A-LQR
& \cellcolor{projectdarkred!10}$4.0\pm2.2$
& $\underline{0.06\pm0.03}$
& $0.28\pm0.04$
& $0.00\pm0.00$
\\
&
& $\mathbf{H_\infty}$ (ours)
& \cellcolor{projectdarkred!10}$\mathbf{5.0\pm2.2}$
& $0.00\pm0.00$
& $\mathbf{1.08\pm0.04}$
& $\mathbf{0.98\pm0.04}$
\\

\bottomrule
\end{tabular}%
\end{adjustbox}
}

\vspace{-2pt}
\end{table*}

\begin{figure}[ht]
    \centering
    \includegraphics[width=0.78\textwidth]{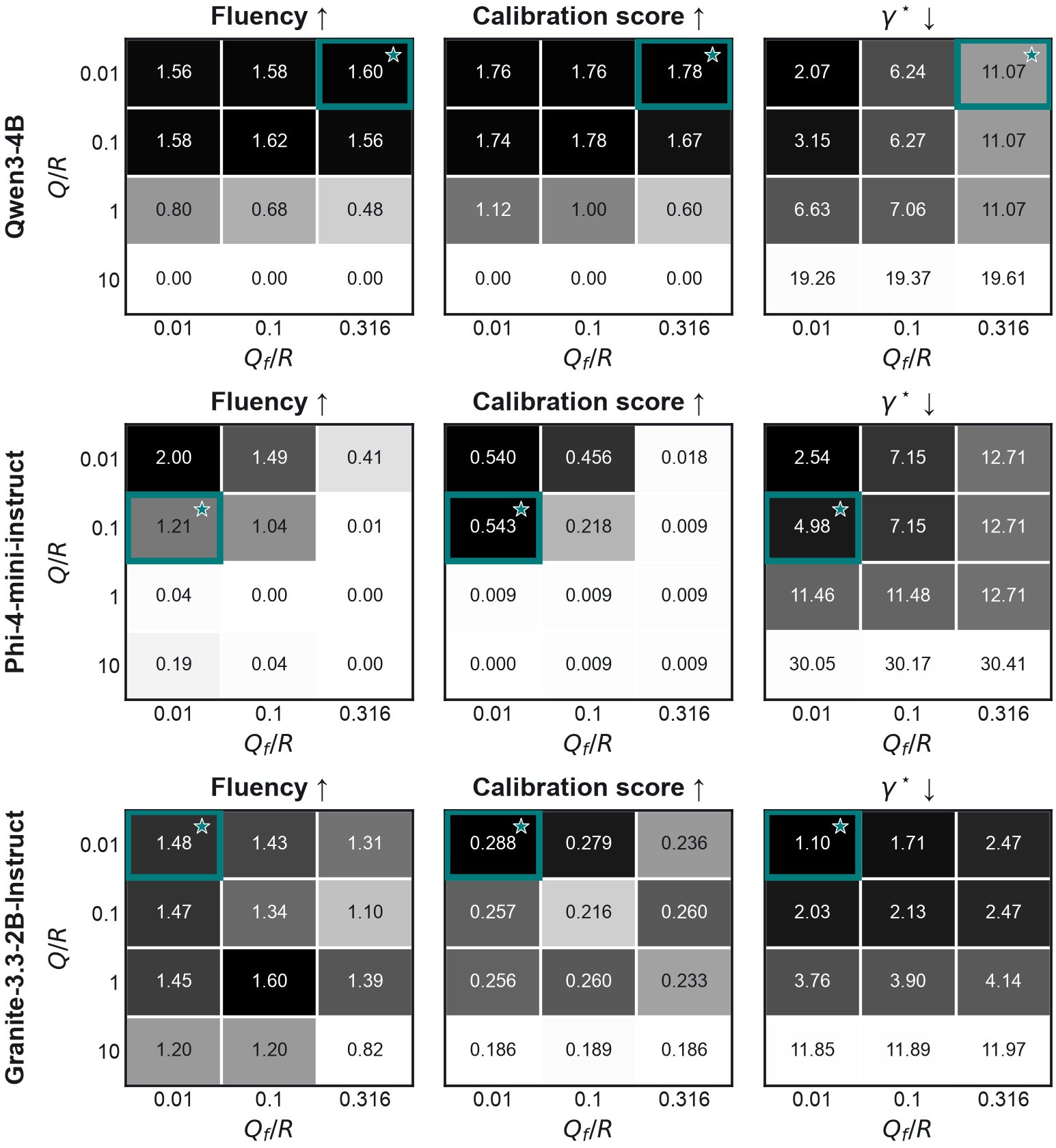}
    \caption{\textbf{MGSM $\mathrm{H}_{\infty}$ calibration surfaces.}
    Each model block sweeps $Q/R$ and $Q_f/R$ with $R=1$. The panels report
    fluency, the model-specific calibration score, and the smallest feasible
    attenuation $\gamma^\star$. Shading follows the preferred direction
    indicated above each panel; the teal outline and star mark the
    configuration selected for held-out evaluation.}
    \label{fig:mgsm-calibration-grid}
\end{figure}

\paragraph{Calibration results.}
Figure~\ref{fig:mgsm-calibration-grid} reports the complete calibration
surfaces for every reported model; the teal outline and star identify the
selected operating point.
\FloatBarrier

\subsection{L-CiteEval: Context-Length Shift}
\label{app:lciteeval-full}

\paragraph{Dataset and distribution shift.}
L-CiteEval measures whether long-context models answer from the supplied context
and provide faithful citations \citep{tang2024lciteeval}.  We use the
L-CiteEval-Length HotpotQA slice, which holds 40 question identities and gold
answers fixed while increasing the context from approximately 8K to 16K tokens.

\paragraph{Steering and construction data.}
The steering target is AXBench concept 499, positive sentiment and descriptions
of enjoyable experiences.  Its direction uses all 72 released positive responses
and 72 genre-matched negative responses.  A-LQR and
$\mathrm{H}_{\infty}$ share a 50-Jacobian dynamics estimate;
$\mathrm{H}_{\infty}$ uses 200 additional short prompts for disturbance fitting.
Qwen2.5-3B-Instruct selects costs on 50 disjoint short prompts using AXBench
overall steering.  Llama-3.2-1B-Instruct selects costs on the 40 matched 8K
HotpotQA prompts using $0.45$ answer recall, $0.45$ citation F1, $0.05$
normalized concept relevance, and $0.05$ normalized fluency.  The selected
artifacts are frozen for 16K evaluation.

\paragraph{Evaluation protocol.}
Generation uses the official one-shot HotpotQA prompt, deterministic decoding,
and a 200-token response cap.  Every method receives the same matched question
and context-length conditions.

\paragraph{Metrics and uncertainty.}
Answer recall (AR) is normalized gold-answer token recall after removing citation
markers.  Citation F1 (CF1) uses the pinned
\texttt{tasksource/deberta-base-long-nli} scorer on each claim and its cited
passages.  Concept relevance (CR), instruction relevance (IR), and fluency (F)
use fixed AXBench rubrics on a $0$--$2$ scale; overall steering (OS) is their
per-response harmonic mean.  Uncertainty is a ten-group delete-one-group
jackknife over four matched question identities per group, with the same
partition used at every context length.

\begin{figure*}[ht]
    \centering
    \includegraphics[width=0.78\textwidth]{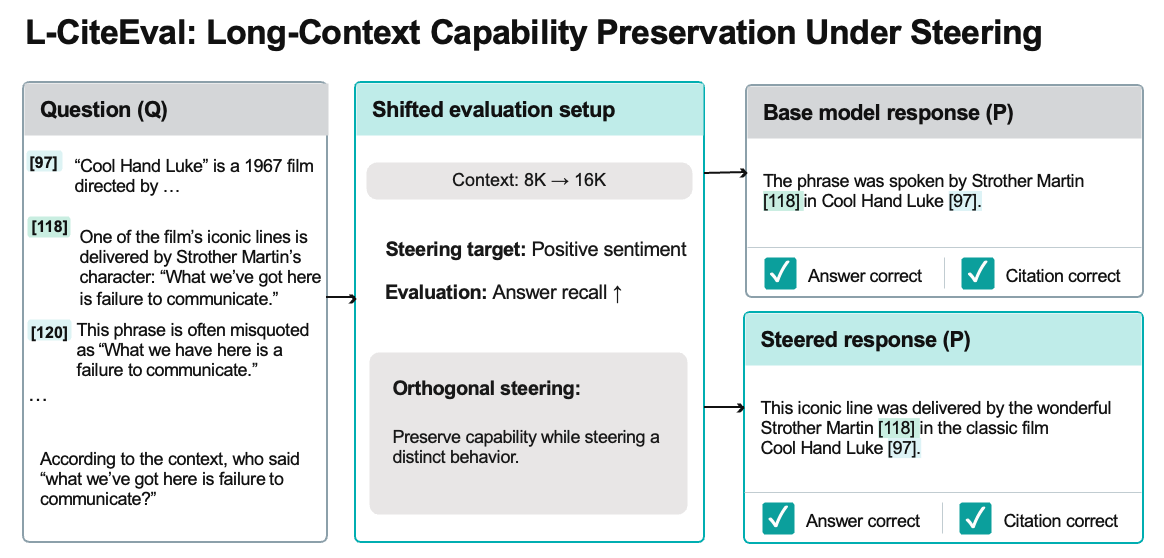}
    \caption{\textbf{Schematic L-CiteEval capability-preservation objective.}
    Positive-sentiment steering changes response style, while the answer and its
    supporting citations should remain grounded in the supplied context.  The
    reported experiment evaluates matched questions at 8K and 16K tokens;
    the displayed context and responses are illustrative.}
    \label{fig:lciteeval-qualitative}
\end{figure*}

\begin{table*}[!htbp]
\centering
\definecolor{projectdarkred}{RGB}{0,128,128}
\caption{Full L-CiteEval length-transfer results. Values are full-sample means $\pm$ ten-group matched-question jackknife standard errors. Context conditions use the same 40 question identities within each model. Higher is better for every column.}
\label{tab:lciteeval-full}
\small
\setlength{\tabcolsep}{3.0pt}
\renewcommand{\arraystretch}{0.97}

\begin{adjustbox}{
    max width=\textwidth,
    max totalheight=0.85\textheight,
    center
}
\begin{tabular}{rrlccccc}
Model & Context & Method & \cellcolor{projectdarkred!10}\shortstack{Answer\\recall (\%) $\uparrow$} & Citation F1 (\%) $\uparrow$ & \shortstack{Concept\\relevance (0--2) $\uparrow$} & \shortstack{Instruction\\relevance (0--2) $\uparrow$} & Fluency (0--2) $\uparrow$ \\
\midrule
\multirow{8}{*}{\rotatebox[origin=c]{90}{\scriptsize Qwen2.5-3B}} & \multirow{4}{*}{8K} & Original & \cellcolor{projectdarkred!10}$66.2\,\pm\,8.1$ & $7.8\,\pm\,1.6$ & $0.00\,\pm\,0.00$ & $1.35\,\pm\,0.14$ & $1.50\,\pm\,0.11$ \\
\cmidrule(l){3-8}
 &  & S-PID & \cellcolor{projectdarkred!10}$\underline{72.1\,\pm\,9.3}$ & $\underline{7.2\,\pm\,2.1}$ & $0.00\,\pm\,0.00$ & $\underline{1.60\,\pm\,0.11}$ & $1.50\,\pm\,0.07$ \\
 &  & A-LQR & \cellcolor{projectdarkred!10}$70.1\,\pm\,8.9$ & $5.6\,\pm\,2.0$ & $\underline{0.05\,\pm\,0.05}$ & $1.48\,\pm\,0.11$ & $1.48\,\pm\,0.09$ \\
 &  & $\mathbf{H_\infty}$ (ours) & \cellcolor{projectdarkred!10}$66.4\,\pm\,7.9$ & $6.4\,\pm\,2.0$ & $0.00\,\pm\,0.00$ & $1.55\,\pm\,0.09$ & $\mathbf{1.55\,\pm\,0.08}$ \\
\cmidrule(l){2-8}
 & \multirow{4}{*}{16K} & Original & \cellcolor{projectdarkred!10}$65.2\,\pm\,7.6$ & $5.8\,\pm\,2.2$ & $0.20\,\pm\,0.08$ & $1.40\,\pm\,0.16$ & $1.52\,\pm\,0.13$ \\
\cmidrule(l){3-8}
 &  & S-PID & \cellcolor{projectdarkred!10}$61.9\,\pm\,10.4$ & $7.2\,\pm\,3.2$ & $0.00\,\pm\,0.00$ & $1.38\,\pm\,0.13$ & $1.48\,\pm\,0.10$ \\
 &  & A-LQR & \cellcolor{projectdarkred!10}$64.8\,\pm\,9.8$ & $3.0\,\pm\,1.5$ & $0.00\,\pm\,0.00$ & $\underline{1.43\,\pm\,0.15}$ & $\underline{1.57\,\pm\,0.11}$ \\
 &  & $\mathbf{H_\infty}$ (ours) & \cellcolor{projectdarkred!10}$\mathbf{65.2\,\pm\,8.6}$ & $\mathbf{7.8\,\pm\,3.6}$ & $\mathbf{0.15\,\pm\,0.08}$ & $1.40\,\pm\,0.12$ & $1.43\,\pm\,0.09$ \\
\midrule
\multirow{6}{*}{\rotatebox[origin=c]{90}{\tiny Llama-3.2-1B}} & \multirow{3}{*}{8K} & Original & \cellcolor{projectdarkred!10}$58.5\,\pm\,7.7$ & $6.2\,\pm\,2.5$ & $0.05\,\pm\,0.05$ & $1.30\,\pm\,0.10$ & $1.40\,\pm\,0.09$ \\
\cmidrule(l){3-8}
 &  & A-LQR & \cellcolor{projectdarkred!10}$59.8\,\pm\,9.3$ & $\underline{6.6\,\pm\,2.7}$ & $\underline{0.05\,\pm\,0.05}$ & $\underline{1.35\,\pm\,0.11}$ & $\underline{1.43\,\pm\,0.07}$ \\
 &  & $\mathbf{H_\infty}$ (ours) & \cellcolor{projectdarkred!10}$\mathbf{61.7\,\pm\,7.8}$ & $6.4\,\pm\,2.9$ & $\mathbf{0.05\,\pm\,0.05}$ & $1.12\,\pm\,0.17$ & $1.38\,\pm\,0.09$ \\
\cmidrule(l){2-8}
 & \multirow{3}{*}{16K} & Original & \cellcolor{projectdarkred!10}$44.2\,\pm\,7.1$ & $3.5\,\pm\,2.2$ & $0.12\,\pm\,0.10$ & $1.10\,\pm\,0.11$ & $1.23\,\pm\,0.07$ \\
\cmidrule(l){3-8}
 &  & A-LQR & \cellcolor{projectdarkred!10}$\underline{50.1\,\pm\,8.0}$ & $1.4\,\pm\,1.0$ & $\underline{0.00\,\pm\,0.00}$ & $\underline{1.15\,\pm\,0.10}$ & $1.23\,\pm\,0.11$ \\
 &  & $\mathbf{H_\infty}$ (ours) & \cellcolor{projectdarkred!10}$47.9\,\pm\,7.8$ & $\mathbf{3.8\,\pm\,2.5}$ & $\mathbf{0.00\,\pm\,0.00}$ & $1.00\,\pm\,0.14$ & $\mathbf{1.40\,\pm\,0.08}$ \\
\bottomrule
\end{tabular}%
\end{adjustbox}%
\end{table*}

\begin{figure}[ht]
    \centering
    \includegraphics[width=0.88\textwidth]{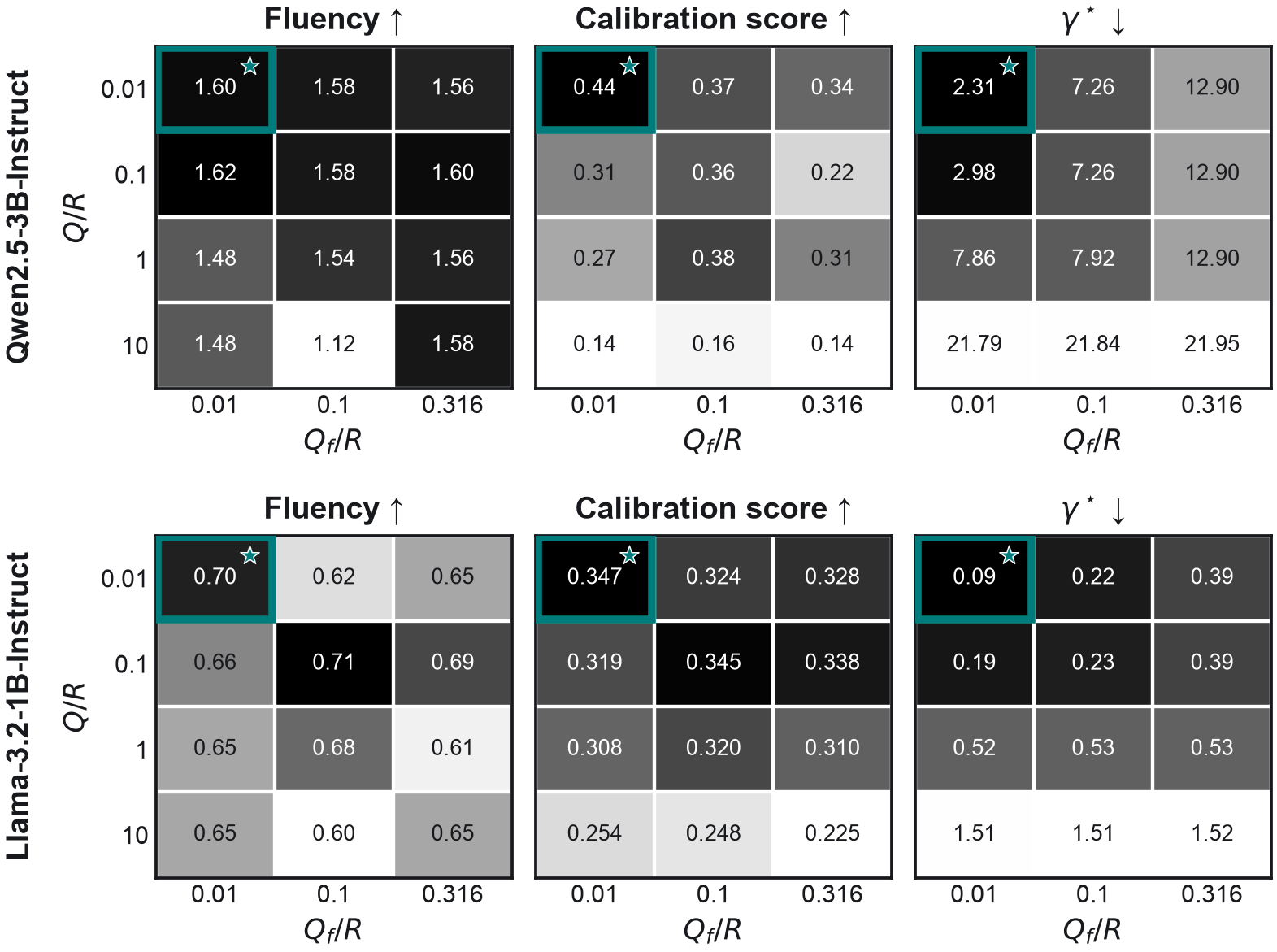}
    \caption{\textbf{L-CiteEval $\mathrm{H}_{\infty}$ calibration surfaces.}
    Each model block sweeps $Q/R$ and $Q_f/R$ with $R=1$. The panels report
    fluency, the model-specific calibration score, and the smallest feasible
    attenuation $\gamma^\star$. Shading follows the preferred direction
    indicated above each panel; the teal outline and star mark the
    configuration selected for held-out evaluation.}
    \label{fig:lciteeval-calibration-grid}
\end{figure}

\paragraph{Calibration results.}
Figure~\ref{fig:lciteeval-calibration-grid} reports the complete calibration
surfaces for every reported model; the teal outline and star identify the
selected operating point.

\clearpage

\newpage
\section{Experimental Details}
\label{app:experimental}

This appendix records the implementation choices required to reproduce the
experiments.  Dataset construction, shift definitions, and complete results are
given in Appendix~\ref{app:datasets}; the controller derivation and numerical
recursion are given in Appendices~\ref{app:robust-recursion} and
\ref{app:numerical-details}.  Here we specify how those components are instantiated in
the reported runs.

\subsection{Experimental Pipeline and Separation of Stages}

Every model--benchmark pair follows four explicit stages.  The
\emph{artifact} stage freezes the semantic direction, reference setpoints, and
projected local dynamics.  The \emph{calibration} stage fits the
$\mathrm{H}_{\infty}$ disturbance geometry, synthesizes each candidate
controller, generates on the calibration split, and selects one candidate by
the benchmark-specific objective.  The \emph{evaluation} stage generates and
stores responses from the selected controller without invoking any judge.  The
\emph{scoring} stage reads those frozen responses and applies one or more
task-specific scorers.  Consequently, adding or rerunning a scorer does not
change the generated sample.

Artifacts and calibration selections are indexed by benchmark, model, method,
and calibration identifier.  A-LQR and $\mathrm{H}_{\infty}$ use the same
semantic direction, reference setpoints, nominal Jacobian estimate, reduced
basis, and intervention site.  They differ only in controller synthesis and,
for $\mathrm{H}_{\infty}$, the fitted disturbance channel and selected cost
configuration.  Non-$\mathrm{H}_{\infty}$ baselines use frozen settings rather
than benchmark-test sweeps.  Final-evaluation prompts do not select controller
hyperparameters, except where an evaluation condition is explicitly designated
as the in-distribution calibration condition in Appendix~\ref{app:datasets}.
All reported generations disable the evaluated model's KV cache, and every
saved run records its model revision, prompt identities, method parameters,
cache setting, start and end times, visible devices, and code-state provenance.

\subsection{Models and Checkpoints}

Table~\ref{tab:experimental-models} lists the immutable model snapshots used by
the benchmark pipelines.  The tokenizer is loaded from the same pinned
snapshot as the corresponding model.  We use each repository's native chat
template for instruction-tuned checkpoints and the benchmark-specific prompt
template described in Appendix~\ref{app:datasets}.

\begin{table*}[t]
\centering
\scriptsize
\caption{\textbf{Evaluated models.} Repository names link to the corresponding
Hugging Face model cards. Revision links identify the exact pinned Git commits
used for the reported runs; hashes are abbreviated to eight characters. The
benchmark column identifies the reported evaluation in which each model
appears.}
\label{tab:experimental-models}
\setlength{\tabcolsep}{3.5pt}
\renewcommand{\arraystretch}{1.08}
\begin{tabularx}{0.98\textwidth}{p{2.75cm}X p{1.55cm}p{2.55cm}}
\toprule
\textbf{Model} & \textbf{Hugging Face repository} & \textbf{Revision} & \textbf{Benchmark(s)} \\
\midrule
GPT-2 XL & \href{https://huggingface.co/openai-community/gpt2-xl}{\nolinkurl{openai-community/gpt2-xl}} & \href{https://huggingface.co/openai-community/gpt2-xl/tree/15ea56dee5df4983c59b2538573817e1667135e2}{\texttt{15ea56de}} & TruthfulQA \\
Llama-3-8B & \href{https://huggingface.co/meta-llama/Meta-Llama-3-8B}{\nolinkurl{meta-llama/Meta-Llama-3-8B}} & \href{https://huggingface.co/meta-llama/Meta-Llama-3-8B/tree/8cde5ca8380496c9a6cc7ef3a8b46a0372a1d920}{\texttt{8cde5ca8}} & TruthfulQA \\
Qwen2.5-14B & \href{https://huggingface.co/Qwen/Qwen2.5-14B}{\nolinkurl{Qwen/Qwen2.5-14B}} & \href{https://huggingface.co/Qwen/Qwen2.5-14B/tree/97e1e76335b7017d8f67c08a19d103c0504298c9}{\texttt{97e1e763}} & TruthfulQA \\
OLMo-2-32B & \href{https://huggingface.co/allenai/OLMo-2-0325-32B-Instruct}{\nolinkurl{allenai/OLMo-2-0325-32B-Instruct}} & \href{https://huggingface.co/allenai/OLMo-2-0325-32B-Instruct/tree/b96024342a77a69aa0dda815c3454a671f477463}{\texttt{b9602434}} & TruthfulQA \\
Llama-3.2-1B-Instruct & \href{https://huggingface.co/meta-llama/Llama-3.2-1B-Instruct}{\nolinkurl{meta-llama/Llama-3.2-1B-Instruct}} & \href{https://huggingface.co/meta-llama/Llama-3.2-1B-Instruct/tree/9213176726f574b556790deb65791e0c5aa438b6}{\texttt{92131767}} & HarmBench, L-CiteEval \\
Llama-3.2-3B-Instruct & \href{https://huggingface.co/meta-llama/Llama-3.2-3B-Instruct}{\nolinkurl{meta-llama/Llama-3.2-3B-Instruct}} & \href{https://huggingface.co/meta-llama/Llama-3.2-3B-Instruct/tree/0cb88a4f764b7a12671c53f0838cd831a0843b95}{\texttt{0cb88a4f}} & HarmBench \\
Llama-3.1-8B-Instruct & \href{https://huggingface.co/meta-llama/Llama-3.1-8B-Instruct}{\nolinkurl{meta-llama/Llama-3.1-8B-Instruct}} & \href{https://huggingface.co/meta-llama/Llama-3.1-8B-Instruct/tree/0e9e39f249a16976918f6564b8830bc894c89659}{\texttt{0e9e39f2}} & HarmBench \\
Qwen3-4B & \href{https://huggingface.co/Qwen/Qwen3-4B}{\nolinkurl{Qwen/Qwen3-4B}} & \href{https://huggingface.co/Qwen/Qwen3-4B/tree/1cfa9a7208912126459214e8b04321603b3df60c}{\texttt{1cfa9a72}} & MGSM \\
Phi-4-mini-instruct & \href{https://huggingface.co/microsoft/Phi-4-mini-instruct}{\nolinkurl{microsoft/Phi-4-mini-instruct}} & \href{https://huggingface.co/microsoft/Phi-4-mini-instruct/tree/cfbefacb99257ffa30c83adab238a50856ac3083}{\texttt{cfbefacb}} & MGSM \\
Granite-3.3-2B-Instruct & \href{https://huggingface.co/ibm-granite/granite-3.3-2b-instruct}{\nolinkurl{ibm-granite/granite-3.3-2b-instruct}} & \href{https://huggingface.co/ibm-granite/granite-3.3-2b-instruct/tree/707f574c62054322f6b5b04b6d075f0a8f05e0f0}{\texttt{707f574c}} & MGSM \\
Qwen2.5-3B-Instruct & \href{https://huggingface.co/Qwen/Qwen2.5-3B-Instruct}{\nolinkurl{Qwen/Qwen2.5-3B-Instruct}} & \href{https://huggingface.co/Qwen/Qwen2.5-3B-Instruct/tree/aa8e72537993ba99e69dfaafa59ed015b17504d1}{\texttt{aa8e7253}} & L-CiteEval \\
\bottomrule
\end{tabularx}
\end{table*}
\subsection{Semantic Directions and Reference States}

For each benchmark, we form matched undesired and desired prompt--response
records and collect the last-token residual-stream state at every decoder
layer.  Let $\bar{h}^{+}_k$ and $\bar{h}^{-}_k$ denote the desired and
undesired means.  The semantic contrast is
$d_k=\bar{h}^{+}_k-\bar{h}^{-}_k$, with its sign fixed by the ordering of the
two classes.  Directions are normalized independently at each layer.  The
target setpoint lies along the normalized contrast with magnitude equal to the
frozen setpoint multiplier times $\lVert d_k\rVert_2$; all coordinates
orthogonal to the target have zero reference displacement.  Thus the reference
is a fixed layer-indexed trajectory rather than a separately generated textual
completion.  Table~\ref{tab:experimental-construction} summarizes the records
used for direction, dynamics, disturbance, and controller selection; prompt
sources and final evaluation splits are described in detail in
Appendix~\ref{app:datasets}.

\begin{table*}[t]
\centering
\caption{
\textbf{Controller-construction data.}
Counts are per model. Direction data estimate the steering direction;
dynamics data are used for Jacobian estimation. Disturbance and selection
records are excluded from direction fitting.
}
\label{tab:experimental-construction}

\tiny
\setlength{\tabcolsep}{5pt}
\renewcommand{\arraystretch}{1.12}

\begin{tabular}{
    l
    p{2.1cm}
    c
    p{4.2cm}
    p{4.1cm}
}
\toprule
\textbf{Benchmark}
& \textbf{Direction}
& \textbf{Dynamics}
& \textbf{Disturbance}
& \textbf{Selection}
\\
\midrule

\textbf{TruthfulQA}
& \textbf{200 + 200} false / true
& \textbf{35}
& \textbf{200} English prompts
& \textbf{50 or 200} disjoint English prompts, depending on model
\\

\textbf{HarmBench}
& \textbf{50 + 50} harmful / safe
& \textbf{50}
& \textbf{50} direct-validation behaviors
& Same \textbf{50} direct-validation behaviors
\\

\textbf{MGSM}
& \textbf{250} matched English--Spanish forms
& \textbf{50}
& Frozen GSM8K prompts; Bengali, German, Russian, and Thai translations
for Phi/Granite
& \textbf{50} prompts for Qwen; \textbf{100} translated prompts for Phi/Granite
\\

\textbf{L-CiteEval}
& \textbf{72 + 72} negative / positive AXBench responses
& \textbf{50}
& \textbf{200} short-context prompts
& \textbf{50} short prompts for Qwen; \textbf{40} matched 8K HotpotQA prompts
for Llama
\\

\bottomrule
\end{tabular}
\end{table*}

The controlled state is the input residual stream of each transformer block.
During generation, a forward hook adds the controller output to the final token
at the corresponding block output.  With KV caching disabled, the complete
prefix is recomputed at each decoding step and the controller is therefore
reevaluated on the current last-token state at every layer.  Original and
controlled methods use identical tokenization and decoding settings.

\subsection{Projected Dynamics and Disturbance Geometry}

For each dynamics prompt, we compute last-token Jacobian--vector products with
the prefix states fixed.  Prompt-wise Jacobians are averaged and then projected
between consecutive layer bases.  We never materialize or retain the full
hidden-dimension Jacobian collection after projection.  The reduced basis is
fitted independently at every layer.  Its first vector is the normalized
semantic contrast; the remaining vectors are the leading right singular
vectors of the calibration activations after removing their component along
that contrast.  We orthonormalize the resulting basis by QR decomposition and
use a reduced state dimension of eight in all reported experiments.  A-LQR and
$\mathrm{H}_{\infty}$ consume the same saved projected transition matrices.

We estimate the disturbance geometry from one-step residuals in these reduced
coordinates.  Residuals are centered separately at each layer, and
$D_kD_k^{\top}$ is set equal to their unbiased empirical covariance
($\mathrm{ddof}=1$).  We obtain $D_k$ from the singular values and right
singular vectors of the centered residual matrix.  Because the state dimension
is eight and every reported disturbance split contains more than nine records,
the retained rank is eight unless the empirical covariance is rank deficient;
no post-hoc explained-variance truncation or disturbance rescaling is applied.
The implementation verifies
$\lVert D_kD_k^{\top}-\widehat{\Sigma}_k\rVert_F/
\lVert\widehat{\Sigma}_k\rVert_F\leq10^{-4}$ before controller synthesis.

All calibration tensors used for synthesis are accumulated in float32.  The
reduced-coordinate construction uses a numerical norm floor of $10^{-4}$.  For
$\mathrm{H}_{\infty}$ synthesis, the attenuation search begins on
$[0,100]$, uses at most 100 bisection iterations with tolerance $10^{-5}$, and
deploys at a relative margin of $0.01$ above the smallest numerically feasible
attenuation.  Positive-definiteness checks use Cholesky factorizations, and the
Riccati matrices are symmetrized after each backward update, as described in
Appendix~\ref{app:numerical-details}.

\subsection{Controllers and Hyperparameter Selection}

Original denotes generation without an activation intervention.  S-PID uses a
fixed proportional--integral--derivative policy in the same semantic
coordinates.  A-LQR uses the fixed model--benchmark configuration inherited
from its source protocol; it is not reswept on our evaluation data.
$\mathrm{H}_{\infty}$ alone is selected through a benchmark-specific grid.
Except for the wider TruthfulQA grid, we evaluate the Cartesian product

\[
Q/R\in\{0.01,0.1,1,10\},
\qquad
Q_f/R\in\{0.01,0.1,10^{-1/2}\},
\qquad R=1.
\]

TruthfulQA uses
$Q/R\in\{0.01,10^{-1.5},0.1,10^{-0.5},1,10^{0.5},10,10^{1.5}\}$
and
$Q_f/R\in\{0.01,10^{-1.5},0.1,10^{-0.5}\}$, again with $R=1$.
The setpoint multiplier is frozen before this cost sweep: it follows the fixed
A-LQR model setting for TruthfulQA and is $1.5$ for HarmBench, MGSM, and
L-CiteEval.  We do not sweep the setpoint multiplier or attenuation level as an
additional model-selection axis.  Each cost candidate nevertheless receives
its own feasible $\gamma^\star$ from the same disturbance-fitted control
problem.

The selection objectives are those reported with the calibration surfaces in
Appendix~\ref{app:datasets}.  TruthfulQA uses the run-specific objective stated
in Appendix~\ref{app:truthfulqa-full}; HarmBench maximizes mean AXBench overall
steering, with validation ASR only as a tie-breaker; MGSM uses either AXBench
overall steering or the reported $0.8/0.2$ accuracy--steering composite,
depending on the frozen model run; and L-CiteEval uses the frozen objective
recorded with its calibration surface.  Exact ties are resolved
deterministically by the stored grid order.  A calibration whose selection
metric is identically zero fails rather than silently choosing a cost pair.

\subsection{Generation and Scoring}

Table~\ref{tab:experimental-generation} records the decoding configurations.
The same prompt identities, order, and decoding parameters are shared by all
methods within a model and condition.  Batch size is a systems parameter and
may differ across devices; it does not change prompt membership or decoding
semantics.

\begin{table*}[t]
\centering
\scriptsize
\caption{\textbf{Generation settings.} All controlled and uncontrolled
methods use the same setting within a benchmark.  The Llama-3.1-8B HarmBench
run uses the resource-reduced cap shown in parentheses.}
\label{tab:experimental-generation}
\setlength{\tabcolsep}{5pt}
\renewcommand{\arraystretch}{1.08}
\begin{tabularx}{0.96\textwidth}{p{2.2cm}p{2.2cm}p{2.1cm}p{2.0cm}X}
\toprule
\textbf{Benchmark} & \textbf{Decoding} & \textbf{Response cap} &
\textbf{Repetitions} & \textbf{Additional settings} \\
\midrule
TruthfulQA & Sampling & 50 tokens & 5; refreshed Llama H$_\infty$, GPT-2 XL,
and OLMo-2-32B: 1 &
$T=1.0$, top-$p=0.3$, repetition penalty $1.2$ \\
HarmBench & Greedy & 512 tokens (8B: 100) & 1 &
Direct and five jailbreak templates share behavior identities \\
MGSM & Greedy & 256 tokens & 1 &
Native eight-shot demonstrations for each input language \\
L-CiteEval & Greedy & 200 tokens & 1 &
Official one-shot prompt; matched 8K and 16K contexts \\
\bottomrule
\end{tabularx}
\end{table*}

Scoring is modular and operates only on completed generation files.  TruthfulQA
truth and informativeness use the pinned
\texttt{allenai/truthfulqa-truth-judge-llama2-7B} and
\texttt{allenai/truthfulqa-info-judge-llama2-7B} checkpoints.  HarmBench ASR
uses the official test classifier
\texttt{cais/HarmBench-Llama-2-13b-cls}.  MGSM accuracy is an exact comparison
between the parsed final Arabic numeral and the stored answer.  L-CiteEval
answer recall uses the released normalized token-overlap rule after citation
markers are removed, and citation precision, recall, and F1 use the released
AutoAIS procedure with the pinned
\texttt{tasksource/deberta-base-long-nli} checkpoint.  AXBench concept
relevance, instruction relevance, and fluency use the released $0$--$2$
rubrics with \texttt{gpt-4o-mini-2024-07-18}; their overall score is the
per-response harmonic mean and is zero when any component is zero.  API scoring
uses temperature zero, batched asynchronous requests, persistent per-response
caches, and bounded retries.  Deterministic scorers and model-based judges store
their raw outputs alongside the aggregate metric so that aggregation can be
reproduced without another generation pass.

\subsection{Statistical Procedures}

All comparisons are paired at the natural benchmark identity: TruthfulQA
question, HarmBench behavior, MGSM problem, or L-CiteEval question.  We first
compute per-response scores, aggregate within each frozen condition, and only
then form cross-condition summaries.  Language-level MGSM summaries are macro
averages over the five evaluation languages; HarmBench template summaries
retain the direct and five adversarial conditions separately.

Full-protocol TruthfulQA rows report the standard error across five independent
generation runs.  The refreshed Llama-3-8B $\mathrm{H}_{\infty}$ row and all
GPT-2 XL and OLMo-2-32B rows use a five-group delete-one-group jackknife over
question identities.  HarmBench, MGSM, and L-CiteEval use ten-group delete-one-group
jackknives over behavior, problem, and matched-question identities,
respectively.  For HarmBench, deleting one behavior removes its direct request
and all five adversarial renderings; for L-CiteEval, the same question grouping
is used at every context length.  The reported intervals therefore measure
sampling variability over benchmark identities.  They do not measure judge
uncertainty and, for deterministic decoding, do not measure decoding
variability.  No unreported significance test is used to determine boldface or
underlining in the result tables.

\subsection{Software, Hardware, and Computational Accounting}
Experiments were implemented in PyTorch and Hugging Face Transformers. Large
full-order Jacobians were averaged and projected after estimation; subsequent
controller synthesis used the saved projected dynamics, reduced bases, and
setpoints.

Table~\ref{tab:experimental-compute} reports allocated accelerator-hours,
defined as stage elapsed time multiplied by the number of accelerators reserved
for that stage. Artifact construction is counted once per model and shared by
the controlled methods. Calibration includes disturbance fitting, controller
synthesis over the complete candidate grid, calibration generation, scoring,
and selection. Generation includes every method, distribution-shift condition,
and repetition used in the reported tables. The accounting includes retained
paper-producing shards from resumed jobs and excludes exploratory runs whose
outputs were discarded; it is therefore a lower bound on total experimental
compute.

Across the reported benchmark suite, these runs account for at least
$229.9$ allocated accelerator-hours: $19.1$ for artifact construction,
$48.7$ for calibration, $159.8$ for final generation, and $2.3$ for scoring.
The campaign used five accelerator classes---RTX 5090, A100, H100, H200, and
RTX PRO 6000 Blackwell---with individual jobs scaling to eight accelerators and
a recorded peak allocation of 12 accelerators concurrently across machines.
The largest model--benchmark allocations were Qwen2.5-3B on L-CiteEval
($52.8$ accelerator-hours), Qwen2.5-14B on TruthfulQA ($34.0$), and
OLMo-2-32B on TruthfulQA ($32.0$).

\begin{table}[H]
\centering
\scriptsize
\caption{\textbf{Accelerator allocation for the reported experiments.}
Entries are allocated accelerator-hours, partitioned by pipeline stage. Total
includes scoring. Hardware lists every accelerator allocation contributing to
the reported model row. Values are lower bounds because discarded exploratory
runs are excluded.}
\label{tab:experimental-compute}
\setlength{\tabcolsep}{2.6pt}
\renewcommand{\arraystretch}{1.08}
\begin{tabular*}{\textwidth}{@{\extracolsep{\fill}}lllrrrrr@{}}
\toprule
\textbf{Benchmark} & \textbf{Model} & \textbf{Accelerator allocation(s)} &
\textbf{Artifacts} & \textbf{Calibration} & \textbf{Generation} &
\textbf{Scoring} & \textbf{Total} \\
\midrule
TruthfulQA & GPT-2 XL & $2\times$ H200 & 0.3 & 0.7 & 0.7 & 0.1 & 1.9 \\
& Llama-3-8B & $4\times$ H100; $8\times$ H200; $2\times$ RTX 5090 &
1.0 & 1.8 & 12.4 & 0.3 & 15.6 \\
& Qwen2.5-14B & $4\times$ H200; $2\times$ RTX 5090 &
2.5 & 1.4 & 29.7 & 0.5 & 34.0 \\
& OLMo-2-32B & $8\times$ H200 & 7.7 & 10.0 & 14.1 & 0.2 & 32.0 \\
\midrule
HarmBench & Llama-3.2-1B & $2\times$ RTX 5090 & 0.2 & 1.4 & 2.9 & 0.4 & 4.9 \\
& Llama-3.2-3B & $2\times$ RTX 5090 & 0.7 & 2.8 & 21.5 & 0.1 & 25.1 \\
& Llama-3.1-8B & $8\times$ RTX PRO 6000 & 2.5 & 7.6 & 9.8 & 0.3 & 20.2 \\
\midrule
MGSM & Qwen3-4B & $4\times$ H100 & 0.3 & $<0.1$ & 10.5 & $<0.1$ & 10.9 \\
& Phi-4-mini & $8\times$ H200 & 3.2 & 3.6 & 2.4 & $<0.1$ & 9.3 \\
& Granite-3.3-2B & $4\times$ RTX PRO 6000 & 0.2 & 7.7 & 6.8 & $<0.1$ & 14.8 \\
\midrule
L-CiteEval & Qwen2.5-3B & $4\times$ H100; $8\times$ H200; $1\times$ A100 &
0.3 & 8.0 & 44.3 & 0.1 & 52.8 \\
& Llama-3.2-1B & $4\times$ H200 & 0.2 & 3.7 & 4.6 & $<0.1$ & 8.5 \\
\midrule
\multicolumn{3}{r}{\textbf{Reported-suite total}} &
\textbf{19.1} & \textbf{48.7} & \textbf{159.8} & \textbf{2.3} &
\textbf{229.9} \\
\bottomrule
\end{tabular*}
\end{table}


\vspace{0.6em}

\noindent
\begin{minipage}[c]{0.22\textwidth}
    \centering
     \includegraphics[width=0.95\linewidth]{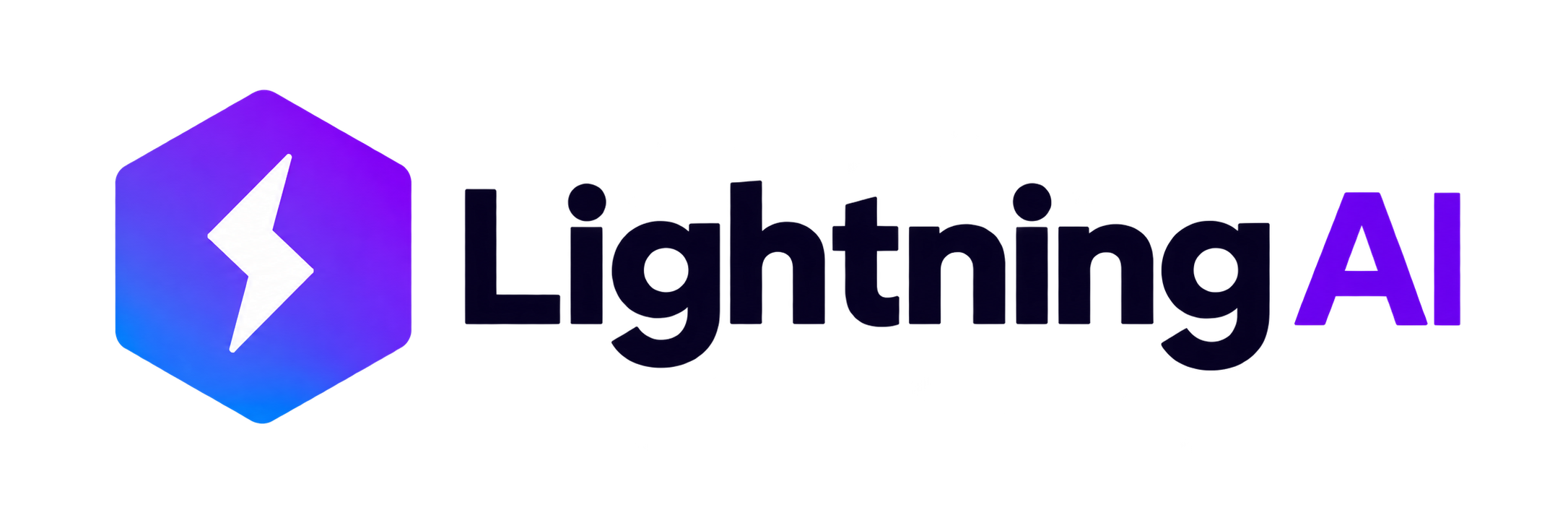}
\end{minipage}
\hfill
\begin{minipage}[R]{0.75\textwidth}
\small
\textbf{Computational support.}
We gratefully acknowledge Lightning AI for providing computational infrastructure and accelerator resources that supported the large-scale experiments reported in this work. Access to these resources enabled us to carry out repeated controller calibration, large-scale generation, and evaluation across multiple model families and accelerator classes at a computational scale that would otherwise have been difficult to achieve in a typical academic research setting.
\end{minipage}

\vspace{0.5em}

\clearpage
\newpage

\section{Robust Steerability and Held-Out Steering Reliability}
\label{app:srob-heldout-reliability}

Robust steerability is computed from calibration dynamics before held-out
generation. We test whether it tracks subsequent adversarial robustness across
the three reported Llama controllers on HarmBench, which fixes the task, target
behavior, calibration objective, and evaluation protocol.

\begin{table}[H]
\centering
\scriptsize
\caption{\textbf{Calibration-time robust steerability and HarmBench
safe-response rates.} Results are for the selected $\mathrm{H}_{\infty}$
controller. Behavioral entries are percentages computed as
$100-\mathrm{ASR}$; higher is better. Mean OOD averages the five frozen
jailbreak templates.}
\label{tab:srob-harmbench-reliability}
\setlength{\tabcolsep}{2.4pt}
\renewcommand{\arraystretch}{1.12}
\begin{tabular}{lcccccccc}
\toprule
\textbf{Model} &
$\boldsymbol{S}_{\mathbf{rob}}$ &
\textbf{Direct} &
\shortstack{\textbf{APM}\\\textbf{Programmer}} &
\shortstack{\textbf{DNE}\\\textbf{Nonresponse}} &
\shortstack{\textbf{Jailbreak}\\\textbf{Bot}} &
\shortstack{\textbf{John}\\\textbf{Persona}} &
\shortstack{\textbf{YOJA/Nona}\\\textbf{Roleplay}} &
\shortstack{\textbf{Mean}\\\textbf{OOD}} \\
\midrule
Llama-3.2-1B & 7.688 & 97.08 & 99.58 & 98.33 & 99.58 & 98.75 & 99.17 & 99.08 \\
Llama-3.2-3B & 2.405 & 91.67 & 97.50 & 96.25 & 96.25 & 95.83 & 95.83 & 96.33 \\
Llama-3.1-8B & 3.487 & 96.67 & 99.17 & 100.00 & 100.00 & 98.33 & 100.00 & 99.50 \\
\bottomrule
\end{tabular}
\end{table}

The relationship between $\Srob$ and safe-response rate has a positive slope
under each held-out jailbreak template. Llama-3.2-3B has the lowest $\Srob$
and the lowest safe-response rate under every attack. The two controllers with
higher $\Srob$ approach ceiling-level refusal, so their ordering varies across
templates. Within HarmBench, calibration-time robust steerability tracks which
model--behavior controller remains reliable under adversarial prompt shift.

The main experiments establish that robust feedback improves adversarial
robustness. Calibration-derived $\Srob$ therefore complements outcome-based
evaluation with a pre-deployment diagnostic of reliability under held-out
attack.

\clearpage
\newpage
\section{Computational Complexity}
\label{app:computational-complexity}

This appendix characterizes controller synthesis and deployment separately.
Synthesis is an offline calibration operation performed once for a selected
model--behavior configuration. Deployment measurements include the complete
autoregressive forward pass, activation hooks, coordinate transformations, and
controller updates. Measurements use Llama-3.2-1B-Instruct on the HarmBench
refusal task.

\begin{table}[H]
\centering
\scriptsize
\caption{\textbf{Controller synthesis and deployment cost.} Medians over 20
synthesis trials and seven generation trials.}
\label{tab:controller-computational-cost}
\setlength{\tabcolsep}{5pt}
\renewcommand{\arraystretch}{1.08}
\begin{tabular}{lccc}
\toprule
\textbf{Method} & \textbf{Offline synthesis} &
\textbf{Latency (ms/token)} & \textbf{Peak memory (GB)} \\
\midrule
A-LQR & $\mathcal{O}(Ld^3)$; 0.2015 s & 5.409 & 2.882 \\
$\mathrm{H}_{\infty}$ (ours) &
$\mathcal{O}(N_{\mathrm{iter}}Lr_x^3)$; 0.1245 s & 5.459 & 2.882 \\
\bottomrule
\end{tabular}
\end{table}

\subsection{Asymptotic Synthesis Complexity}
\label{app:complexity-asymptotic}

Let $L$ denote the number of controlled layers, $d$ the hidden dimension,
$r_x$ the reduced controller-state dimension, and $N_{\mathrm{iter}}$ the
number of attenuation-search iterations. A-LQR performs one finite-horizon
Riccati sweep with $d\times d$ matrices, giving synthesis time
$\mathcal{O}(Ld^3)$ and gain storage $\mathcal{O}(Ld^2)$. The measured
configuration uses $L=16$ and $d=2048$.

$\mathrm{H}_{\infty}$ synthesis operates in the reduced state space. Each
feasibility sweep costs $\mathcal{O}(Lr_x^3)$, and bisection gives total time
$\mathcal{O}(N_{\mathrm{iter}}Lr_x^3)$ with gain storage
$\mathcal{O}(Lr_x^2)$. The deployed configuration uses $r_x=8$.

The production implementations therefore differ in both recursion and state
dimension. To isolate the additional work introduced by robust feasibility
search, we also time one A-LQR Riccati sweep on the same $r_x=8$ reduced
problem used by $\mathrm{H}_{\infty}$. This matched-core measurement separates
the cost of repeated robust solves from the dimensional difference between the
deployed controllers.

\subsection{Empirical Synthesis Runtime}
\label{app:complexity-synthesis}

Offline synthesis was profiled on one NVIDIA GeForce RTX 5090. Garbage
collection and the CUDA allocator cache were cleared before each measurement,
CUDA was synchronized around the timed region, and controller order was
alternated across 20 trials. Synthesis used float32 tensors.

\begin{table}[H]
\centering
\scriptsize
\caption{\textbf{Offline controller-synthesis measurements.}
Statistics summarize 20 trials. The matched A-LQR row uses the same
eight-dimensional state as $\mathrm{H}_{\infty}$.}
\label{tab:controller-synthesis-detail}
\setlength{\tabcolsep}{5pt}
\renewcommand{\arraystretch}{1.08}
\begin{tabular}{lcccc}
\toprule
\textbf{Method} & \textbf{State dim.} & \textbf{Median (s)} &
\textbf{Mean $\pm$ SD (s)} & \textbf{Range (s)} \\
\midrule
A-LQR (deployed) & 2048 & 0.201472 & $0.200567\pm0.002489$ &
$[0.194700,\,0.203315]$ \\
A-LQR (matched reduced core) & 8 & 0.003073 &
$0.003085\pm0.000041$ & $[0.003034,\,0.003164]$ \\
$\mathrm{H}_{\infty}$ (deployed) & 8 & 0.124515 &
$0.124512\pm0.000214$ & $[0.124153,\,0.124880]$ \\
\bottomrule
\end{tabular}
\end{table}

The deployed $\mathrm{H}_{\infty}$ synthesis completed in a median of
$0.124515$ seconds. At the matched eight-dimensional state size, this was
$40.52\times$ the time required for one A-LQR Riccati sweep, reflecting the
repeated feasibility solves. The deployed full-state A-LQR implementation took
$0.201472$ seconds because its recursion operates at $d=2048$ rather than
$r_x=8$. Peak synthesis memory was $0.160$~GB for deployed A-LQR and below
$0.009$~GB for both reduced-state controllers.

The production attenuation search converged in 24 bisection iterations to
$\gamma^\star=0.13007522$. Its final infeasible--feasible bracket was
$[0.13006926,\,0.13007522]$, with width $5.96\times10^{-6}$. We assessed
sensitivity to the fitted disturbance geometry by drawing 100 seeded bootstrap
samples of size 50, with replacement, from the 50 prompt-level calibration
residual records. For each resample, the layer-wise disturbance covariance was
refitted and the unchanged synthesis procedure was rerun. All 100 fits
converged in 24 iterations. The resulting $\gamma^\star$ values had mean
$0.12935$ and standard deviation $0.00602$.

Each bootstrap resample refits the empirical disturbance geometry before
synthesis; the reported iteration count is therefore defined per fitted
control problem.

\subsection{Inference-Time Deployment}
\label{app:complexity-inference}

For one final-token state at one controlled layer, the deployed A-LQR policy
constructs a full-dimensional semantic tracking error and multiplies it by a
$d\times d$ gain, requiring $\mathcal{O}(d^2)$ controller operations. The
$\mathrm{H}_{\infty}$ policy projects the hidden state into $r_x$
coordinates, applies the reduced feedback and control-channel matrices, and
lifts the intervention back to the hidden dimension. Its controller cost is
$\mathcal{O}(dr_x+r_x^2)$ per layer. Across the controlled horizon, the
controller-only costs are $\mathcal{O}(Ld^2)$ for A-LQR and
$\mathcal{O}(L(dr_x+r_x^2))$ for $\mathrm{H}_{\infty}$.
Because these operations are embedded in the transformer forward pass, we also
measure end-to-end decoding latency and memory.

Deployment was measured on the same 16 frozen HarmBench direct prompts for all
methods. Each prompt generated exactly 50 new tokens, giving 800 generated
tokens per method and trial. Generation was greedy, batch size was eight, the
evaluated-model KV cache was disabled, and method order was rotated across
seven synchronized trials. The model used bfloat16 parameters with PyTorch
scaled dot-product attention. Peak allocated memory statistics were reset
before every trial.

\begin{table}[H]
\centering
\scriptsize
\caption{\textbf{End-to-end deployment cost.}
Latency is synchronized wall time per generated token over seven trials.
$\Delta$ is relative to Original.}
\label{tab:controller-inference-detail}
\setlength{\tabcolsep}{4pt}
\renewcommand{\arraystretch}{1.08}
\begin{tabular}{lcccc}
\toprule
\textbf{Method} & \textbf{Median latency} & \textbf{Mean $\pm$ SD} &
\textbf{$\Delta$ vs. Original} & \textbf{Peak / incremental GB} \\
& \textbf{(ms/token)} & \textbf{(ms/token)} & \textbf{(\%)} & \\
\midrule
Original & 5.347 & $5.346\pm0.005$ & 0.00 & 2.882 / 0.265 \\
A-LQR & 5.409 & $5.403\pm0.014$ & +1.16 & 2.882 / 0.265 \\
$\mathrm{H}_{\infty}$ (ours) & 5.459 & $5.454\pm0.013$ & +2.09 &
2.882 / 0.265 \\
\bottomrule
\end{tabular}
\end{table}

The median $\mathrm{H}_{\infty}$ latency was $0.050$ ms/token above A-LQR
and $2.09\%$ above Original. Median peak allocated memory was $2.882$ GB for
all methods, with $0.265$ GB allocated above the resident model. Measurements
used Python 3.11.16, PyTorch 2.9.1 with CUDA 12.8, Transformers 4.57.1, and
bfloat16 scaled dot-product attention on an NVIDIA GeForce RTX 5090.

\end{document}